\documentclass{article}

\def\NotArxiv#1{}
\def\ArxivOnly#1{#1}

\usepackage{booktabs}

\usepackage{tikz}  \usetikzlibrary{intersections}
\usetikzlibrary{decorations.pathreplacing}
\usetikzlibrary{calc}

\usepackage{xspace}

\usepackage{etoolbox}
\usepackage{stmaryrd}

\usepackage{subcaption}

\usepackage{amsmath}  \usepackage{amssymb}  \usepackage{amsthm}  
\usepackage{graphicx}
\graphicspath{{./Pictures/}}
\DeclareGraphicsExtensions{.pdf}

\usepackage{tabularx}
\usepackage{longtable}

\usepackage{hyperref}
\usepackage{varwidth}

\ArxivOnly{
  \usepackage{natbib}
  \usepackage{fullpage}
}

\theoremstyle{plain}
\newtheorem{theorem}{Theorem}
\newtheorem{lemma}[theorem]{Lemma}

\theoremstyle{definition}
\newtheorem{definition}[theorem]{Definition}
\newtheorem{example}[theorem]{Example}
\theoremstyle{remark}

\usepackage{h_vocabulary}
\usepackage{h_agc} \usepackage{h_color}

\usepackage{Add}

{
  \theoremstyle{definition}
  \newtheorem{Definition}[theorem]{Definition}
}

\OFF{
  \newtheorem{Theorem}{Theorem}

  {\theorembodyfont{\rmfamily}
    \newtheorem{Definition}[Theorem]{Definition}
  }

}

\newcommand{\RefFigure}[1]{Figure\,\ref{#1}}
\newcommand{\RefFig}[1]{Fig.\,\ref{#1}}

\newcommand{\RefLem}[1]{Lem.\,\ref{#1}}

\newcommand{\RefEq}[1]{(\ref{#1})}

\newcommand{\RefSection}[1]{Section\,\ref{#1}}
\newcommand{\RefSec}[1]{Sect.\,\ref{#1}}

\providecommand{\SetUnitlength}[1]{\setlength{\unitlength}{#1}\ifx\tikzpicture\undefined\relax\else\tikzset{x=#1}\tikzset{y=#1}\fi\ifx\PSTricksLoaded\undefined\relax\else\psset{unit=#1}\fi}

\newif\ifDebugPicture
\DebugPicturefalse
\newcommand{\CoorNode}[1]{ coordinate (#1) \ifDebugPicture node[opacity=0.5]  {\scriptsize #1} \fi }

\newcommand{\SetIntersect}[3]{\path [name intersections={of=#1 and #2,name=tmp}] ; \path (tmp-1) \CoorNode{#3} ;
}

\newcommand{\nameunder}[3]{\draw [decorate,decoration={brace,amplitude=5pt,mirror},xshift=0pt,yshift=-4pt]
  (#1-0.4,0.5) -- (#2+0.2,0.5) node [black,midway,below=.6em,] {\parbox{1.5cm}{\centering #3}};
}\newcommand{\nameright}[5]{\begin{scope}[shift={(.7,-.7)}]
    \draw [decorate,decoration={brace,amplitude=3pt,mirror}]
    (#1-0.5,#2-0.5) -- (#3+0.6,#4+0.6);
    \path ({(#1+#3)/2+.7},{(#2+#4)/2-.7}) -- node [black,pos=0,anchor=west,sloped] {{\parbox{2cm}{\raggedright #5}}} +(5,-5);
  \end{scope}}

\newcommand{\SafeZonePictureParameter}{\newcommand{\SpeedS}{\ensuremath{s}}\newcommand{\SpeedIval}{.6}\newcommand{\SpeedKval}{.1}\newcommand{\SpeedJval}{-.4}\newcommand{\SpeedSval}{4}\newcommand{\EpsilonVal}{.215}}

\newenvironment{DisplayRuleInP}{
  \begin{equation*}
  }{\end{equation*}
  }

\newenvironment{DisplayRuleNoI}{

\begin{equation*}
  }{\end{equation*}
}

\newenvironment{ParameterList}{\begingroup\small\footnotesize\begin{tabular}{r@{ = }>{$\displaystyle}l<{$}}\multicolumn{1}{r}{\bf Parameter\quad\mbox{}} & \multicolumn{1}{@{}l}{\bf\ Value} \\
    }{
  \end{tabular}\endgroup}

\providecommand{\IntegerInterval}[2]{\JDLvocabularyMathXspace{{\llbracket}#1,#2{\rrbracket}}}

\newcommand{\ForAllSpeedE}{\ensuremath{\forall \SpeedSubset{\subseteq}\IntegerInterval{1}{\SpeedSetNumber}}}
\newcommand{\ForAllSpeedI}{\ensuremath{\forall \SpeedIndexI{\in}\IntegerInterval{1}{\SpeedSetNumber}}}
\newcommand{\ForAllSpeedIE}{\ensuremath{\forall \SpeedIndexI{\in}\IntegerInterval{1}{\SpeedSetNumber}}, \ForAllSpeedE}
\newcommand{\ForAllSpeedIEFltI}{\raggedright\ensuremath{\ForAllSpeedI, \ForAllSpeedE,\SpeedSubsetTwo{\subseteq}\IntegerInterval{1}{i-1}, |\SpeedSubsetTwo| \neq 0}} 
\newcommand{\ForAllSpeedIL}{\ensuremath{\forall \SpeedIndexI,\SpeedIndexL{\in}\IntegerInterval{1}{\SpeedSetNumber}}}
\newcommand{\ForAllSpeedILE}{\ensuremath{\forall \SpeedIndexI,\SpeedIndexL{\in}\IntegerInterval{1}{\SpeedSetNumber}}, \ForAllSpeedE}
\newcommand{\ForAllSpeedIME}{\ensuremath{\forall \SpeedIndexI,\SpeedIndexM{\in}\IntegerInterval{1}{\SpeedSetNumber}}, \ForAllSpeedE}
\newcommand{\ForAllSpeedILinE}{\ensuremath{\forall \SpeedIndexI,\SpeedIndexL{\in}\IntegerInterval{1}{\SpeedSetNumber}}, \ForAllSpeedE, \ensuremath{\SpeedIndexL{\in}\SpeedSubset}}
\newcommand{\ForAllSpeedJltI}{\ensuremath{\forall \SpeedIndexI,\SpeedIndexJ{\in}\IntegerInterval{1}{\SpeedSetNumber},\,\SpeedIndexJ {<} \SpeedIndexI}}

\newcommand{\ForAllSpeedJteKltI}{\ensuremath{\forall \SpeedIndexI,\SpeedIndexJ,\SpeedIndexK{\in}\IntegerInterval{1}{\SpeedSetNumber},\,\SpeedIndexJ {\le} \SpeedIndexK {<} \SpeedIndexI}}

\newcommand{\ForAllSpeedJleKltI}{\ensuremath{\forall \SpeedIndexI,\SpeedIndexJ,\SpeedIndexK{\in}\IntegerInterval{1}{\SpeedSetNumber},\,\SpeedIndexJ {\leq} \SpeedIndexK {<} \SpeedIndexI}}
\newcommand{\ForAllSpeedJleKltIL}{\ensuremath{\forall \SpeedIndexI,\SpeedIndexJ,\SpeedIndexK,\SpeedIndexL{\in}\IntegerInterval{1}{\SpeedSetNumber},\,\SpeedIndexJ {\leq} \SpeedIndexK {<} \SpeedIndexI}}

\newcommand{\ForAllSpeedJltLltI}{\ensuremath{\forall \SpeedIndexI,\SpeedIndexJ,\SpeedIndexL{\in}\IntegerInterval{1}{\SpeedSetNumber},\,\SpeedIndexJ {<} \SpeedIndexL {<} \SpeedIndexI}}
\newcommand{\ForAllSpeedJltLltKltI}{\ensuremath{\forall \SpeedIndexI,\SpeedIndexJ,\SpeedIndexK,\SpeedIndexL{\in}\IntegerInterval{1}{\SpeedSetNumber},\,\SpeedIndexJ {<} \SpeedIndexL {<} \SpeedIndexK {<} \SpeedIndexI}}
\newcommand{\ForAllSpeedJleLltKltIMneL}{\ensuremath{\forall \SpeedIndexI,\SpeedIndexJ,\SpeedIndexK,\SpeedIndexL,\SpeedIndexM{\in}\IntegerInterval{1}{\SpeedSetNumber},\,\SpeedIndexJ {\leq} \SpeedIndexL {<} \SpeedIndexK {<} \SpeedIndexI,\SpeedIndexL\neq\SpeedIndexM}}
\newcommand{\ForAllSpeedKltI}{\ensuremath{\forall \SpeedIndexI,\SpeedIndexK{\in}\IntegerInterval{1}{\SpeedSetNumber},\, \SpeedIndexK {<} \SpeedIndexI}}

\newcommand{\ForAllSpeedKltILleK}{\ensuremath{\forall \SpeedIndexI,\SpeedIndexK,\SpeedIndexL{\in}\IntegerInterval{1}{\SpeedSetNumber},\, \SpeedIndexK {<} \SpeedIndexI,\, \SpeedIndexK {\leq} \SpeedIndexL}}

\newcommand{\ForAllSpeedLltJleKltI}{\ensuremath{\forall \SpeedIndexI,\SpeedIndexJ, \SpeedIndexK,\SpeedIndexL{\in}\IntegerInterval{1}{\SpeedSetNumber},\, \SpeedIndexL {<} \SpeedIndexJ {\le} \SpeedIndexK {\leq} \SpeedIndexI}}

\newcommand{\MakeCaseMSCR}[2]{\expandafter\newcommand\csname MSForAllSpeed#1\endcsname[2][\BaseSpeed{\SpeedIndexI}]{\ensuremath{#2},&##2&##1\\}
  \expandafter\newcommand\csname CRForAllSpeed#1\endcsname[2]{\ensuremath{#2},&\AGCruleDefArray{##1}{##2}\\}
}

\MakeCaseMSCR{JltKltI}{\forall \SpeedIndexI,\SpeedIndexJ, \SpeedIndexK{\in}\IntegerInterval{1}{\SpeedSetNumber},\, \SpeedIndexJ {<} \SpeedIndexK {<} \SpeedIndexI}

\newcommand{\MS}[2][\BaseSpeed{\SpeedIndexI}]{&#2&#1\\}
\newcommand{\MSForAllSpeedE}[2][\BaseSpeed{\SpeedIndexI}]{\ForAllSpeedE,&#2&#1\\}
\newcommand{\MSForAllSpeedI}[2][\BaseSpeed{\SpeedIndexI}]{\ForAllSpeedI,&#2&#1\\}
\newcommand{\MSForAllSpeedIE}[2][\BaseSpeed{\SpeedIndexI}]{\ForAllSpeedIE,&#2&#1\\}
\newcommand{\MSForAllSpeedIL}[2][\BaseSpeed{\SpeedIndexI}]{\ForAllSpeedIL,&#2&#1\\}
\newcommand{\MSForAllSpeedJltI}[2][\SpeedRapid]{\ForAllSpeedJltI,&#2&#1\\}

\newcommand{\MSForAllSpeedJteKltI}[2][\SpeedRapid]{\ForAllSpeedJteKltI,&#2&#1\\}
\newcommand{\MSForAllSpeedKltI}[2][\SpeedRapid]{\ForAllSpeedKltI,&#2&#1\\}

\newenvironment{MSlist}{\begingroup\small\footnotesize\begin{tabular}{@{}r@{\,}c@{\,}>{\begin{math}}c<{\end{math}}@{}}
    &\,\bf Meta-signal\,& \text{\bf speed\!\!} \\
  }{
  \end{tabular}\endgroup}

\newcommand{\CRForAllSpeedI}[2]{\ForAllSpeedI,&\AGCruleDefArray{#1}{#2}\\}
\newcommand{\CRForAllSpeedIE}[3][]{\ForAllSpeedIE,&\AGCruleDefArray{#2}{#3}#1\\}
\newcommand{\CRForAllSpeedIEFltI}[2]{\ForAllSpeedIEFltI,&\AGCruleDefArray{#1}{#2}\\} 

\newcommand{\CRForAllSpeedIL}[2]{\ForAllSpeedIL,&\AGCruleDefArray{#1}{#2}\\}
\newcommand{\CRForAllSpeedILE}[3][]{\ForAllSpeedILE,&\AGCruleDefArray{#2}{#3}#1\\}
\newcommand{\CRForAllSpeedILinE}[2]{\ForAllSpeedILinE,&\AGCruleDefArray{#1}{#2}\\}
\newcommand{\CRForAllSpeedIME}[3][]{\ForAllSpeedIME,&\AGCruleDefArray{#2}{#3}#1\\}
\newcommand{\CRForAllSpeedJltI}[2]{\ForAllSpeedJltI,&\AGCruleDefArray{#1}{#2}\\}

\newcommand{\CRForAllSpeedJleKltI}[2]{\ForAllSpeedJleKltI,&\AGCruleDefArray{#1}{#2}\\}
\newcommand{\CRForAllSpeedJleKltIL}[2]{\ForAllSpeedJleKltIL,&\AGCruleDefArray{#1}{#2}\\}
\newcommand{\CRForAllSpeedJltLltKltI}[2]{\ForAllSpeedJltLltKltI,&\AGCruleDefArray{#1}{#2}\\}
\newcommand{\CRForAllSpeedJleLltKltIMneL}[2]{\ForAllSpeedJleLltKltIMneL,&\AGCruleDefArray{#1}{#2}\\}
\newcommand{\CRForAllSpeedJltLltI}[2]{\ForAllSpeedJltLltI,&\AGCruleDefArray{#1}{#2}\\}
\newcommand{\CRForAllSpeedKltI}[2]{\ForAllSpeedKltI,&\AGCruleDefArray{#1}{#2}\\}

\newcommand{\CRForAllSpeedKltILleK}[2]{\ForAllSpeedKltILleK,&\AGCruleDefArray{#1}{#2}\\}
\newcommand{\CRForAllSpeedLltJleKltI}[2]{\ForAllSpeedLltJleKltI,&\AGCruleDefArray{#1}{#2}\\}

\newcommand{\ShiftR}[2]{\CRForAllSpeedI{\SigShrinkBottomBothRi,#1}{#2,\SigShrinkBottomBothRi} \CRForAllSpeedI{\SigShrinkBottomRi,#1}{#2,\SigShrinkBottomRi} \CRForAllSpeedI{#2,\SigShrinkBackRi}{\SigShrinkBackRi,#1} }

\renewcommand{\AGCruleDefArray}[2]{{#1}&{#2}}

\newenvironment{CRlist}{\begingroup\small\footnotesize\scriptsize\begin{tabular}{@{}r@{\,}>{\raggedleft\arraybackslash\{}r<{\,\}}@{\,$\to$\,\{\,}>{\raggedright\arraybackslash}l<{\}}}
    }{
  \end{tabular}\endgroup}

\newcounter{Example}
\newcommand{\PicExample}[2]{\begin{figure}[hbt]
    \centering
    \mbox{}\hfill\includegraphics[scale=#2]{test_#1}\hfill\hfill\includegraphics[scale=#2]{test_#1_sim}\hfill\hfill\mbox{}\stepcounter{Example}
    \caption{Example \theExample.}
    \label{fig:annexe:example:#1}
  \end{figure}
}

\providecommand{\mathxspace}[1]{\ensuremath{#1}\xspace}

\providecommand{\mathCalxspace}[1]{\mathxspace{\mathcal{#1}}}

\JDLvocabulary{\NaturalSet}{\JDLvocabularyMathBBXspace{N}}{}{Set of all natural integers}
\JDLvocabulary{\IntegerSet}{\JDLvocabularyMathBBXspace{Z}}{}{Set of all integers}
\JDLvocabulary{\RealSet}{\JDLvocabularyMathBBXspace{R}}{}{Set of all real numbers}

\newcommand{\SpeedA}{\JDLvocabularyMathXspace{\alpha}}
\newcommand{\SpeedB}{\JDLvocabularyMathXspace{\beta}}
\newcommand{\SpeedSS}{\JDLvocabularyMathXspace{s}}
\newcommand{\TZero}{\JDLvocabularyMathXspace{-\frac{\SpeedSS-\SpeedA}{\SpeedSS-\SpeedB}}}
\newcommand{\XZero}{\JDLvocabularyMathXspace{-\SpeedB\frac{\SpeedSS-\SpeedA}{\SpeedSS-\SpeedB}}}

\JDLvocabulary{\SymbUndefined}{\JDLvocabularyMathXspace{\bot}}{}{Symbol used to complete the definition of the bi-infinite word associated to a configuration and a position}

\JDLvocabulary{\SpeedSet}{\mathCalxspace{S}}{}{(finite) Set of speeds}
\JDLvocabulary{\SpeedSetNumber}{\JDLvocabularyMathXspace{\mathfrak{n}}}{}{Number of speeds in \string\SpeedSet}
\JDLvocabulary{\USpeedSet}{\mathCalxspace{S_U}}{}{(finite) Set of speeds used for simulating all machines with a given speed set}
\JDLvocabulary{\UMetaSignalSet}{\JDLvocabularyMathXspace{M_U}}{}{(finite) Set of meta-signals used for simulating all machines with a given speed set}
\JDLvocabulary{\URuleSet}{\JDLvocabularyMathXspace{M_U}}{}{(finite) Set of collision-rules used for simulating all machines with a given speed set}
\JDLvocabulary{\UniversalMSSpeed}{\mathCalxspace{U_S}}{}{Signal machine capable of simulating all signal machines using only speed in \string\SpeedSet}

\JDLvocabulary{\UniversalMSSpeedCheckedColl}{\JDLvocabularyMathXspace{\UniversalMSSpeed^{\operatorname{checked}}}}{}{Submachine of \UniversalMSSpeed for dealing with checked configurations}
\JDLvocabulary{\OtherUniversalMSSpeed}{\JDLvocabularyMathXspace{\UniversalMSSpeed'}}{}{Some variant of \UniversalMSSpeed}
\JDLvocabulary{\SpeedIndexI}{\JDLvocabularyMathXspace{i}}{}{Index for speeds}
\JDLvocabulary{\SpeedIndexJ}{\JDLvocabularyMathXspace{j}}{}{Index for speeds}
\JDLvocabulary{\SpeedIndexK}{\JDLvocabularyMathXspace{k}}{}{Index for speeds}
\JDLvocabulary{\SpeedIndexL}{\JDLvocabularyMathXspace{l}}{}{Index for speeds}
\JDLvocabulary{\SpeedIndexM}{\JDLvocabularyMathXspace{m}}{}{Index for speeds}
\JDLvocabulary{\SpeedIndexSome}{\JDLvocabularyMathXspace{-}}{}{Unspecified index for speeds}

\newcommand{\BaseSpeed}[1]{\JDLvocabularyMathXspace{\AGCspeed_{#1}}}
\newcommand{\SpeedI}{\BaseSpeed{\SpeedIndexI}}\newcommand{\SpeedK}{\BaseSpeed{\SpeedIndexK}}\newcommand{\SpeedL}{\BaseSpeed{\SpeedIndexL}}
\JDLvocabulary{\SpaceCoordinateX}{\JDLvocabularyMathXspace{x}}{}{Some spatial coordinate}
\JDLvocabulary{\Duration}{\JDLvocabularyMathXspace{\Delta t}}{}{Some duration}

\JDLvocabulary{\SpeedSubset}{\JDLvocabularyMathXspace{E}}{}{Some subset of {1}..{\SpeedSetNumber}}
\JDLvocabulary{\SpeedSubsetTwo}{\JDLvocabularyMathXspace{F}}{}{Some (other) subset of {1}..{\SpeedSetNumber}}

\JDLvocabulary{\SafeTop}{\JDLvocabularyTextXspace{Z$_T$}}{}{Point on top of the safety zone}
\JDLvocabulary{\SafeLeft}{\JDLvocabularyTextXspace{Z$_L$}}{}{Point on left of the safety zone}
\JDLvocabulary{\SafeRight}{\JDLvocabularyTextXspace{Z$_R$}}{}{Point on right of the safety zone}
\JDLvocabulary{\SafeBot}{\JDLvocabularyTextXspace{Z$_B$}}{}{Point at the bottom of the safety zone}
\JDLvocabulary{\SafeCm}{\JDLvocabularyTextXspace{C$_{\SpeedIndexM}$}}{}{Point of intersection of macro-collision on output}
\JDLvocabulary{\TestEpsilon}{\JDLvocabularyMathXspace{\varepsilon}}{}{Parameter for ensuring a large enough safety zone}
\JDLvocabulary{\CoefCheck}{\JDLvocabularyMathXspace{\tau_{\text{check}}}}{}{[check] Maximal relative height between middle and  encounter from  start}
\JDLvocabulary{\CoefTest}{\JDLvocabularyMathXspace{\tau_{U}}}{}{[Test] To position $U$, must be in $(2/3,1)$}

\newcommand{\GenericMSSpeedNbr}[2]{\JDLvocabularyMathXspace{{_{#1}{\mu}^{#2}}}}
\JDLvocabulary{\MetaSignalIndexK}{\JDLvocabularyMathXspace{\sigma}}{}{Index for meta-signal of \AGCmachine}
\JDLvocabulary{\GenericMSSpeedNbrik}{\GenericMSSpeedNbr{\SpeedIndexI}{\MetaSignalIndexK}}{}{Meta-signal \MetaSignalIndexK of speed \JDLvocabularyMathXspace{\BaseSpeed{\SpeedIndexI}} of \AGCmachine}

\tikzAGCmakeSignal{MOne}{$\AGCmetaSignal_1$}{Black,thick,dashed}\tikzAGCmakeSignal{MTwo}{$\AGCmetaSignal_2$}{DarkBlue,,thick,loosely dotted}\tikzAGCmakeSignal{MThree}{$\AGCmetaSignal_3$}{Red,solid,thick}\tikzAGCmakeSignal{MFour}{$\AGCmetaSignal_4$}{DarkGreen,thick,densely dotted}

\newcommand{\IdBaseSpeedOtherSpeedParamParam}[5]{\JDLvocabularyMathXspace{{_{\text{\AGCsigPolice{#2}}}^{\text{\AGCsigPolice{#3}}}}\AGCsigPolice{#1}_{\text{\AGCsigPolice{#4}}}^{\text{\AGCsigPolice{#5}}}}}
\newcommand{\IdBaseSpeed}[2]{\IdBaseSpeedOtherSpeedParamParam{#1}{#2}{}{}{}}

\tikzstyle{StyleFail}=[Red,densely dashed]

\JDLvocabulary{\SymbolBorderLeft}{\AGCsigPolice{border-left}}{}{To encode }
\newcommand{\IdBorderLeft}[1][\SpeedIndexI]{\IdBaseSpeed{\SymbolBorderLeft}{#1}}
\tikzstyle{StyleBorder}=[DarkRed,solid,thick]
\tikzAGCmakeSignal{BorderLeft}{\IdBorderLeft[\SpeedIndexSome]}{StyleBorder}
\tikzAGCmakeSignal{BorderLefti}{\IdBorderLeft}{StyleBorder}
\tikzAGCmakeSignal{BorderLeftj}{\IdBorderLeft[\SpeedIndexJ]}{StyleBorder}
\tikzAGCmakeSignal{BorderLeftk}{\IdBorderLeft[\SpeedIndexK]}{StyleBorder}
\tikzAGCmakeSignal{BorderLeftl}{\IdBorderLeft[\SpeedIndexL]}{StyleBorder}
\tikzAGCmakeSignal{BorderLeftm}{\IdBorderLeft[\SpeedIndexM]}{StyleBorder}
\tikzAGCmakeSignal{BorderLeftmpo}{\IdBorderLeft[\SpeedIndexM{+}1]}{StyleBorder}
\tikzAGCmakeSignal{BorderLeftOne}{\IdBorderLeft[1]}{StyleBorder}
\tikzAGCmakeSignal{BorderLeftTwo}{\IdBorderLeft[2]}{StyleBorder}
\tikzAGCmakeSignal{BorderLeftMax}{\IdBorderLeft[\SpeedSetNumber]}{StyleBorder}
\JDLvocabulary{\SymbolBorderRight}{\AGCsigPolice{border-right}}{}{To encode }
\newcommand{\IdBorderRight}[1][\SpeedIndexI]{\IdBaseSpeed{\SymbolBorderRight}{#1}}
\tikzAGCmakeSignal{BorderRight}{\IdBorderRight{-}}{StyleBorder}
\tikzAGCmakeSignal{BorderRighti}{\IdBorderRight}{StyleBorder}
\tikzAGCmakeSignal{BorderRightj}{\IdBorderRight[\SpeedIndexJ]}{StyleBorder}
\tikzAGCmakeSignal{BorderRightk}{\IdBorderRight[\SpeedIndexK]}{StyleBorder}
\tikzAGCmakeSignal{BorderRightl}{\IdBorderRight[\SpeedIndexL]}{StyleBorder}
\tikzAGCmakeSignal{BorderRightm}{\IdBorderRight[\SpeedIndexM]}{StyleBorder}
\tikzAGCmakeSignal{BorderRightmpo}{\IdBorderRight[\SpeedIndexM{+}1]}{StyleBorder}
\tikzAGCmakeSignal{BorderRightOne}{\IdBorderRight[1]}{StyleBorder}
\tikzAGCmakeSignal{BorderRightMax}{\IdBorderRight[\SpeedSetNumber]}{StyleBorder}
\newcommand{\SymbolId}{\AGCsigPolice{id}}
\newcommand{\IdId}[1][]{\IdBaseSpeedOtherSpeedParamParam{\SymbolId}{#1}{}{}{}}
\tikzstyle{StyleId}=[DarkPurple,solid]
\tikzAGCmakeSignal{ID}{\IdId[\_]}{StyleId}
\tikzAGCmakeSignal{IDi}{\IdId[\SpeedIndexI]}{StyleId}
\tikzAGCmakeSignal{IDk}{\IdId[\SpeedIndexK]}{StyleId}
\tikzAGCmakeSignal{IDl}{\IdId[\SpeedIndexL]}{StyleId}
\tikzAGCmakeSignal{IDTT}{\IdId[2]{2}}{Black,dashed,thick}\JDLvocabulary{\SymbolMain}{\IdBaseSpeed{\AGCsigPolice{main}}{}}{}{Base for signal places exactly where simulated ones are}
\newcommand{\IdMain}[2][\ensuremath{\emptyset}]{\IdBaseSpeedOtherSpeedParamParam{\SymbolMain}{#2}{}{}{#1}}
\tikzstyle{StyleMain}=[Black,solid,ultra thick]
\tikzAGCmakeSignal{Main}{\IdMain{\SpeedIndexSome}}{StyleMain}
\tikzAGCmakeSignal{MainSome}{\IdMain[\ensuremath{-}]{\SpeedIndexSome}}{StyleMain}
\tikzAGCmakeSignal{Maini}{\IdMain{\SpeedIndexI}}{StyleMain}
\tikzAGCmakeSignal{MainiE}{\IdMain[\SpeedSubset]{\SpeedIndexI}}{StyleMain}
\tikzAGCmakeSignal{Mainj}{\IdMain{\SpeedIndexJ}}{StyleMain}
\tikzAGCmakeSignal{Maink}{\IdMain{\SpeedIndexK}}{StyleMain}
\tikzAGCmakeSignal{Mainl}{\IdMain{\SpeedIndexL}}{StyleMain}
\tikzAGCmakeSignal{Mainm}{\IdMain{\SpeedIndexM}}{StyleMain}
\tikzAGCmakeSignal{Mainmpo}{\IdMain{\SpeedIndexM{+}1}}{StyleMain}
\tikzAGCmakeSignal{MainOne}{\IdMain{1}}{StyleMain}
\tikzAGCmakeSignal{MainTwo}{\IdMain{2}}{StyleMain}
\tikzAGCmakeSignal{MainMax}{\IdMain{\SpeedSetNumber}}{StyleMain}
\newcommand{\SymbolRuleBound}{\AGCsigPolice{rule-bound}}
\newcommand{\IdRuleBound}[1][\SpeedIndexI]{\IdBaseSpeed{\SymbolRuleBound}{#1}}
\tikzstyle{StyleRuleBound}=[DarkBlue,solid,ultra thick]
\tikzAGCmakeSignal{RuleBoundi}{\IdRuleBound}{StyleRuleBound}
\tikzAGCmakeSignal{RuleBoundk}{\IdRuleBound[\SpeedIndexK]}{StyleRuleBound}
\tikzAGCmakeSignal{RuleBoundl}{\IdRuleBound[\SpeedIndexL]}{StyleRuleBound}
\newcommand{\SymbolRuleMiddle}{\AGCsigPolice{rule-middle}}
\newcommand{\IdRuleMiddle}[1][\SpeedIndexI]{\IdBaseSpeed{\SymbolRuleMiddle}{#1}}
\tikzstyle{StyleRuleMiddle}=[StyleRuleBound]
\tikzAGCmakeSignal{RuleMiddlei}{\IdRuleMiddle}{StyleRuleMiddle}
\tikzAGCmakeSignal{RuleMiddlek}{\IdRuleMiddle[\SpeedIndexK]}{StyleRuleMiddle}
\tikzAGCmakeSignal{RuleMiddlel}{\IdRuleMiddle[\SpeedIndexL]}{StyleRuleMiddle}
\newcommand{\SymbolIf}{\AGCsigPolice{if}}
\newcommand{\IdIf}[2][\SpeedIndexI]{\IdBaseSpeedOtherSpeedParamParam{\SymbolIf}{#1}{}{#2}{}}
\tikzstyle{StyleIf}=[DarkRed,solid,thick]
\tikzAGCmakeSignal{Ifi}{\IdIf[]{\SpeedIndexI}}{StyleIf}
\tikzAGCmakeSignal{Ifik}{\IdIf{\SpeedIndexK}}{StyleIf}
\tikzAGCmakeSignal{Ifil}{\IdIf{\SpeedIndexL}}{StyleIf}
\tikzAGCmakeSignal{Ifim}{\IdIf{\SpeedIndexM}}{StyleIf}
\tikzAGCmakeSignal{IfiOne}{\IdIf{1}}{DarkGreen,solid,thick}
\tikzAGCmakeSignal{IfiThree}{\IdIf{3}}{DarkGreen,solid,thick}
\tikzAGCmakeSignal{IfiFour}{\IdIf{4}}{Red,solid,thick}
\tikzAGCmakeSignal{IfiFive}{\IdIf{5}}{Blue,solid,thick}
\tikzAGCmakeSignal{IfiSix}{\IdIf{6}}{DarkGreen,solid,thick}
\tikzAGCmakeSignal{IfiSeven}{\IdIf{7}}{DarkOrange,solid,thick}
\tikzAGCmakeSignal{IfiEight}{\IdIf{8}}{Blue,solid,thick}
\tikzAGCmakeSignal{IfiMax}{\IdIf{\SpeedSetNumber}}{Blue,solid,thick}
\tikzAGCmakeSignal{Ifkl}{\IdIf[\SpeedIndexK]{\SpeedIndexL}}{StyleIf}
\tikzAGCmakeSignal{Iflm}{\IdIf[\SpeedIndexL]{\SpeedIndexM}}{StyleIf}
\newcommand{\SymbolThen}{\AGCsigPolice{then}}
\newcommand{\IdThen}[2][\SpeedIndexI]{\IdBaseSpeedOtherSpeedParamParam{\SymbolThen}{#1}{}{#2}{}}
\tikzstyle{StyleOne}=[DarkOrange,solid]
\tikzstyle{StyleTwo}=[DarkGreen,solid]
\tikzstyle{StyleThen}=[DarkBrown]
\tikzAGCmakeSignal{TheniOne}{{\IdThen{1}}}{StyleOne}
\tikzAGCmakeSignal{TheniTwo}{{\IdThen{2}}}{StyleTwo}
\tikzAGCmakeSignal{TheniFour}{{\IdThen{4}}}{Red,solid,thick}
\tikzAGCmakeSignal{TheniMax}{{\IdThen{\SpeedSetNumber}}}{StyleOne}
\tikzAGCmakeSignal{Thenik}{{\IdThen{\SpeedIndexK}}}{StyleOne}
\tikzAGCmakeSignal{Thenil}{{\IdThen{\SpeedIndexL}}}{StyleOne}
\tikzAGCmakeSignal{Thenim}{{\IdThen[\SpeedIndexL]{\SpeedIndexM}}}{StyleOne}
\tikzAGCmakeSignal{Thenkl}{{\IdThen[\SpeedIndexK]{\SpeedIndexL}}}{StyleOne}
\tikzAGCmakeSignal{Thenlm}{{\IdThen[\SpeedIndexL]{\SpeedIndexM}}}{StyleOne}

\JDLvocabulary{\SymbolShrinkBottom}{\AGCsigPolice{shrink-bottom}}{}{To encode the bottom part of shrinking}
\newcommand{\IdShrinkBottom}[2]{\IdBaseSpeedOtherSpeedParamParam{\SymbolShrinkBottom}{}{#1}{#2}{}}
\tikzstyle{StyleShrinkBottom}=[Black,thick,densely dotted]
\tikzAGCmakeSignal{ShrinkBottomRi}{{\IdShrinkBottom{\SpeedIndexI}{R}}}{StyleShrinkBottom}
\tikzAGCmakeSignal{ShrinkBottomLi}{{\IdShrinkBottom{\SpeedIndexI}{L}}}{StyleShrinkBottom}
\tikzAGCmakeSignal{ShrinkBottomRk}{{\IdShrinkBottom{\SpeedIndexK}{R}}}{StyleShrinkBottom}
\tikzAGCmakeSignal{ShrinkBottomLk}{{\IdShrinkBottom{\SpeedIndexK}{L}}}{StyleShrinkBottom}
\newcommand{\IdShrinkBottomBoth}[2]{\IdBaseSpeedOtherSpeedParamParam{\SymbolShrinkBottom}{}{#1}{#2}{both}}
\tikzAGCmakeSignal{ShrinkBottomBothLi}{{\IdShrinkBottomBoth{\SpeedIndexI}{L}}}{StyleShrinkBottom}
\tikzAGCmakeSignal{ShrinkBottomBothLj}{{\IdShrinkBottomBoth{\SpeedIndexJ}{L}}}{StyleShrinkBottom}
\tikzAGCmakeSignal{ShrinkBottomBothLk}{{\IdShrinkBottomBoth{\SpeedIndexK}{L}}}{StyleShrinkBottom}
\tikzAGCmakeSignal{ShrinkBottomBothLl}{{\IdShrinkBottomBoth{\SpeedIndexL}{L}}}{StyleShrinkBottom}
\tikzAGCmakeSignal{ShrinkBottomBothRi}{{\IdShrinkBottomBoth{\SpeedIndexI}{R}}}{StyleShrinkBottom}
\JDLvocabulary{\SymbolShrinkTop}{\AGCsigPolice{shrink-top}}{}{To encode the bottom part of shrinking}
\newcommand{\IdShrinkTop}[2]{\IdBaseSpeedOtherSpeedParamParam{\SymbolShrinkTop}{}{#1}{#2}{}}
\tikzstyle{StyleShrinkTop}=[StyleBorder]
\tikzAGCmakeSignal{ShrinkTopLi}{{\IdShrinkTop{\SpeedIndexI}{L}}}{StyleShrinkTop}
\tikzAGCmakeSignal{ShrinkTopLj}{{\IdShrinkTop{\SpeedIndexJ}{L}}}{StyleShrinkTop}
\tikzAGCmakeSignal{ShrinkTopLk}{{\IdShrinkTop{\SpeedIndexK}{L}}}{StyleShrinkTop}
\tikzAGCmakeSignal{ShrinkTopLl}{{\IdShrinkTop{\SpeedIndexL}{L}}}{StyleShrinkTop}
\tikzAGCmakeSignal{ShrinkTopRi}{{\IdShrinkTop{\SpeedIndexI}{R}}}{StyleShrinkTop}
\tikzAGCmakeSignal{ShrinkTopRk}{{\IdShrinkTop{\SpeedIndexK}{R}}}{StyleShrinkTop}
\tikzAGCmakeSignal{ShrinkTopRl}{{\IdShrinkTop{\SpeedIndexL}{R}}}{StyleShrinkTop}
\newcommand{\IdShrinkTopTest}[2]{\IdBaseSpeedOtherSpeedParamParam{\SymbolShrinkTop}{}{#1}{#2}{test}}
\tikzAGCmakeSignal{ShrinkTopTestLk}{{\IdShrinkTopTest{\SpeedIndexK}{L}}}{StyleShrinkTop}
\tikzAGCmakeSignal{ShrinkTopTestLi}{{\IdShrinkTopTest{\SpeedIndexI}{L}}}{StyleShrinkTop}
\tikzAGCmakeSignal{ShrinkTopTestRi}{{\IdShrinkTopTest{\SpeedIndexI}{R}}}{StyleShrinkTop}
\JDLvocabulary{\SymbolShrinkBack}{\AGCsigPolice{shrink-back}}{}{To encode the bottom part of shrinking}
\newcommand{\IdShrinkBack}[2]{\IdBaseSpeedOtherSpeedParamParam{\SymbolShrinkBack}{}{#1}{#2}{}}
\tikzstyle{StyleShrinkBack}=[StyleShrinkBottom]
\tikzAGCmakeSignal{ShrinkBackLi}{{\IdShrinkBack{\SpeedIndexI}{L}}}{StyleShrinkBack}
\tikzAGCmakeSignal{ShrinkBackRi}{{\IdShrinkBack{\SpeedIndexI}{R}}}{StyleShrinkBack}
\tikzAGCmakeSignal{ShrinkBackLk}{{\IdShrinkBack{\SpeedIndexK}{L}}}{StyleShrinkBack}
\tikzAGCmakeSignal{ShrinkBackRk}{{\IdShrinkBack{\SpeedIndexK}{R}}}{StyleShrinkBack}
\JDLvocabulary{\SymbolShrinkTest}{\AGCsigPolice{shrink-test}}{}{To encode the bottom part of shrinking}
\newcommand{\IdShrinkTest}[2]{\IdBaseSpeedOtherSpeedParamParam{\SymbolShrinkTest}{}{#1}{}{#2}}
\tikzstyle{StyleShrinkTest}=[thick,solid,DarkOrange]
\tikzAGCmakeSignal{ShrinkTestik}{{\IdShrinkTest{\SpeedIndexI,\SpeedIndexK}{}}}{StyleShrinkTest}
\tikzAGCmakeSignal{ShrinkTestiR}{{\IdShrinkTest{\SpeedIndexI}{R}}}{StyleShrinkTest}
\tikzAGCmakeSignal{ShrinkTestiL}{{\IdShrinkTest{\SpeedIndexI}{L}}}{StyleShrinkTest}
\tikzAGCmakeSignal{ShrinkTestkL}{{\IdShrinkTest{\SpeedIndexK}{L}}}{StyleShrinkTest}
\JDLvocabulary{\SymbolShrinkTestOK}{\AGCsigPolice{shrink-test-ok}}{}{To encode the success of the size comparison}
\newcommand{\IdShrinkTestOK}[1]{\IdBaseSpeedOtherSpeedParamParam{\SymbolShrinkTestOK}{#1}{}{}{}}
\tikzstyle{StyleShrinkTestOK}=[thick,solid,DarkOrange]
\tikzAGCmakeSignal{ShrinkTestOKi}{\IdShrinkTestOK{\SpeedIndexI}}{StyleShrinkTestOK}
\tikzAGCmakeSignal{ShrinkTestOKk}{\IdShrinkTestOK{\SpeedIndexK}}{StyleShrinkTestOK}
\JDLvocabulary{\SymbolShrinkTestFail}{\AGCsigPolice{shrink-test-fail}}{}{To encode the failure of the size comparison}
\newcommand{\IdShrinkTestFail}[1]{\IdBaseSpeedOtherSpeedParamParam{\SymbolShrinkTestFail}{#1}{}{}{}}
\tikzstyle{StyleShrinkTestFail}=[StyleFail,thick]
\tikzAGCmakeSignal{ShrinkTestFaili}{\IdShrinkTestFail{\SpeedIndexI}}{StyleShrinkTestFail}
\JDLvocabulary{\SymbolShrinkOrder}{\AGCsigPolice{shrink-order}}{}{To encode the failure of the size comparison}
\newcommand{\IdShrinkOrder}[1]{\IdBaseSpeedOtherSpeedParamParam{\SymbolShrinkOrder}{}{#1}{}{}}
\tikzstyle{StyleShrinkOrder}=[StyleShrinkTestFail]
\tikzAGCmakeSignal{ShrinkOrderi}{\IdShrinkOrder{\SpeedIndexI}}{StyleShrinkOrder}
\tikzAGCmakeSignal{ShrinkOrderk}{\IdShrinkOrder{\SpeedIndexK}}{StyleShrinkOrder}
\JDLvocabulary{\SymbolShrinkId}{\AGCsigPolice{shrink-id}}{}{To encode }
\newcommand{\IdShrinkId}[1]{\IdBaseSpeedOtherSpeedParamParam{\SymbolShrinkId}{}{#1}{}{}}
\tikzAGCmakeSignal{ShrinkIDi}{\IdShrinkId{\SpeedIndexI}}{StyleId}
\tikzAGCmakeSignal{ShrinkIDk}{\IdShrinkId{\SpeedIndexK}}{StyleId}
\JDLvocabulary{\SymbolShrinkRuleBound}{\AGCsigPolice{shrink-rule-bound}}{}{To encode }
\newcommand{\IdShrinkRuleBound}[1][]{\IdBaseSpeedOtherSpeedParamParam{\SymbolShrinkRuleBound}{}{#1}{}{}}
\tikzAGCmakeSignal{ShrinkRuleBoundi}{\IdShrinkRuleBound[\SpeedIndexI]}{StyleRuleBound}
\JDLvocabulary{\SymbolShrinkRuleMiddle}{\AGCsigPolice{shrink-rule-middle}}{}{To encode }
\newcommand{\IdShrinkRuleMiddle}[1][]{\IdBaseSpeedOtherSpeedParamParam{\SymbolShrinkRuleMiddle}{}{#1}{}{}}
\tikzAGCmakeSignal{ShrinkRuleMiddlei}{\IdShrinkRuleMiddle[\SpeedIndexI]}{StyleRuleMiddle}
\JDLvocabulary{\SymbolShrinkIf}{\AGCsigPolice{shrink-if}}{}{To encode the if part of rules during shrinking}
\newcommand{\IdShrinkIf}[2][]{\IdBaseSpeedOtherSpeedParamParam{\SymbolShrinkIf}{}{#1}{}{#2}}
\tikzAGCmakeSignal{ShrinkIfil}{\IdShrinkIf[\SpeedIndexI]{\SpeedIndexL}}{StyleIf}
\JDLvocabulary{\SymbolShrinkThen}{\AGCsigPolice{shrink-then}}{}{To encode the then part of rules during shrinking}
\newcommand{\IdShrinkThen}[2][]{\IdBaseSpeedOtherSpeedParamParam{\SymbolShrinkThen}{}{#1}{}{#2}}
\tikzAGCmakeSignal{ShrinkThenil}{{\IdShrinkThen[\SpeedIndexI]{\SpeedIndexL}}}{StyleThen}
\JDLvocabulary{\SymbolShrinkDelayed}{\AGCsigPolice{shrink-delayed}}{}{To encode a shrinking that is waiting for the previous one to finish}
\newcommand{\IdShrinkDelayed}[2]{\IdBaseSpeedOtherSpeedParamParam{\SymbolShrinkDelayed}{}{#1}{#2}{}}
\tikzstyle{StyleShrinkDelayed}=[StyleShrinkBottom]
\tikzAGCmakeSignal{ShrinkDelayedLi}{{\IdShrinkDelayed{\SpeedIndexI}{L}}}{StyleShrinkDelayed}
\tikzAGCmakeSignal{ShrinkDelayedRi}{{\IdShrinkDelayed{\SpeedIndexI}{R}}}{StyleShrinkDelayed}
\tikzAGCmakeSignal{ShrinkDelayedLk}{{\IdShrinkDelayed{\SpeedIndexK}{L}}}{StyleShrinkDelayed}
\tikzAGCmakeSignal{ShrinkDelayedRk}{{\IdShrinkDelayed{\SpeedIndexK}{R}}}{StyleShrinkDelayed}

\JDLvocabulary{\SymbolTestStart}{\AGCsigPolice{test-start}}{}{To encode }
\newcommand{\IdTestStart}[1]{\IdBaseSpeedOtherSpeedParamParam{\SymbolTestStart}{}{#1}{}{}}
\tikzAGCmakeSignal{TestStarti}{\IdTestStart{\SpeedIndexI}}{Magenta,solid,thick}
\JDLvocabulary{\SymbolTestLeft}{\AGCsigPolice{test-left}}{}{To encode }
\newcommand{\IdTestLeft}[1]{\IdBaseSpeedOtherSpeedParamParam{\SymbolTestLeft}{}{#1}{}{}}
\tikzAGCmakeSignal{TestLefti}{\IdTestLeft{\SpeedIndexI}}{Magenta,solid,thick}
\JDLvocabulary{\SymbolTestRightOk}{\AGCsigPolice{test-right-ok}}{}{To encode }
\newcommand{\IdTestRightOk}[2]{\IdBaseSpeedOtherSpeedParamParam{\SymbolTestRightOk}{}{#1}{}{#2}}
\tikzAGCmakeSignal{TestRightOKij}{\IdTestRightOk{\SpeedIndexI}{\SpeedIndexJ}}{Magenta,solid,thick}
\JDLvocabulary{\SymbolTestRight}{\AGCsigPolice{test-right}}{}{To encode }
\newcommand{\IdTestRight}[3]{\IdBaseSpeedOtherSpeedParamParam{\SymbolTestRight}{}{#1}{#2}{#3}}
\tikzAGCmakeSignal{TestRighti}{\IdTestRight{\SpeedIndexI}{}{}}{Magenta,solid,thick}
\tikzAGCmakeSignal{TestRightij}{\IdTestRight{\SpeedIndexI,\SpeedIndexJ}{}{}}{Magenta,solid,thick}
\tikzAGCmakeSignal{TestRightijk}{\IdTestRight{\SpeedIndexI,\SpeedIndexJ}{}{\SpeedIndexK}}{Magenta,solid,thick}
\tikzAGCmakeSignal{TestRightijl}{\IdTestRight{\SpeedIndexI,\SpeedIndexJ}{}{\SpeedIndexL}}{Magenta,solid,thick}
\tikzAGCmakeSignal{TestRightik}{\IdTestRight{\SpeedIndexI,\SpeedIndexK}{}{\SpeedIndexK}}{Magenta,solid,thick}
\tikzAGCmakeSignal{TestRightikj}{\IdTestRight{\SpeedIndexI,\SpeedIndexK}{}{\SpeedIndexJ}}{Magenta,solid,thick}
\tikzAGCmakeSignal{TestRightikk}{\IdTestRight{\SpeedIndexI,\SpeedIndexK}{}{\SpeedIndexK}}{Magenta,solid,thick}
\tikzAGCmakeSignal{TestRightikl}{\IdTestRight{\SpeedIndexI,\SpeedIndexK}{}{\SpeedIndexL}}{Magenta,solid,thick}

\JDLvocabulary{\SymbolTestRightWait}{\AGCsigPolice{test-right-wait}}{}{To encode }
\newcommand{\IdTestRightWait}[3]{\IdBaseSpeedOtherSpeedParamParam{\SymbolTestRightWait}{}{#1}{#2}{#3}}
\tikzAGCmakeSignal{TestRightWaitij}{\IdTestRightWait{\SpeedIndexI,\SpeedIndexJ}{}{\SpeedIndexJ}}{Magenta,solid,thick}
\tikzAGCmakeSignal{TestRightWaitijk}{\IdTestRightWait{\SpeedIndexI,\SpeedIndexK}{}{\SpeedIndexJ}}{Magenta,solid,thick}
\tikzAGCmakeSignal{TestRightWaitijl}{\IdTestRightWait{\SpeedIndexI,\SpeedIndexJ}{}{\SpeedIndexL}}{Magenta,solid,thick}
\tikzAGCmakeSignal{TestRightWaitikj}{\IdTestRightWait{\SpeedIndexI,\SpeedIndexK}{}{\SpeedIndexJ}}{Magenta,solid,thick}
\tikzAGCmakeSignal{TestRightWaitikl}{\IdTestRightWait{\SpeedIndexI,\SpeedIndexK}{}{\SpeedIndexL}}{Magenta,solid,thick}
\JDLvocabulary{\SymbolTestLeftUp}{\AGCsigPolice{test-left-up}}{}{To encode }
\newcommand{\IdTestLeftUp}[2]{\IdBaseSpeedOtherSpeedParamParam{\SymbolTestLeftUp}{}{#1}{}{#2}}
\tikzAGCmakeSignal{TestLeftUpik}{\IdTestLeftUp{\SpeedIndexI,\SpeedIndexK}{}}{Magenta,solid}
\JDLvocabulary{\SymbolTestRightUp}{\AGCsigPolice{test-right-up}}{}{To encode }
\newcommand{\IdTestRightUp}[2]{\IdBaseSpeedOtherSpeedParamParam{\SymbolTestRightUp}{}{#1}{}{#2}}
\tikzAGCmakeSignal{TestRightUpi}{\IdTestRightUp{\SpeedIndexI}{}}{Magenta,solid}
\JDLvocabulary{\SymbolTestLeftOk}{\AGCsigPolice{test-left-ok}}{}{To encode }
\newcommand{\IdTestLeftOk}[1]{\IdBaseSpeedOtherSpeedParamParam{\SymbolTestLeftOk}{}{#1}{}{}}
\tikzAGCmakeSignal{TestLeftOKi}{\IdTestLeftOk{\SpeedIndexI}}{Magenta,solid,thick}
\JDLvocabulary{\SymbolMainTestOK}{\IdBaseSpeedOtherSpeedParamParam{\SymbolMain}{}{}{test}{ok}}{}{To encode the success of the test on main}
\newcommand{\IdMainTestOK}[1][]{\IdBaseSpeed{\SymbolMainTestOK}{#1}}
\tikzAGCmakeSignal{MainTestOKi}{\IdMainTestOK[\SpeedIndexI]}{Magenta,solid,ultra thick}
\JDLvocabulary{\SymbolMainTestFailL}{\IdBaseSpeedOtherSpeedParamParam{\SymbolMain}{}{}{test}{fail-l}}{}{To encode a left failure of the test on main}
\newcommand{\IdMainTestFailL}[1][]{\IdBaseSpeed{\SymbolMainTestFailL}{#1}}
\tikzAGCmakeSignal{MainTestFailLi}{\IdMainTestFailL[\SpeedIndexI]}{StyleFail,ultra thick}
\JDLvocabulary{\SymbolMainTestFailR}{\IdBaseSpeedOtherSpeedParamParam{\SymbolMain}{}{}{test}{fail-r}}{}{To encode a right failure of the test on main}
\newcommand{\IdMainTestFailR}[1][]{\IdBaseSpeed{\SymbolMainTestFailR}{#1}}
\tikzAGCmakeSignal{MainTestFailRi}{\IdMainTestFailR[\SpeedIndexI]}{StyleFail,ultra thick}
\JDLvocabulary{\SymbolTestLeftFail}{\AGCsigPolice{test-left-fail}}{}{To encode }
\newcommand{\IdTestLeftFail}[1]{\IdBaseSpeedOtherSpeedParamParam{\SymbolTestLeftFail}{}{#1}{}{}}
\tikzAGCmakeSignal{TestLeftFaili}{\IdTestLeftFail{\SpeedIndexI}}{StyleFail,thick}
\JDLvocabulary{\SymbolTestRightFail}{\AGCsigPolice{test-right-fail}}{}{To encode }
\newcommand{\IdTestRightFail}[1]{\IdBaseSpeedOtherSpeedParamParam{\SymbolTestRightFail}{}{#1}{}{}}
\tikzAGCmakeSignal{TestRightFaili}{\IdTestRightFail{\SpeedIndexI}}{StyleFail,thick}

\JDLvocabulary{\SymbolCheck}{\AGCsigPolice{check}}{}{To check the right positioning of \SigMain}
\newcommand{\IdCheck}[2]{\IdBaseSpeedOtherSpeedParamParam{\SymbolCheck}{}{#1,#2}{}{}}
\tikzstyle{StyleCheck}=[DarkOrange,solid,thick] 
\tikzAGCmakeSignal{Checkij}{\IdCheck{\SpeedIndexI}{\SpeedIndexJ}}{StyleCheck}
\JDLvocabulary{\SymbolCheckMaybe}{\AGCsigPolice{check-?}}{}{To check the right positioning of \SigMain}
\newcommand{\IdCheckMaybe}[2]{\IdBaseSpeedOtherSpeedParamParam{\SymbolCheckMaybe}{}{#1,#2}{}{}}
\tikzAGCmakeSignal{CheckMaybeij}{\IdCheckMaybe{\SpeedIndexI}{\SpeedIndexJ}}{StyleCheck}
\JDLvocabulary{\SymbolCheckUp}{\AGCsigPolice{check-up}}{}{To check the right positioning of \SigMain}
\newcommand{\IdCheckUp}[2]{\IdBaseSpeedOtherSpeedParamParam{\SymbolCheckUp}{}{#1,#2}{}{}}
\tikzAGCmakeSignal{CheckUpij}{\IdCheckUp{\SpeedIndexI}{\SpeedIndexJ}}{StyleCheck}
\JDLvocabulary{\SymbolCheckIntercept}{\AGCsigPolice{check-intersect}}{}{To check the right positioning of \SigMain}
\newcommand{\IdCheckIntercept}[3]{\IdBaseSpeedOtherSpeedParamParam{\SymbolCheckIntercept}{}{#1,#2}{}{#3}}
\tikzAGCmakeSignal{CheckInterceptikj}{\IdCheckIntercept{\SpeedIndexI}{\SpeedIndexK}{\SpeedIndexJ}}{StyleCheck}
\tikzAGCmakeSignal{CheckInterceptikl}{\IdCheckIntercept{\SpeedIndexI}{\SpeedIndexK}{\SpeedIndexL}}{StyleCheck}
\tikzAGCmakeSignal{CheckInterceptikm}{\IdCheckIntercept{\SpeedIndexI}{\SpeedIndexK}{\SpeedIndexM}}{StyleCheck}
\JDLvocabulary{\SymbolCheckOK}{\AGCsigPolice{check-ok}}{}{To encode }
\newcommand{\IdCheckOK}[2]{\IdBaseSpeedOtherSpeedParamParam{\SymbolCheckOK}{}{#1,#2}{}{}}
\tikzAGCmakeSignal{CheckOKij}{\IdCheckOK{\SpeedIndexI}{\SpeedIndexJ}}{StyleCheck}
\JDLvocabulary{\SymbolCheckFail}{\AGCsigPolice{check-fail}}{}{To encode }
\newcommand{\IdCheckFail}[2]{\IdBaseSpeedOtherSpeedParamParam{\SymbolCheckFail}{}{#1,#2}{}{}}
\tikzAGCmakeSignal{CheckFailij}{\IdCheckFail{\SpeedIndexI}{\SpeedIndexJ}}{StyleFail,thick}

\tikzstyle{HYPOTETIC}=[DarkGrey,ultra thick,densely dotted,double]

\JDLvocabulary{\SymbolRuleBoundFail}{\AGCsigPolice{rule-bound-fail}}{}{To encode }
\newcommand{\IdRuleBoundFail}[1][]{\IdBaseSpeed{\SymbolRuleBoundFail}{#1}}
\tikzstyle{StyleRuleBoundFail}=[StyleFail,thick]
\tikzAGCmakeSignal{RuleBoundFaili}{\IdRuleBoundFail[\SpeedIndexI]}{StyleRuleFail}
\JDLvocabulary{\SymbolRuleMiddleFail}{\AGCsigPolice{rule-middle-fail}}{}{To encode }
\newcommand{\IdRuleMiddleFail}[1][]{\IdBaseSpeed{\SymbolRuleMiddleFail}{#1}}
\tikzstyle{StyleRuleFail}=[StyleRuleBoundFail]
\tikzAGCmakeSignal{RuleMiddleFaili}{\IdRuleMiddleFail[\SpeedIndexI]}{StyleRuleFail}
\JDLvocabulary{\SymbolIfOK}{\AGCsigPolice{if-ok}}{}{To encode the validation if part of rules}
\newcommand{\IdIfOK}[2][]{\IdBaseSpeedOtherSpeedParamParam{\SymbolIfOK}{#1}{}{#2}{}}
\tikzstyle{StyleIfOK}=[DarkGreen,densely dotted,thick]
\tikzAGCmakeSignal{IfOKi}{\IdIfOK[\SpeedIndexI]{\SpeedIndexI}}{StyleIfOK}
\tikzAGCmakeSignal{IfOKik}{\IdIfOK[\SpeedIndexI]{\SpeedIndexK}}{StyleIfOK}
\tikzAGCmakeSignal{IfOKim}{\IdIfOK[\SpeedIndexI]{\SpeedIndexM}}{StyleIfOK}
\tikzAGCmakeSignal{IfOKiFour}{\IdIfOK[\SpeedIndexI]{4}}{Red,dotted,thick}
\tikzAGCmakeSignal{IfOKiFive}{\IdIfOK[\SpeedIndexI]{5}}{Blue,dotted,thick}
\tikzAGCmakeSignal{IfOKiSix}{\IdIfOK[\SpeedIndexI]{6}}{DarkGreen,dotted,thick}
\JDLvocabulary{\IdCrossSymbol}{\AGCsigPolice{cross}}{}{To bring the id of colliding macro-signals onto the rules}
\newcommand{\IdCross}[1][]{\IdBaseSpeedOtherSpeedParamParam{\IdCrossSymbol}{}{}{#1}{}}
\tikzstyle{StyleCross}=[DarkOrange,solid,thick]
\tikzAGCmakeSignal{Crossi}{\IdCross[\SpeedIndexI]}{StyleCross}
\tikzAGCmakeSignal{Crossk}{\IdCross[\SpeedIndexK]}{StyleCross}
\tikzAGCmakeSignal{CrossFour}{\IdCross[4]}{Red,solid,thick}
\tikzAGCmakeSignal{CrossFive}{\IdCross[5]}{Blue,solid,thick}
\tikzAGCmakeSignal{CrossSix}{\IdCross[6]}{DarkGreen,solid,thick}
\JDLvocabulary{\IdCrossOkSymbol}{cross-ok}{}{When an id is inactivated after colliding on an if part of a rule}
\newcommand{\IdCrossOK}[1][]{\IdBaseSpeedOtherSpeedParamParam{\IdCrossOkSymbol}{}{}{#1}{}}
\tikzstyle{StyleCrossOK}=[densely dotted,DarkOrange,thick]
\tikzAGCmakeSignal{CrossOKi}{\IdCrossOK[\SpeedIndexI]}{StyleCrossOK}
\tikzAGCmakeSignal{CrossOKk}{\IdCrossOK[\SpeedIndexK]}{StyleCrossOK}
\tikzAGCmakeSignal{CrossOKFour}{\IdCrossOK[4]}{Red,dotted,thick}
\tikzAGCmakeSignal{CrossOKFive}{\IdCrossOK[5]}{Blue,dotted,thick}
\tikzAGCmakeSignal{CrossOKSix}{\IdCrossOK[6]}{DarkGreen,dotted,thick}
\JDLvocabulary{\IdCrossBackSymbol}{cross-back}{}{To bring the id of colliding macro-signals onto the rules}
\newcommand{\IdCrossBack}[1][]{\IdBaseSpeedOtherSpeedParamParam{\IdCrossBackSymbol}{}{}{#1}{}}
\tikzAGCmakeSignal{CrossBacki}{\IdCrossBack[\SpeedIndexI]}{StyleCross}
\JDLvocabulary{\IdCrossBackOkSymbol}{cross-back-ok}{}{When an id is inactivated after colliding on an if part of a rule}
\newcommand{\IdCrossBackOK}[1][]{\IdBaseSpeedOtherSpeedParamParam{\IdCrossBackOkSymbol}{}{}{#1}{}}
\tikzAGCmakeSignal{CrossBackOKi}{\IdCrossBackOK[\SpeedIndexI]}{StyleCrossOK}
\JDLvocabulary{\SymbolCollect}{\AGCsigPolice{collect}}{}{To encode }
\newcommand{\IdCollect}[2]{\IdBaseSpeedOtherSpeedParamParam{\SymbolCollect}{}{}{#1}{#2}}
\tikzAGCmakeSignal{Collectij}{\IdCollect{\SpeedIndexI}{\SpeedIndexJ}}{Magenta,solid,thick}

\JDLvocabulary{\SymbolIdCopy}{\AGCsigPolice{id-copy}}{}{To make a copy of the selected rule}
\newcommand{\IdCopy}[2][\SpeedIndexI]{\IdBaseSpeedOtherSpeedParamParam{\SymbolIdCopy}{}{}{#2}{#1}}
\tikzAGCmakeSignal{IdCopyOne}{{\IdCopy{1}}}{StyleOne}
\tikzAGCmakeSignal{IdCopyTwo}{{\IdCopy{2}}}{StyleTwo}
\tikzAGCmakeSignal{IdCopyl}{{\IdCopy{\SpeedIndexL}}}{StyleOne}
\JDLvocabulary{\SymbolIdSelected}{\AGCsigPolice{id-select}}{}{Copy of the selected rule awaiting dispatching}
\newcommand{\IdSelected}[2][\SpeedIndexI]{\IdBaseSpeedOtherSpeedParamParam{\SymbolIdSelected}{#1}{}{#2}{}}
\tikzAGCmakeSignal{IdSelectedl}{{\IdSelected{\SpeedIndexL}}}{StyleOne}
\tikzAGCmakeSignal{IdSelectedOne}{{\IdSelected{1}}}{StyleOne}
\tikzAGCmakeSignal{IdSelectedTwo}{{\IdSelected{2}}}{StyleTwo}
\JDLvocabulary{\SymbolReady}{\AGCsigPolice{ready}}{}{To encode }
\newcommand{\IdReady}[2][\ensuremath{\emptyset}]{\IdBaseSpeedOtherSpeedParamParam{\SymbolReady}{}{}{#2}{#1}}
\tikzstyle{StyleReady}=[DarkPink,solid,thick]
\tikzAGCmakeSignal{Readyi}{\IdReady{\SpeedIndexI}}{StyleReady}
\tikzAGCmakeSignal{Readyl}{\IdReady{\SpeedIndexL}}{StyleReady}
\tikzAGCmakeSignal{ReadyiE}{\IdReady[\SpeedSubset]{\SpeedIndexI}}{StyleReady}
\tikzAGCmakeSignal{ReadyiEl}{\IdReady[\ensuremath{\SpeedSubset\cup\{\SpeedIndexL\}}]{\SpeedIndexI}}{StyleReady}
\JDLvocabulary{\SymbolReadyNo}{\AGCsigPolice{ready-no}}{}{To encode }
\newcommand{\IdReadyNo}[2][\ensuremath{\emptyset}]{\IdBaseSpeedOtherSpeedParamParam{\SymbolReadyNo}{}{}{#2}{#1}}
\tikzstyle{StyleReadyNo}=[Red,dashed,thick]
\tikzAGCmakeSignal{ReadyNoi}{\IdReadyNo{\SpeedIndexI}}{StyleReadyNo}
\tikzAGCmakeSignal{ReadyNoiE}{\IdReadyNo[\SpeedSubset]{\SpeedIndexI}}{StyleReadyNo}
\newcommand{\SymbolReadyU}{\IdBaseSpeedOtherSpeedParamParam{ready}{}{}{}{1}}
\tikzstyle{StyleReadyU}=[StyleReady]
\tikzAGCmakeSignal{ReadyU}{\SymbolReadyU}{StyleReadyU}
\newcommand{\SymbolReadyUU}{\IdBaseSpeedOtherSpeedParamParam{ready}{}{}{}{2}}
\tikzstyle{StyleReadyUU}=[StyleReadyU]
\tikzAGCmakeSignal{ReadyUU}{\SymbolReadyUU}{StyleReadyUU}

\JDLvocabulary{\SymbolFastLeft}{\AGCsigPolice{fast-left}}{}{To encode }
\tikzAGCmakeSignal{FastLeft}{\IdBaseSpeedOtherSpeedParamParam{\SymbolFastLeft}{}{}{}{\SpeedSubset}}{Black,densely dashed,thick}
\JDLvocabulary{\SymbolFastRight}{\AGCsigPolice{fast-right}}{}{To encode }
\tikzAGCmakeSignal{FastRight}{\IdBaseSpeedOtherSpeedParamParam{\SymbolFastRight}{}{}{}{\SpeedSubset}}{Black,densely dashed,thick}

\JDLvocabulary{\OutputRight}{\JDLvocabularyTextXspace{\textsf{O}$_R$}}{}{Right limit of outputting signals}
\JDLvocabulary{\OutputLeft}{\JDLvocabularyTextXspace{\textsf{O}$_L$}}{}{Left limit of outputting signals}

\JDLvocabulary{\AGCconfigurationSetZeroNonEmpty}{\JDLvocabularyMathXspace{\mathcal{C}^{*}}}{}{The set of all configurations of \AGCmachine with $\AGCconfiguration(0) \neq \AGCextendedValueVoid$}
\JDLvocabulary{\SymbConfFun}{\JDLvocabularyMathXspace{\operatorname{symb}}}{}{Function yielding the symbolic representation of a configuration}
\JDLvocabulary{\AGCsimGroup}{\JDLvocabularyMathXspace{\operatorname{interp}}}{}{Local function used for decoding signals in simulation}
\JDLvocabulary{\AGCsimGroupConf}{\JDLvocabularyMathXspace{\overline{\AGCsimGroup}}}{}{Local function used for decoding whole configurations in simulation}
\JDLvocabulary{\AGCsimClean}{\JDLvocabularyMathXspace{\operatorname{clean}}}{}{Local predicate indicating configurations that can be used as the initial configuration of the simulator}
\JDLvocabulary{\AGCsimRepr}{\JDLvocabularyMathXspace{\operatorname{repr}}}{}{Function used to find a representative of a signal, i.e. encoding it}
\JDLvocabulary{\AGCsimReprConf}{\JDLvocabularyMathXspace{\overline{\AGCsimRepr}}}{}{Extension of \AGCsimRepr to whole configurations}
\JDLvocabulary{\AGCsimStartingMSwidth}{\JDLvocabularyMathXspace{\delta}}{}{Maximum width of macro-signals entering a checked macro-collision}
\JDLvocabulary{\AGCsimMSwidth}{\JDLvocabularyMathXspace{\operatorname{width}}}{}{Initial width of a macro-signal}
\JDLvocabulary{\AGCsimOutputDelay}{\JDLvocabularyMathXspace{\delta_{o}}}{}{Delay between a (macro-)collision and the time when all its outputs are ready and clean}
\JDLvocabulary{\AGCsimMaxSignalsInMacroSignal}{\JDLvocabularyMathXspace{\mathfrak{w}}}{}{Maximum of signals in any value of $\AGCsimRepr(\cdot)$}
\JDLvocabulary{\AGCruleMax}{\JDLvocabularyMathXspace{\AGCrule_{\operatorname{max}}}}{}{Collision with an input signal of each speed, each with maximal id.}
\providecommand{\AGCruleMaxIn}{\mathxspace{\AGCruleIn_{\operatorname{max}}}}

\JDLvocabulary{\AGCconfigurationSMSymbol}{\JDLvocabularyMathXspace{a}}{}{Some configuration of \AGCmachine}
\JDLvocabulary{\AGCconfigurationSMotherSymbol}{\JDLvocabularyMathXspace{b}}{}{Some configuration of \AGCmachineOther}

\providecommand{\AGCconfigurationSM}[1][]{\JDLvocabularyMathXspace{\AGCconfigurationSMSymbol_{#1}}}{}{}
{}{}
\JDLvocabulary{\MimicEncodingFunction}{\JDLvocabularyMathXspace{\kappa}}{}{Function encoding for mimicking}

\JDLvocabulary{\SMbaseSet}{\JDLvocabularyMathXspace{\Gamma}}{}{Base meta-signal set for mimicking}
\JDLvocabulary{\SMbase}{\JDLvocabularyMathXspace{\gamma}}{}{Base meta-signal set for mimicking}
\JDLvocabulary{\SMclean}{\JDLvocabularyMathXspace{\chi}}{}{Cleaning function for mimicking}
\JDLvocabulary{\SMmimickCollision}{\JDLvocabularyMathXspace{\times}}{}{Yet unlabelled collision for mimicking}
\JDLvocabulary{\SMprefixSet}{\JDLvocabularyMathXspace{P}}{}{Prefix-free set of  sequences of \AGCconfigurationSM-meta-signals for mimicking}
\JDLvocabulary{\SMprefix}{\JDLvocabularyMathXspace{p}}{}{Sequences of \AGCmachine-meta-signals for mimicking}
\JDLvocabulary{\SMrelabeling}{\JDLvocabularyMathXspace{\lambda}}{}{Relabelling function for mimicking}
\JDLvocabulary{\SMquerySpeed}{\JDLvocabularyMathXspace{\nu}}{}{Speed query function for mimicking}
\JDLvocabulary{\SMrelabelSignal}{\JDLvocabularyMathXspace{\varphi}}{}{Signal relabeling for mimicking}
\JDLvocabulary{\SMrelabelCollision}{\JDLvocabularyMathXspace{\psi}}{}{Collision relabelling for mimicking}

\newcommand{\SpeedSpecial}[2][]{\JDLvocabularyMathXspace{\AGCspeed_{\text{\sf #1}}^{\text{\sf #2}}}}
\JDLvocabulary{\SpeedMaxAbs}{\SpeedSpecial{max}}{}{Maximum absolute value of speed}
\JDLvocabulary{\SpeedRapid}{\SpeedSpecial{rapid}}{}{Base speed for Check signals}
\JDLvocabulary{\SpeedShrink}{\SpeedSpecial{shrink}}{}{Base speed for shrinking signals}
\JDLvocabulary{\SpeedTestUpi}{\SpeedSpecial[\SpeedIndexI]{test-up}}{}{Speed for testing}
\JDLvocabulary{\SpeedTestRightik}{\SpeedSpecial[\SpeedIndexI,\SpeedIndexK]{test-right}}{}{Speed for testing}
\JDLvocabulary{\SpeedTestRighti}{\SpeedSpecial[\SpeedIndexI]{test-right}}{}{Speed for testing}
\JDLvocabulary{\SpeedTestRightUpi}{\SpeedSpecial[\SpeedIndexI]{test-right-up}}{}{Speed for testing}
\JDLvocabulary{\SpeedTestBackLefti}{\SpeedSpecial[\SpeedIndexI]{test-left-back}}{}{Speed for testing}
\JDLvocabulary{\SpeedTestBackRighti}{\SpeedSpecial[\SpeedIndexI]{test-right-back}}{}{Speed for testing}
\JDLvocabulary{\SpeedCheckUp}{\SpeedSpecial{chk-up}}{}{Speed for checking \SigMain positions}

\JDLvocabulary{\Uik}{\JDLvocabularyMathXspace{U^{i,k}}}{}{Point used for testing}

\begin{document}

\title{Abstract Geometrical Computation 10: \\
  An Intrinsically Universal Family of Signal Machines\footnote{The authors are thankful to the Franco-Iranian PHC Gundishapur 2017 number 38071PC ``Dynamique des machines \`a signaux'' for funding this research.}}

\author{Florent Becker$^1$
\and Tom Besson$^1$
\and J{\'e}r{\^o}me Durand-Lose$^{1,2}$
\and Aur{\'e}lien Emmanuel$^1$
\and Mohammad-Hadi Foroughmand-Araabi $^3$
\and Sama Goliaei$^4$
\and Shahrzad Heydarshahi$^1$}

\maketitle

{\renewcommand{\abstractname}{}
  \begin{abstract}
    \parindent 0cm
    
    \noindent$^{1}$Univ. Orléans, INSA Centre Val de Loire, LIFO, France
    \texttt{\{florent.becker, jerome.durand-lose\}@univ-orleans.fr}
    
$^{2}${LIX, CNRS-Inria-École Polytechnique, France}

$^{3}${Department of Mathematical Sciences, Sharif University of Technology, Tehran, Iran
  \texttt{foroughmand@sharif.ir}}

$^{4}${Faculty of New Sciences and Technologies, University of Tehran, Tehran, Iran
  \texttt{sgoliaei@ut.ac.ir}}
\end{abstract}
}

\begin{abstract}\parindent 1.25 emSignal machines form an abstract and idealised model of collision computing. Based on dimensionless signals moving on the real line, they model particle/signal dynamics in Cellular Automata.
  Each particle, or \emph{signal}, moves at constant speed in continuous time and space.
  When signals meet, they get replaced by other signals.
  A signal machine defines the types of available signals, their speeds and the rules for replacement in collision.

  A signal machine \AGCmachine simulates another one \AGCmachineOther if all the space-time diagrams of \AGCmachineOther can be generated from space-time diagrams of \AGCmachine by removing some signals and renaming other signals according to local information. 
  Given any finite set of speeds \SpeedSet, we construct a signal machine that is able to simulate any signal machine whose speeds belong to \SpeedSet.
  Each signal is simulated by a \emph{macro-signal}, a ray of parallel signals.
  Each macro-signal has a main signal located exactly where the simulated signal would be, as well as auxiliary signals which encode its id and the collision rules of the simulated machine.

  The simulation of a collision, a \emph{macro-collision}, consists of two phases.
  In the first phase, macro-signals are shrunk, then the macro-signals involved in the collision are identified and it is ensured that no other macro-signal comes too close.
  If some do, the process is aborted and the macro-signals are shrunk, so that the correct macro-collision will eventually be restarted and successfully initiated.
  Otherwise, the second phase starts: the appropriate collision rule is found and new macro-signals are generated accordingly.

  Considering all finite set of speeds \SpeedSet and their corresponding simulators provides an intrinsically universal family of signal machines.
\end{abstract}

{\renewcommand{\abstractname}{Key-Words}
\begin{abstract}
  Abstract Geometrical Computation;
  Collision computing;
  Intrinsic universality;
  Signal machine;
  Simulation  
\end{abstract}
}

\section{Introduction}
\label{sec:intro}

\emph{Signal Machines} (SM) arose as a continuous abstraction of Cellular Automata (CA) \citep{durand-lose08jac}.
In dimension one, the dynamics of Cellular Automata is often described as signals interacting in collisions resulting in the generation of new signals.
Signals allow to store and transmit information, to start a process, to synchronise, etc.
The use of signals in the context of CA is widespread in the literature:
collision computing \citep{adamatzky02book}, 
gliders \cite{jin+chen16}, 
solitons \citep{jakubowski+steiglitz+squier96cs,jakubowski+steiglitz+squier97,jakubowski+steiglitz+squier17,siwak01ip}, 
particles \citep{boccara+nasser+roger91phys-rev-a,mitchell96,hordijk+crutchfield+mitchell98},
Turing-computation \citep{lindgren+nordahl90,cook04cs},
synchronisation \citep{varshavsky+marakhovsky+peschansky70mst,yunes07mcu},
geometrical constructions \citep{cook04cs},
signals \citep{mazoyer+terrier99tcs,delorme+mazoyer02book-adamatzky}, etc.

In signal machines, signals are dimention-less points moving on a 1-dimensional Euclidean space in continuous time.
They have uniform movement and thus draw line segments on space-time diagrams.
Each signal is an instance of a \emph{meta-signal} among a finite given set of meta-signals.
As soon as two or more signals meet, a collision happens: incoming signals are instantly replaced by outgoing signals according to \emph{collision rules}, depending on the meta-signals of the incoming signals.
In-between collisions, signals propagate at some uniform speed depending on their meta-signal.

A signal machine is defined by
a finite set of meta-signals,
a function assigning a speed to each meta-signal (negative for leftward), and
a set of collision rules.
A collision rule associates a set of at least two meta-signals of different speeds (\emph{incoming}) with another set of meta-signals of different speeds (\emph{outgoing}).
Collision rules are deterministic: a set appears at most once as the incoming part of a collision rule.

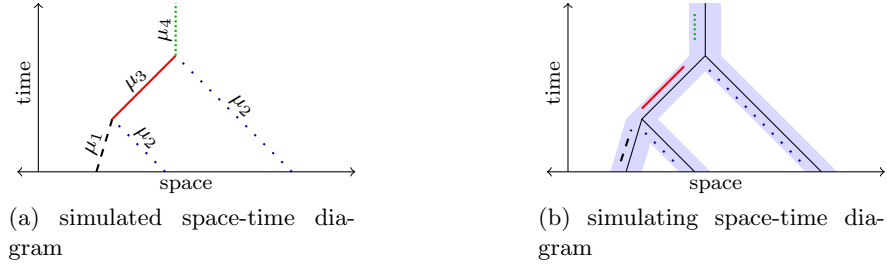
\begin{figure}\newcommand{\TOP}{16}
  \newcommand{\PicSim}[1]{\footnotesize\SetUnitlength{.47em}\begin{tikzpicture}[y=\unitlength]
      \coordinate (AA) at (7.5,0);
      \coordinate (BB) at (14,0);
      \coordinate (CC) at (26,0);
      \coordinate (A) at (9,5);
      \coordinate (B) at (15,11);
      \coordinate (C) at (15,\TOP);
      #1
      \draw[<->] (0, 0) -- node[below]{space} +(32,0);
      \draw[->] (2, 0) -- node[sloped,left=-.2em,anchor=south]{time} +(0,\TOP);
    \end{tikzpicture}\vspace*{-.5em}}\mbox{}\hfill\SubFigure[simulated space-time diagram\label{fig:schematic-simulation:ed}]{\PicSim{\DrawSigMOneLUAbove(AA)(A);\DrawSigMTwoLUAbove(BB)(A);\DrawSigMThreeLUAbove(A)(B);\DrawSigMTwoLUAbove(CC)(B);\DrawSigMFourLUAbove(B)(C);}}\hfill\SubFigure[simulating space-time diagram\label{fig:schematic-simulation:ion}]{\PicSim{\fill[data-fill] ([shift={(1.5,0)}]AA) -- ([shift={(-1.5,0)}]AA) -- ([shift={(-1.5,0)}]A) -- ([shift={(1.5,0)}]A) --cycle;
      \fill[data-fill] ([shift={(1.5,0)}]BB) --([shift={(-1.5,0)}]BB) -- ([shift={(-1.5,0)}]A) -- ([shift={(1.5,0)}]A) -- cycle;
      \fill[data-fill] ([shift={(1.5,0)}]CC) -- ([shift={(-1.5,0)}]CC) -- ([shift={(-1.5,0)}]B) -- ([shift={(1.5,0)}]B) -- cycle;		
      \fill[data-fill] ([shift={(1.5,0)}]A) -- ([shift={(-1.5,0)}]A) -- ([shift={(-1.5,0)}]B) -- ([shift={(1.5,0)}]B) -- cycle;		
      \fill[data-fill] ([shift={(1.5,0)}]B) -- ([shift={(-1.5,0)}]B) -- ([shift={(-1.5,0)}]C) -- ([shift={(1.5,0)}]C) -- cycle;
      \draw (AA) -- (A);
      \draw (BB) -- (A) -- (B);
      \draw (CC) -- (B);
      \draw (B) -- (C);
      \DrawSigMOne([shift={(-.5,1)}]AA)([shift={(-1,-1)}]A);\DrawSigMTwo([shift={(-2,1)}]BB)([shift={(0,-1)}]A);\DrawSigMThree([shift={(0,1)}]A)([shift={(-2,-1)}]B);\DrawSigMTwo([shift={(-2,1)}]CC)([shift={(0,-1)}]B);\DrawSigMFour([shift={(-1,1.5)}]B)([shift={(-1,-1)}]C);}}\hfill\mbox{}
  \caption{Simulation scheme.}
  \label{fig:schematic-simulation}
\end{figure}

In any configuration, there are finitely many signals and collisions located at distinct places on the real line.
The aggregation of the configurations reachable from some (initial) configuration forms a two dimensional space-time diagram like the one in \RefFig{fig:schematic-simulation:ed} in which the traces of signals are line segments.
Signals corresponding to the same meta-signal have the same speed: their traces are parallel segments (like the dotted \SigMTwo).
Collisions provide a discrete time scale and a directed acyclic graph structure inside each space-time diagram.
This emphasises the hybrid aspect of SM: continuous steps separated by discrete steps.

Signal machines are known to be able to compute by simulating Turing machine and even to hyper-compute \citep{durand-lose12ijuc-uc-hypercomp}.
As an analog model of computation they correspond exactly to the linear BSS model \citep{blum+shub+smale89bams,durand-lose07cie}.

As with any computing dynamical system, it is natural to ask whether there is a signal machine which is able to simulate all signal machines.
Intrinsic universality (being able to simulate any device of its own kind) is an important property, since it means to represent all machines and to exhibit all the behaviours available in the class.
In computer science, the existence of (intrinsically) universal Turing machines is the cornerstone of computability theory.
Many computing systems have intrinsically universal instances: the (full) BSS model, Cellular Automata (CA) \citep{albert+culik87,mazoyer+rapaport98stac,ollinger01fct,ollinger03stacs,goles+meunier+rapaport+theyssier11tcs}, reversible CA \citep{durand-lose95latin}, quantum CA \citep{arrigh+grattage12naco}, some tile assembly models at temperature $2$ \citep{doty+lutz+patitz+summers+woods10stacs,woods13mcu}, etc.
Some tile assembly models at temperature $1$ \citep{meunier+patitz+summers+theyssier+winslow+woods14soda} or causal graph dynamics \citep{martiel+martin15mcu} admit infinite intrinsically universal families but no single intrinsically universal instance.

One key characteristic of intrinsic universality is that it is expected to \emph{simulate according to the model}.
Transitive simulations across models are not enough: simulating a TM that can simulate any rational signal machine totally discards relevant aspects of the models such as directed acyclic graph representation, relative location, spatial positioning, energy levels, etc.

It should be noted that although instances of signal-based systems with Turing-computation capability are very common in the literature \citep{lindgren+nordahl90,cook04cs}, to our knowledge, the present paper provides the first result about intrinsic universality in a purely signal-based continuous system.

For Cellular Automata, simulation and intrinsic universality can be defined with an operation of \emph{grouping} on space-time diagrams \cite{mazoyer+rapaport98stac,ollinger03stacs}.
This operation consists of creating a space-time diagram from another one by applying a local function on blocks of the former.
Because Cellular Automata are discrete, it is possible to consider the domain of this local function to be finite.

With signal machines, because space is continuous and there is no canonical scale within a diagram, decoding a space-time diagram is done by uniformly applying a \emph{local} decoding function on each point of each configuration of the diagram, rather than having grouping and blocks.
The notion of locality for the decoding function of the space-time diagram has to be defined by stating that the decoding function should only look at a uniformly-bounded amount of signals around a collision.
Because signal machines lack the discrete time-steps of Cellular Automata, a special handling of the initial configuration is necessary, somewhat like with self-assembling systems \cite{doty+lutz+patitz+schweller+summers+woods12focs}.

Having defined a fitting concept of simulation, the present paper provides, for any finite set of speeds \SpeedSet, a signal machine capable of simulating all signal machines which only use speeds in \SpeedSet.

In a simulation by one of our universal signal machines, each signal of the simulated SM is replaced by a ray of signals (shaded in \RefFig{fig:schematic-simulation:ion}) called \emph{macro-signal}.
Each macro-signal has a non-zero width and contains a \SigMain signal (black in the middle) which is exactly positioned as the simulated signal.
The meta-signal (dot, dash or thick in \RefFig{fig:schematic-simulation:ed}) is encoded within the macro-signal (greyed zone), as illustrated in \RefFig{fig:schematic-simulation:ion}, which is then used by the decoding function to recover the meta-signal.

Each macro-signal encodes its identity in unary together with the list of all the collision rules of the simulated signal machine.
Notice that at any time, the amount of information in the macro-signal is bounded.
Macro-collisions are handled locally.

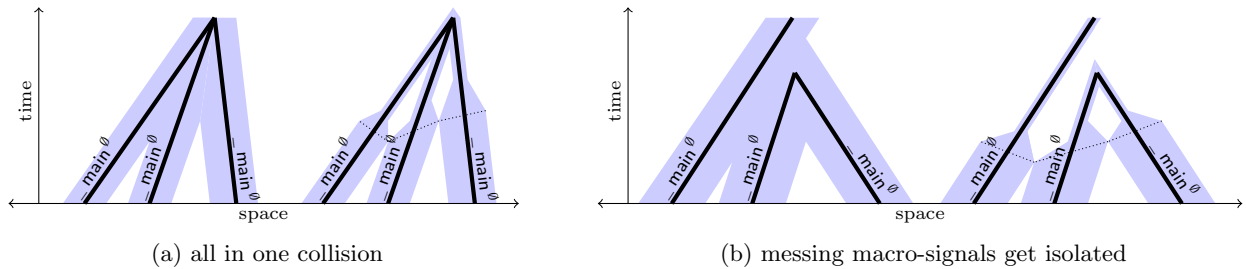
\begin{figure}[hbt]
  \footnotesize\scriptsize\SetUnitlength{.49em}\newcommand{\PicSim}[2]{\begin{tikzpicture}[x=.65\unitlength]\begin{scope}[inner sep=-.4ex]
    #1      
    \end{scope}
    \draw[<->] (-7, 0) -- node[below]{space} +(51+#2,0);
    \draw[->] (-2.5, 0) -- node[sloped,left=-.2em,anchor=south]{time} +(0,19);
  \end{tikzpicture}}\centering\SubFigure[all in one collision\label{fig:shrink-role:collision}]{\PicSim{
  \begin{scope}[x=-.7\unitlength,shift={(-25,0)}]
    \fill[blue!20!white] (-3,0) -- (0,18) -- (3,18) -- (5,8) -- (3.66,0) -- cycle;
    \fill[blue!20!white] (9,0) -- (5,8) -- (3,18) -- (6,18) -- (12,6) -- (15,0) -- cycle;
    \fill[blue!20!white] (18,0) -- (12,6) -- (3,18) -- (6,18) -- (24,0) -- cycle;
    \DrawSigMainLUAboveRight(3,18)(0,0);
    \DrawSigMainLUAboveLeft(12,0)(3,18);
    \DrawSigMainLUAboveLeft(21,0)(3,18);
  \end{scope}
  \begin{scope}[x=-.7\unitlength,shift={(-58,0)}]
    \coordinate (ll) at (-1.5,9) ;    
    \coordinate (l) at (5,8) ;    
    \coordinate (r) at (12,6) ;    
    \coordinate (rr) at (16,8) ;    
    \fill[blue!20!white] (-3,0) -- (ll) -- (1,12) -- (2,18) -- (3,19) -- (3,18) -- (0,0) -- cycle;
    \fill[blue!20!white] (0,0) -- (3,18) -- (3.5,15) -- (3,12) -- (l) -- (3.66,0) -- cycle;
    \fill[blue!20!white] (9,0) -- (l) -- (6,10) -- (3.5,15) -- (3,18) -- (3,19) -- (6,14) -- (9,8) -- (12,6) -- (15,0) -- cycle;
    \fill[blue!20!white] (18,0) -- (12,6) -- (12,8) -- (6,14) -- (3,18) -- (3,19) -- (13,9) -- (16,8) -- (24,0) -- cycle;
    \draw[densely dotted] (ll) -- (l) ;
    \draw[densely dotted] (l) -- (r) ;
    \draw[densely dotted] (r) -- (rr) ;
    \DrawSigMainLUAboveRight(3,18)(0,0);
    \DrawSigMainLUAboveLeft(12,0)(3,18)(12,0);
    \DrawSigMainLUAboveLeft(21,0)(3,18);
  \end{scope}
}{25}
\vspace*{-.25em}}\hfill\SubFigure[messing macro-signals get isolated\label{fig:shrink-role:non-collision}]{\PicSim{
  \begin{scope}[x=-.65\unitlength,shift={(-40,0)}]
    \fill[blue!20!white] (0,0) -- (16,16)  -- (14,18) -- (22,18) -- (41,0) -- (31,0) -- (27,4) -- (29,0) -- (19,0) --
    (16,6) -- (10,0) -- cycle;
    \DrawSigMainLUAboveRight(17.66,12.66)(5,0);
    \DrawSigMainLUAboveLeft(24,0)(17.66,12.66);
    \DrawSigMainLUAboveLeft(36,0)(18,18);
  \end{scope}
  \begin{scope}[x=-.65\unitlength,shift={(-85,0)}]
    \coordinate (ll) at (8,8) ;    
    \coordinate (l) at (16,6) ;    
    \coordinate (r) at (27,4) ;    
    \coordinate (rr) at (35,6) ;    
    \fill[blue!20!white] (0,0) -- (ll) -- (13,9) -- (18,14) -- (17.66,12.66) -- (5,0) -- cycle;
    \fill[blue!20!white] (5,0) -- (17.66,12.66) -- (17.33,11.33) -- (14,8) -- (l) -- (10,0) -- cycle;
    \fill[blue!20!white] (19,0) -- (16,6) -- (19,8) -- (17.33,11.33) -- (17.66,12.66) -- (24,0) -- cycle;
    \fill[blue!20!white] (24,0) -- (17.66,12.66) -- (18,14) -- (22,6) -- (27,4) -- (29,0) -- cycle;
    \fill[blue!20!white] (31,0) -- (27,4) -- (28,7) -- (18,17) -- (18,18) -- (36,0) -- cycle;
    \fill[blue!20!white] (31,0) -- (27,4) -- (28,7) -- (18,17) -- (18,18) -- (36,0) -- cycle;
    \fill[blue!20!white] (36,0) -- (17,18) -- (19,18) -- (30,7) -- (rr) -- (41,0) -- cycle;
    \draw[densely dotted] (ll) -- (l) ;
    \draw[densely dotted] (l) -- (r) ;
    \draw[densely dotted] (r) -- (rr) ;
    \DrawSigMainLUAboveRight(17.66,12.66)(5,0) ;
    \DrawSigMainLUAboveLeft(24,0)(17.66,12.66) ;  
    \DrawSigMainLUAboveLeft(36,0)(18,18) ;
  \end{scope}
}{45}
\vspace*{-.25em}}
  \caption{The effect of shrinking (on right of each case).}
  \label{fig:shrink-role}
\end{figure}

The main challenge is that macro-signals and macro-collisions have non-zero width and might overlap and disturb one another.
\RefFigure{fig:shrink-role} illustrates this problem.
In \RefFig{fig:shrink-role:collision} all three present macro-signals rightfully interact whereas in \RefFig{fig:shrink-role:non-collision} the leftmost one should not participate.
In the same spirit, once a collision resolution is started, other signals should be far away enough not to intersect the zone needed for its resolution.

To cope with this, as soon as the borders of two macro-signals touch, both are shrunk in order to ``delay'' the macro-collision resolution (right part in Figs. \ref{fig:shrink-role:collision} and \ref{fig:shrink-role:non-collision}).
This delay is to be understood relative to the width of the input macro-signals: the time of the collision is not changed, but after shrinking the input signals, it becomes a larger multiple of the width of each input signal.
This delay is used to check which macro-signals exactly enter the macro-collision and to ensure that non-participating macro-signals are far away enough. 
This checking identifies macro-signals participating in the ongoing collision and ensures that no other signal may collide with control signals, not before the collision nor after a while after collision. 
This zone in which no other signal might enter is called the \emph{safety zone}.

If any constrain is not satisfied, the macro-collision aborts; nothing happens but the macro-signals have been shrunk and thus relatively spaced.
Later on, testing will be restarted as these thinner macro-signals touch again.
Eventually all correct macro-collisions will happen.

If all constraints are satisfied, the macro-collision is resolved.
This is done by gathering information of id's of all participating macro-signals and finding the appropriate collision rule.
After actual collision between \SigMainSome signals, according to the selected collision rule, macro-signals are replaced by new macro-signals representing output signals of the simulated SM.

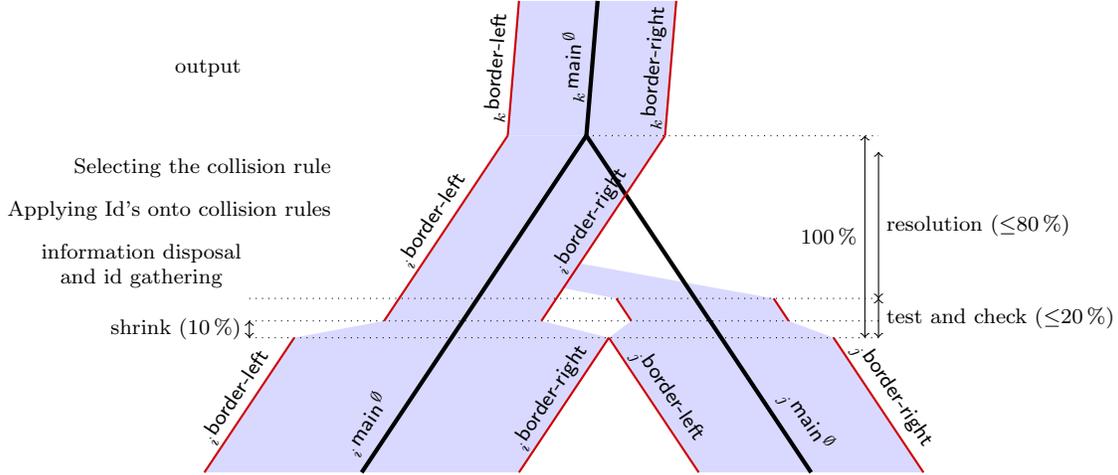
\begin{figure}[hbt]
  \centerline{\scriptsize\footnotesize\SetUnitlength{1em}\newcommand{\WidI}{7}\newcommand{\WidJ}{5}\newcommand{\XMin}{-15}\newcommand{\XMax}{15}\newcommand{\YShrink}{.75}\newcommand{\YDisposal}{1}\newcommand{\YOutput}{6}\newcommand{\BOT}{-15}\begin{tikzpicture}\path
      (0,0) \CoorNode{O}
      +(\XMin,0) \CoorNode{Ol} +(\XMax-2,0) \CoorNode{Or}
      +(-\WidI,0) \CoorNode{OLi} +(\WidI,0) \CoorNode{ORi}
      +(-\WidJ,0) \CoorNode{OLj} +(\WidJ,0) \CoorNode{ORj}
      (-10,\BOT) \CoorNode{I} +(-\WidI,0) \CoorNode{Li} +(\WidI,0) \CoorNode{Ri}
      (10,\BOT) \CoorNode{J} +(-\WidJ,0) \CoorNode{Lj} +(\WidJ,0) \CoorNode{Rj}
      ;
      \path[name path=oi] (O) -- (I) ;
      \path[name path=oj] (O) -- (J) ;
      \path[name path=oli] (Li) -- (OLi) ;
      \path[name path=ori] (Ri) -- (ORi) ;
      \path[name path=olj] (Lj) -- (OLj) ;
      \path[name path=orj] (Rj) -- (ORj) ;
      \SetIntersect{ori}{olj}{A}    
      \path let \p1=(A) in
      (\XMin,\y1) \CoorNode{Al}
      (\XMax-2,\y1) \CoorNode{Ar}
      (\XMin,\y1+\YShrink\unitlength) \CoorNode{Bl}
      (\XMax-2,\y1+\YShrink\unitlength) \CoorNode{Br}
      (\XMin,\y1+\YShrink\unitlength+\YDisposal\unitlength) \CoorNode{Cl}
      (\XMax-2,\y1+\YShrink\unitlength+\YDisposal\unitlength) \CoorNode{Cr} ;
      \path[name path=A] (Ar) -- (Al) ;
      \path[name path=B] (Br) -- (Bl) ;
      \path[name path=C] (Cr) -- (Cl) ;
      \SetIntersect{oli}{A}{ALi}
      \SetIntersect{orj}{A}{ARj}
      \SetIntersect{oi}{B}{Bi}
      \SetIntersect{oj}{B}{Bj}
      \SetIntersect{oi}{C}{Ci}
      \SetIntersect{oj}{C}{Cj}
      \path
      (O) +(-\WidI/2,0) \CoorNode{Li'}  +(\WidI/2,0) \CoorNode{Ri'} 
      ++(.5,\YOutput) \CoorNode{K} +(-\WidI/2,0) \CoorNode{Lk}  +(\WidI/2,0) \CoorNode{Rk} 
      (Bi) +(-\WidI/2,0) \CoorNode{BLi}  +(\WidI/2,0) \CoorNode{BRi} 
      (Bj) +(-\WidI/2,0) \CoorNode{BLj}  +(\WidI/2,0) \CoorNode{BRj} 
      (Cj) +(-\WidI/2,0) \CoorNode{CLj}  +(\WidI/2,0) \CoorNode{CRj} ; 
      \path (\XMin,\YOutput) \CoorNode{Dl} ;
      \fill[data-fill] (Li) -- (ALi) -- (BLi) -- (Li') -- (Ri') --(BRi) -- (A) -- (Ri) -- cycle;
      \fill[data-fill] (Lj) -- (A) -- (BLj) -- (CLj) -- ([shift={(1,1)}]Ci) -- ([shift={(1,2.2)}]Ci) -- (CRj) -- (BRj) -- (ARj) -- (Rj) -- cycle;
      \fill[data-fill] (Li') -- (Lk) -- (Lk)  -- (Rk) -- (Ri') -- cycle ;
      \DrawSigMainiLUAboveLeft(I)(O)
      \DrawSigMainjLUAboveRight(O)(J)
      \DrawSigMainkLUAbove(O)(K)
      \DrawSigBorderLeftkLUAbove(Lk)(Li')
      \DrawSigBorderRightkLUAbove(Rk)(Ri')
      \draw[dotted] (O) -- (Or) ;
      \draw[dotted] (Ar) -- (Al) ;
      \draw[dotted] (Br) -- (Bl) ;
      \draw[dotted] (Cr) -- (Cl) ;
      \DrawSigBorderLeftiLUAbove(Li)(ALi)
      \DrawSigBorderRightiLUAbove(Ri)(A)
      \DrawSigBorderLeftjLUAbove(Lj)(A)
      \DrawSigBorderRightjLUAbove(Rj)(ARj)
      \DrawSigBorderLeftiLUAbove(Li')(BLi)
      \DrawSigBorderRightiLUAbove(Ri')(BRi)
      \DrawSigBorderLeftj(CLj)(BLj)
      \DrawSigBorderRightj(CRj)(BRj)
      \path ($.5*(Dl)+.5*(Ol)$) node[left] {output} ;
      \path ($.2*(Cl)+.8*(Ol)$) ++(4,0) node[left] {Selecting the collision rule} ;
      \path ($.46*(Cl)+.54*(Ol)$) ++(4,0) node[left] {Applying Id's onto collision rules} ;
      \path ($.8*(Cl)+.2*(Ol)$) node[left] {\begin{tabular}[c]{@{}c@{}}
          information disposal\\ and id gathering
        \end{tabular}} ;
      \begin{scope}[<->]
        \draw (Al) -- node[left] {shrink (10\,\%)}  (Bl) ;
        \draw ([xshift=-1.5ex]Ar) -- node[left] {$100\,\%$} ([xshift=-1.5ex]Or) ;
        \draw (Ar) -- node[right] {test and check ($\leq$20\,\%)} (Cr);
        \draw (Cr) -- node[right] {resolution ($\leq$80\,\%)} ($.1*(Cr)+.9*(Or)$) ;
      \end{scope}
      \path[use as bounding box] (\XMin,\BOT) rectangle (\XMax,\YOutput)  ;
    \end{tikzpicture}}\caption{Phases of a successful macro-collision.}
  \label{fig:phases}
\end{figure}

The different phases and their relative duration in a successful macro-collision are presented in \RefFig{fig:phases} where percentages are taken relative to the duration from the collision of \SigBorderRighti and \SigBorderLeftj to the exact location of the collision, i.e. the meeting of \SigMaini and \SigMainj.
These proportions are arbitrary, the only condition is that the macro-collision resolution is started before the signals \SigBorderRighti and \SigBorderLeftj met again.
The duration of the shrinking phase (10\,\%) is fixed.
The duration of the test and check phase is at most 20\,\%, for success as well as failure.
It is ensured that aborting or disposal is carried out before any two present macro-signals meet again (and initiate a different macro-collision).

This way the constructed signal machine is able to simulate any signal machine with speeds included in a given set.
By varying this set, an intrinsically universal family of signal machines is obtained.

All definitions are gathered in \RefSec{sec:def}.
The encoding is detailed in \RefSec{sec:encoding}.
Macro-collision resolution is explained in \RefSec{sec:collisions}; the testing prior to it is found in \RefSec{sec:preparing}.
\RefSection{sec:example} provides some simulation examples.
Conclusion, remarks and perspectives are gathered in \RefSec{sec:conc}.

\section{Definitions}
\label{sec:def}

A signal machine regroups the definitions of its meta-signals and their dynamics: rewriting rules at collisions and constant speed in-between.

\begin{Definition}
  A \emph{signal machine} (SM) \AGCmachine is a triplet $(\AGCmetaSignalSet,\AGCspeedFun,\AGCruleSet)$ such that: 
  \AGCmetaSignalSet is a finite set of \emph{meta-signals};
  $\AGCspeedFun:\AGCmetaSignalSet\to\RealSet$ is the \emph{speed function} (each meta-signal has a constant speed); and
  \AGCruleSet is a finite set of \emph{collision rules} which are denoted by $\AGCruleIn\to\AGCruleOut$ where \AGCruleIn and \AGCruleOut are sets of meta-signals of distinct speeds. 
  Each \AGCruleIn must have at least two meta-signals.
  \AGCruleSet is deterministic: all \AGCruleIn are different.
\end{Definition}

Let $\AGCextendedValueSet$ be the set $\AGCmetaSignalSet \cup \AGCruleSet \cup \{\AGCextendedValueVoid\}$.
A \emph{(\AGCmachine-)configuration} \AGCconfiguration, is a map from \RealSet to \AGCextendedValueSet, that is from the points of the real line to either a meta-signal, a rule or the value \AGCextendedValueVoid (indicating that there is nothing there), with only finitely many non-\AGCextendedValueVoid locations in any configuration.

A signal machine evolution is defined in terms of dynamics.
If there is a signal of speed \AGCspeed at \SpaceCoordinateX, then after a duration \Duration its position is $\SpaceCoordinateX+\AGCspeed{\cdot}\Duration$, unless it enters a collision before.
At a collision, all incoming signals are instantly replaced according to rules by outgoing signals in the following configurations.
This is formalised as follows.

To simplify notations, the relation \emph{issued from}, $\AGCatOrOut\subset\AGCmetaSignalSet\times\AGCextendedValueSet$, is defined to be true only in the following cases:
\begin{itemize}
\item $\AGCmetaSignal\AGCatOrOut\AGCmetaSignal$, $\forall\AGCmetaSignal\in\AGCmetaSignalSet$ and
\item $\AGCmetaSignal\AGCatOrOut\AGCrule$, $\forall\AGCrule\in\AGCmetaSignalSet$ such that $\AGCmetaSignal\in\AGCruleOut$.
\end{itemize}
The relation $\AGCatOrOut$ means ``is equal to (some meta-signal) or belongs to the output of (a collision)''.

\begin{definition}[Dynamics]\label{def:dynamics}
  Considering a configuration \AGCconfiguration, the \emph{time to the next collision}, 
  $\AGCtimeToNextCollision(\AGCconfiguration)$, is equal to the minimum of the positive real numbers $d$ such that:
  \[
    \exists \AGCspacialPosition_1,\AGCspacialPosition_2 \in\RealSet,
    \exists \AGCmetaSignal_1, \AGCmetaSignal_2\in\AGCmetaSignalSet
    \left\{
      \begin{array}{@{\,}c@{\,}l@{}}
        & \AGCspacialPosition_1+d{\cdot}\AGCspeedFun(\AGCmetaSignal_1)= \AGCspacialPosition_2+d{\cdot}\AGCspeedFun(\AGCmetaSignal_2)\\
        \wedge & \AGCmetaSignal_1 \AGCatOrOut\AGCconfiguration(\AGCspacialPosition_1) \\
        \wedge & \AGCmetaSignal_2 \AGCatOrOut\AGCconfiguration(\AGCspacialPosition_2) 
      \end{array}
    \right.
    \enspace .
  \]
  It is $+\infty$ if there is no such a $d$.
  
  Let $\AGCconfiguration[\AGCtemporalPosition]$ be the configuration at time $\AGCtemporalPosition$.
  
  For \AGCtemporalPositionOther ($\AGCtemporalPosition<\AGCtemporalPositionOther<\AGCtemporalPosition{+}\AGCtimeToNextCollision(\AGCconfiguration[\AGCtemporalPosition])$), the configuration at $\AGCtemporalPositionOther$ is defined as follows.
  First, signals are set according to $\AGCconfiguration[\AGCtemporalPositionOther](\AGCspacialPosition)=\AGCmetaSignal$ iff
  $\AGCmetaSignal\AGCatOrOut\AGCconfiguration[\AGCtemporalPosition]\left(\AGCspacialPosition+(\AGCtemporalPosition{-}\AGCtemporalPositionOther){\cdot}\AGCspeedFun(\AGCmetaSignal)\right)$.
  There is no collision to set ($\AGCtemporalPositionOther$ is before the next collision) thus no ambiguity.
  The rest is \AGCextendedValueVoid.

  For the configuration at $\AGCtemporalPosition+\AGCtimeToNextCollision(\AGCconfiguration[\AGCtemporalPosition])$, collisions are set first: 
  $\AGCconfiguration[{\AGCtemporalPosition+\AGCtimeToNextCollision(\AGCconfiguration[\AGCtemporalPosition])}](\AGCspacialPosition)=\AGCrule$
  where
  $\AGCruleIn=\{
  \AGCmetaSignal\in\AGCmetaSignalSet\,|\,\AGCmetaSignal\AGCatOrOut\AGCconfiguration[\AGCtemporalPosition](\AGCspacialPosition-\AGCtimeToNextCollision(\AGCconfiguration[\AGCtemporalPosition]){\cdot}\AGCspeedFun(\AGCmetaSignal))\}$.
  Then meta-signals are set (with above condition) where there is not already a collision, and finally \AGCextendedValueVoid everywhere else.
\end{definition}

The dynamics is uniform in both space and time.
Since configurations are finite, the infimum is non-zero and is reached.

A \emph{space-time diagram} of \AGCmachine is the aggregation of configurations as times elapses, that is, a function from $\RealSet^+$ into the set of configurations of \AGCmachine.
It forms a two dimensional picture (time is always elapsing upwards in the figures).
It is denoted \AGCspaceTimeDiagramSM or $(\AGCmachine,\AGCconfiguration)$ to emphasis on the signal machine and the initial configuration.

\subsection{Simulations among Signal Machines}

In this subsection, we define formally what the sentence ``signal machine $\AGCmachine$ simulates signal machine $\AGCmachineOther$'' entails.

\paragraph{Local functions on configurations}
First, let us define how to recover a configuration of a signal machine from a configuration of another one\,--a putative simulator.

Let \AGCmachine be a signal machine, and \AGCconfigurationSet be the set of all its configurations.
Let $\AGCconfigurationSetZeroNonEmpty = \{ \AGCconfiguration \in \AGCconfigurationSet | \AGCconfiguration(0) \neq \AGCextendedValueVoid \}$.
Let $\AGCconfiguration \in \AGCconfigurationSetZeroNonEmpty$, $\SymbConfFun(\AGCconfiguration)$ is the bi-infinite word defined by: for all $n\in\IntegerSet$, $\SymbConfFun(\AGCconfiguration)_n$ is the $n$-th non-$\AGCextendedValueVoid$ value in \AGCconfiguration, counting from position $0$.
Pose $\SymbConfFun(\AGCconfiguration)_n = \SymbUndefined$ if $\AGCconfiguration$ does not have enough non-$\AGCextendedValueVoid$ values.

Let $f$ be a function from $\AGCconfigurationSet$ to some set $F$.
The function $f$ is \emph{local} if the followings hold:
\begin{itemize}
\item there is $f_{\AGCextendedValueVoid}\in F$ such that when $\AGCconfiguration(0) = \AGCextendedValueVoid$, $f(\AGCconfiguration) = f_{\AGCextendedValueVoid}$,
\item there is a function $\hat{f}$ on bi-infinite words such that $f(\AGCconfiguration) = \hat{f}(\SymbConfFun(\AGCconfiguration))$ when $c\in\AGCconfigurationSetZeroNonEmpty$, and
\item there is $n \in \NaturalSet$ such that $\hat{f}$ only depends on the $n$ symbols around $0$ in its input word.
\end{itemize}
In the analogy with Cellular Automata, local functions will play a role similar to that of grouping functions.

\paragraph{Signal machine simulation}

Let $\AGCmachine$ and \AGCmachineOther be two signal machines; subscripts are used to identify the machine an elements belongs to.
Let \AGCsimGroup be a local function from the configurations of \AGCmachineOther into $\AGCextendedValueSet_{\AGCmachine}$ such that $\AGCsimGroup_{\AGCextendedValueVoid}=\AGCextendedValueVoid_{\AGCmachine}$.
For a configuration $\AGCconfigurationSMotherSymbol$ of \AGCmachineOther, we define the configuration $\AGCsimGroupConf(\AGCconfigurationSMotherSymbol)$ by $\AGCsimGroupConf(\AGCconfigurationSMotherSymbol)(x) = \AGCsimGroup(y \mapsto \AGCconfigurationSMotherSymbol(y + x))$.
Schematically, \AGCsimGroup interprets sequences of signal centered around position $0$ and $\AGCsimGroupConf$ uses it (composed with translations) to rebuild whole configurations of the simulated signal machine.

A \emph{representation} function is a function \AGCsimRepr from $(\AGCmetaSignalSet_{\AGCmachine} \cup \AGCruleSet_{\AGCmachine})$ to $\AGCconfigurationSet[\AGCmachineOther]$, such that the support of $\AGCsimRepr(\cdot)$ is always included in the interval $[-1, 1]$.
One such function will be used for encoding the signals and collisions making up the initial configuration.
This function yields a configuration for each signal or collision.
For a configuration $\AGCconfigurationSMSymbol$, let $d_{\AGCconfigurationSMSymbol}(x) = min \{ |d| | \AGCconfigurationSMSymbol(x + d) \neq \AGCextendedValueVoid \wedge d \neq 0\}$.
Then, we note $\AGCsimReprConf(\AGCconfigurationSMSymbol)$ for the union of the $\AGCsimRepr(\AGCconfigurationSMSymbol(x), d_{\AGCconfigurationSMSymbol}(x))$ for $\AGCconfigurationSMSymbol(x) \neq \AGCextendedValueVoid$, each translated by $x$, and scaled by $d_{\AGCconfigurationSMSymbol}(x) / 3$. Note that because of this translations and scaling, there are no collisions between the different $\AGCsimRepr(\cdot)$.

\begin{definition}[Simulation between signal machines]
  $\AGCmachine = (\AGCmetaSignalSet_{\AGCmachine}, \AGCspeedFun_{\AGCmachine}, \AGCruleSet_{\AGCmachine})$ and $\AGCmachineOther = (\AGCmetaSignalSet_{\AGCmachineOther}, \AGCspeedFun_{\AGCmachineOther}, \AGCruleSet_{\AGCmachineOther})$ be two signal machines.
  Let \AGCsimGroup be a local function from the configurations of \AGCmachineOther into $\AGCextendedValueSet_{\AGCmachine}$.
  Let \AGCsimRepr be a representation function.

  For a diagram $\AGCspaceTimeDiagramSM = (\AGCmachine, \AGCconfigurationSM)$ of $\AGCmachine$, let $\AGCspaceTimeDiagramSMother_{\AGCconfiguration} = (\AGCmachineOther, \AGCsimReprConf(\AGCconfigurationSM))$ be the diagram of $\AGCmachineOther$ on initial configuration \AGCsimReprConf(\AGCconfigurationSM).
  Then \AGCmachineOther \emph{simulates} \AGCmachine if:
  \begin{equation*}
    \forall \AGCconfigurationSM, \forall t \geq 0, \AGCsimGroupConf(\AGCspaceTimeDiagramSMother_{\AGCconfigurationSM}(t)) = \AGCspaceTimeDiagramSM(t)
    \enspace.
  \end{equation*}
\end{definition}

The reader familiar with simulations and intrinsic universality in Cellular Automata may wonder what purpose the \AGCsimRepr function serves.
In Cellular Automata, there is a time grouping as well as a space grouping: when some CA $B$ simulates a CA $A$, there is a $k$ such that for all $i$, the configuration of $B$ at time $k \cdot i$ represents the configuration of $A$ at time $i$, and indeed, any configuration of $B$ seen at a time multiple of $k$ can be used as an initial configuration for simulating the corresponding configuration of $A$.
In a signal machine, having such a periodicity would require the spacing of any auxiliary signals in \AGCconfigurationSMotherSymbol to be uniform.
This in turn would require that there be a positive lower bound to the distance between two signals, uniformly over configurations of \AGCmachine, which cannot be the case.
Instead, a simulator uses \AGCsimRepr to get an encoding of each signal and collision at time $0$.
A similar distinction is necessary in the definition of simulation for self-assembling systems \cite{doty+lutz+patitz+schweller+summers+woods12focs}, for the same reason of asynchronicity.

\begin{lemma}
  For any signal machine $M$, $M$ simulates $M$.
\end{lemma}

\begin{proof}
  Let $\AGCsimGroup(\AGCconfigurationSMotherSymbol)= \AGCconfigurationSMotherSymbol(0)$.
  For any $x$ and $\AGCconfigurationSMotherSymbol$ we have $\AGCsimGroupConf(\AGCconfigurationSMotherSymbol)(x) = \AGCsimGroup(y \mapsto \AGCconfigurationSMotherSymbol(y + x))$.
  So we have $\AGCsimGroupConf(\AGCconfigurationSMotherSymbol)(x) = (y \mapsto \AGCconfigurationSMotherSymbol(y + x))(0) = \AGCconfigurationSMotherSymbol(x)$.
  Then $\AGCsimGroupConf(\AGCconfigurationSMotherSymbol)= \AGCconfigurationSMotherSymbol$.
  For any $x$, if $\AGCconfigurationSMSymbol(x)=\mu\neq \AGCextendedValueVoid $, then $\AGCsimReprConf(\AGCconfigurationSM)(x)=\AGCsimRepr (\mu)(0) =\mu = \AGCconfigurationSMSymbol(x)$, otherwise $\AGCsimReprConf(\AGCconfigurationSM)(x)=\AGCextendedValueVoid$.
  So $\AGCsimReprConf(\AGCconfigurationSM)(x) = \AGCconfigurationSMSymbol(x)$.
  
  Let $\AGCspaceTimeDiagramSM_{\AGCconfigurationSMSymbol} = (\AGCmachine, \AGCsimReprConf(\AGCconfigurationSM))$.
  Then for each $t\geq 0$ and each $\AGCconfigurationSMSymbol$ we have $\AGCsimGroupConf(\AGCspaceTimeDiagramSM_{\AGCconfigurationSMSymbol}(t))=\AGCspaceTimeDiagramSM_{\AGCconfigurationSMSymbol}(t)= (\AGCmachine, \AGCsimReprConf(\AGCconfigurationSM))=(\AGCmachine, \AGCconfigurationSMSymbol)$. 
\end{proof}

\begin{example}
  Let $\AGCmachine = (\AGCmetaSignalSet_{\AGCmachine}, \AGCspeedFun_{\AGCmachine}, \AGCruleSet_{\AGCmachine})$ and $\AGCmachineOther = (\AGCmetaSignalSet_{\AGCmachineOther}, \AGCspeedFun_{\AGCmachineOther}, \AGCruleSet_{\AGCmachineOther})$ be two signal machine such that  $\AGCmetaSignalSet_{\AGCmachine}= (\{ \mu_a\ , \mu_b\}, \{S(\mu_a)=-1, S(\mu_b)=1\}, R_A: \{\rho_c:\{\mu_a , \mu_b\}\rightarrow\{\mu_b\}\})$ and $\AGCmetaSignalSet_{\AGCmachineOther}= (\{ \mu_1\ , \mu_2 , \mu_3\}, \{S(\mu_1)=-1, S(\mu_2)=S(\mu_3)=1, \},R_B: \{\rho_4:\{\mu_1 , \mu_2\}\rightarrow\{\mu_3\}\}, \{\rho_5:\{\mu_1 , \mu_3\}\rightarrow\{\mu_2\}\})$.
  We want to show that $\AGCmachine $ simulates $\AGCmachineOther$.
  We define two functions \AGCsimGroup and \AGCsimRepr. 

  We want that for all diagram $\AGCspaceTimeDiagramSMother$ of $\AGCmachineOther$ , $\AGCsimGroupConf(\AGCspaceTimeDiagramSMother)(x,t)= \mu_a$ if $ \AGCspaceTimeDiagramSMother(x,t)=\mu_1$, 
  $\AGCsimGroupConf(\AGCspaceTimeDiagramSMother)(x,t)= \mu_b$ if $ \AGCspaceTimeDiagramSMother(x,t)=\mu_2 $ or $ \mu_3$, 
  and
  $\AGCsimGroupConf(\AGCspaceTimeDiagramSMother)(x,t)= \rho_c$ if $ \AGCspaceTimeDiagramSMother(x,t)\in \{\rho_4, \rho_5\}$.

  We take $\AGCsimRepr(\mu_a)(x)= \mu_1$ if $x=0$, otherwise $\AGCextendedValueVoid$; $\AGCsimRepr(\mu_b) (x)= \mu_2$ if $x=0$, otherwise $\AGCextendedValueVoid$; $\AGCsimRepr(\rho_c) (x)= \rho_4$ if $x=0$, otherwise $\AGCextendedValueVoid$.

  Also, $\AGCsimGroup(\AGCconfigurationSMotherSymbol)=\AGCextendedValueVoid$ if $\AGCconfigurationSMotherSymbol(0)=\AGCextendedValueVoid$,
  $\AGCsimGroup(\AGCconfigurationSMotherSymbol)=\mu_a$ if $\AGCconfigurationSMotherSymbol(0)=\mu_1$, 
  $\AGCsimGroup(\AGCconfigurationSMotherSymbol)=\mu_b$ if $\AGCconfigurationSMotherSymbol(0)=\mu_2$,
  $\AGCsimGroup(\AGCconfigurationSMotherSymbol)=\mu_b$ if $\AGCconfigurationSMotherSymbol(0)=\mu_3$,
  $\AGCsimGroup(\AGCconfigurationSMotherSymbol)=\rho_c$ if $\AGCconfigurationSMotherSymbol(0)=\rho_4$ and
  $\AGCsimGroup(\AGCconfigurationSMotherSymbol)=\rho_c$ if $\AGCconfigurationSMotherSymbol(0)=\rho_4$.

  If  $\AGCspaceTimeDiagramSMother$ is a space-time diagram of 
  $\AGCmachineOther $, then $\AGCsimGroupConf(\AGCspaceTimeDiagramSMother)$ is a space-time diagram of $\AGCmachine$, because the collisions are preserved by $\AGCsimGroupConf$. 

  Also, for each $s$, $\AGCsimGroupConf(\AGCspaceTimeDiagramSMother)$ is a space-time diagram of $\AGCmachine$ and has initial configuration $s$.

  So $\AGCsimGroupConf(\AGCspaceTimeDiagramSMother_s)=\AGCspaceTimeDiagramSM_s $.
\end{example}

\subsection{Intrinsically universal machines}

\begin{definition}[\SpeedSet-(intrinsic) universality]
  Let $\SpeedSet$ be a set of speeds.
  A signal machine $\UniversalMSSpeed = (\UMetaSignalSet, \URuleSet, \USpeedSet)$, is \emph{\SpeedSet-universal} if, for any \AGCmachine with speeds in \SpeedSet, there is a local function $\AGCsimGroup_{\AGCmachine}$ and a representation function $\AGCsimRepr_{\AGCmachine}$ such that \UniversalMSSpeed simulates \AGCmachine through $\AGCsimGroup_{\AGCmachine}$ and $\AGCsimRepr_{\AGCmachine}$.
\end{definition}

For the rest of the paper, we fix $\SpeedSet = \{\BaseSpeed{1}, \ldots, \BaseSpeed{\SpeedSetNumber}\} \subset \RealSet$ a set of speeds, with $\BaseSpeed{1} < \ldots < \BaseSpeed{\SpeedSetNumber}$ and provide the construction of an \SpeedSet-universal machine \UniversalMSSpeed.
Its set of speeds, its set of meta-signals and its rules only depend on \SpeedSet.
The functions $\AGCsimGroup_{\AGCmachine}$ and \AGCsimRepr will be defined along with the construction of \UniversalMSSpeed.
By default, $\AGCsimGroup_{\AGCmachine}(x)$ is undefined; in the following, we will list the cases where $\AGCsimGroup_{\AGCmachine}(x)$ is defined.
The number of signals that $\AGCsimGroup_{\AGCmachine}$ actually reads is bounded for each \AGCmachine.
The definition of \AGCsimRepr will be given in \RefSec{sec:encoding_repr}.

In the rest of the paper, not all collision rules of intrinsically undefined SM are explicitly defined.
They can be found online in the simulation.
For a set of meta-signals $s$ which is not explicitly defined as the input of a collision rule, a collision rule $\AGCrule_s$ is implicitly defined thus:
\begin{itemize}
\item if $|s| = 2$, then $\AGCruleIn\!\!_s = \AGCruleOut\!\!_s = s$
\item if there is a unique explicit collision rule \AGCrule such that $\AGCruleIn \subset s$, then $\AGCruleIn\!\!_s = s$, and $\AGCruleOut\!\!_s = \AGCruleOut \cup (s \setminus \AGCruleIn)$,
\item if, for all set of rules $\{ \AGCrule_1, \ldots, \AGCrule_k \}$ such that $\bigcup_{1 \leq i \leq k} \AGCruleIn\!\!_i = s$, $\bigcup_{1 \leq i \leq k} \AGCruleOut\!\!_i$ is the same set $\AGCruleOut$, then $\AGCruleIn\!\!_s = s$ and $\AGCruleOut\!\!_s = \AGCruleOut$,
\item otherwise $\AGCrule_s$ is undefined.
\end{itemize}
In other words, unlisted rules with two inputs are blank, they output the same signal as in input.
Unlisted rules can also be the ``superposition'' of one meta-signal in both input and output of a defined collision rule (this cover the case when a defined collision happens exactly on some irrelevant signal).
It might also happen that two or more unrelated collisions happen at the same position and share some input and output signal or that two consecutive collisions become synchronous.
The definition of the corresponding collision rules is straightforward from the defined collision rules and can be considered as limit cases, in the sense that a small perturbation of the input configuration gets rid of them.
They are not listed to avoid unnecessary listings of collision rules.

\section{Encoding of Signals and Signal Machines}
\label{sec:encoding}

\subsection{Meta-Signal Notation}

In the rest of the paper the names of \UniversalMSSpeed-meta-signals are organised around a \AGCsigPolice{base} name---in \AGCsigPolice{sans-serif font}---decorated with parameters:
\begin{equation*}
  \IdBaseSpeedOtherSpeedParamParam{base}{a}{}{c}{d}
  \text{\qquad
    or 
    \qquad}
  \IdBaseSpeedOtherSpeedParamParam{base}{}{b}{c}{d}
  \enspace .
\end{equation*}

A signal noted $\IdBaseSpeedOtherSpeedParamParam{base}{a}{}{c}{d}$, is instantiated for speed $\BaseSpeed{\text{\AGCsigPolice{a}}} \in \SpeedSet$, and its actual speed is $\BaseSpeed{\text{\AGCsigPolice{a}}}$.
A signal noted $\IdBaseSpeedOtherSpeedParamParam{base}{}{b}{c}{d}$, is instantiated for speed $\BaseSpeed{\text{\AGCsigPolice{b}}} \in \SpeedSet$, but its actual speed is not $\BaseSpeed{\text{\AGCsigPolice{b}}}$, but some other speed, generally computed from \AGCsigPolice{b}, \AGCsigPolice{c} and \AGCsigPolice{d}.
We use $\IdBaseSpeedOtherSpeedParamParam{base}{\SpeedIndexSome}{}{c}{d}$ or $\IdBaseSpeedOtherSpeedParamParam{base}{}{\SpeedIndexSome}{c}{d}$ when the actual value of the parameter is not relevant.
A signal noted $\IdBaseSpeedOtherSpeedParamParam{base}{}{}{c}{d}$ belongs to a family that is not parametrized by speeds in \SpeedSet.
For example, \SigMainOne and \SigMainTwo are different meta-signals of respective speeds \BaseSpeed{1} and \BaseSpeed{2} but with the same meaning, with respect to \BaseSpeed{1} and \BaseSpeed{2} respectively.

Parameters \AGCsigPolice{c} and \AGCsigPolice{d} are used to hold a finite amount of information.

\subsection{Encoding of signals and the \AGCsimRepr and \AGCsimRepr and \AGCsimGroup functions}
\label{sec:encoding_repr}

Let \GenericMSSpeedNbrik be any meta-signal of \AGCmachine.
The integer \SpeedIndexI indicates that the speed of \GenericMSSpeedNbrik speed is \BaseSpeed{\SpeedIndexI} and \MetaSignalIndexK ($0<\MetaSignalIndexK$) is its index in some numbering of the meta-signals of \AGCmachine of speed \BaseSpeed{\SpeedIndexI}.

We define $\AGCsimRepr(\GenericMSSpeedNbrik)$ as a so-called \emph{macro-signal}, i.e. a configuration with finite support, delimited by two parallel signals, here \SigBorderLefti and \SigBorderRighti. The space used by a macro-signal is called its \emph{support zone.}

The configuration $\AGCsimRepr(\GenericMSSpeedNbrik)$ contains the following signals in the following order:
\begin{DisplayRuleInP}
  \SigBorderLefti \quad (\SigIDi)^{\MetaSignalIndexK} \quad \SigMaini \quad _{\SpeedIndexI}{<}\text{rules encoding}{>} \quad \SigBorderRighti
  \enspace .
\end{DisplayRuleInP}
The left part is thus made of $\MetaSignalIndexK$ parallel signals of speed \BaseSpeed{i} encoding \MetaSignalIndexK in unary.

The relative position of signals of $\AGCsimRepr(\GenericMSSpeedNbrik)$ are defined with \SigBorderLefti at $-1 - \frac{\BaseSpeed{\SpeedIndexI} + \SpeedMaxAbs}{\SpeedRapid}$, \SigBorderRighti at $1 - \frac{\BaseSpeed{\SpeedIndexI} + \SpeedMaxAbs}{\SpeedRapid}$, \SigMaini at $0$, and other signals regularly spaced between them (before the rescaling and translation done by $\AGCsimReprConf$; the value of \AGCsimStartingMSwidth is given in section~\ref{sec:collisions}); 
\SpeedMaxAbs is the maximum absolute value of speeds of \SpeedSet and \SpeedRapid is a very fast speed defined in section \ref{sub:apply-in-rules}.
The reason for this choice is explained in section \ref{sub:output}.

The value of $\AGCsimRepr$ on collisions (rather than signals) is defined in \RefSec{sub:output}.

\smallskip

All the rules of \AGCmachine are encoded between \SigMaini and \SigBorderRighti, one after the other.
Each rule (to be read from the right) is encoded as a \emph{then-part} followed by an \emph{if-part}:
\begin{DisplayRuleNoI}
  \SigRuleBoundi
  \quad 
  (\SigTheniOne)^* \ \cdots \ (\SigTheniMax)^*
  \quad
  \SigRuleMiddlei
  \quad
  (\SigIfiOne)^* \ \cdots \  (\SigIfiMax)^*
  \quad 
  \SigRuleBoundi
  \enspace .
\end{DisplayRuleNoI}

Let $\AGCruleMax$ be the collision rule of \AGCmachine where $\AGCruleMaxIn$ is the set of all meta-signals with maximum id for each speed.
The encoding of \AGCruleMax has to appear in the rules.
This is a technicality to ensures a correct decoding later one.

The number of \SigIfil signals between \SigRuleMiddlei and \SigRuleBoundi corresponds to the index of the \AGCmachine-meta-signal of speed \BaseSpeed{\SpeedIndexL} which is expected as input by this rule (or zero for no \BaseSpeed{\SpeedIndexL}-speed meta-signal). 
Then, the number of \SigThenil between \SigRuleBoundi and \SigRuleMiddlei corresponds to the index of the \AGCmachine-meta-signal of speed \BaseSpeed{\SpeedIndexL} which is output by this rule.
\RefFigure{fig:encoding-rule} provides an example of a rule encoding. 

\begin{figure}[hbt]
  \centering\scriptsize\SetUnitlength{3.4em}\newcommand{\TOP}{4}
  \begin{tikzpicture}[y=.45\unitlength]
    \DrawSigRuleBoundiLUAbove(0,0)(1,\TOP) ;
    \DrawSigTheniTwoLUAbove(1,0)(2,\TOP) ;
    \DrawSigTheniTwoLUAbove(1.5,0)(2.5,\TOP) ;
    \DrawSigTheniTwoLUAbove(2,0)(3,\TOP) ;
    \DrawSigTheniFourLUAbove(3,0)(4,\TOP) ;
    \DrawSigRuleMiddleiLUAbove(4,0)(5,\TOP) ;
    \DrawSigIfiThreeLUAbove(5,0)(6,\TOP) ;
    \begin{scope}[shift={(1,0)}]
      \DrawSigIfiSevenLUAbove(5,0)(6,\TOP) ;
      \DrawSigIfiSevenLUAbove(5.5,0)(6.5,\TOP) ;
      \DrawSigIfiSevenLUAbove(6,0)(7,\TOP) ;
      \DrawSigIfiSevenLUAbove(6.5,0)(7.5,\TOP) ;
      \DrawSigIfiEightLUAbove(7.5,0)(8.5,\TOP) ;
      \DrawSigIfiEightLUAbove(8,0)(9,\TOP) ;
      \DrawSigIfiEightLUAbove(8.5,0)(9.5,\TOP) ;
      \DrawSigIfiEightLUAbove(9,0)(10,\TOP) ;
      \DrawSigIfiEightLUAbove(9.5,0)(10.5,\TOP) ;
      \DrawSigRuleBoundiLUAbove(10.5,0)(11.5,\TOP) ;
    \end{scope}\end{tikzpicture}\caption{Encoding of the rule $\{\,
    \GenericMSSpeedNbr{3}{1},\,
    \GenericMSSpeedNbr{7}{4},\,
    \GenericMSSpeedNbr{8}{5}\,
    \}\rightarrow\{\,
    \GenericMSSpeedNbr{2}{3},\,
    \GenericMSSpeedNbr{4}{1}\,
    \}$ in the direction \SpeedIndexI.}
  \label{fig:encoding-rule}
\end{figure}

All the needed meta-signals are defined in \RefFig{fig:ms:encod}.
Later in the construction, the empty set of \SigMaini is replaced by a subset of \IntegerInterval{1}{\SpeedSetNumber} to store the directions to output after a collision.

\begin{figure}[hbt]
  \centering
  \begin{MSlist}
    \MSForAllSpeedI{\SigBorderLefti}\MSForAllSpeedI{\SigIDi}\MSForAllSpeedIE{\SigMainiE}\MSForAllSpeedI{\SigBorderRighti}\end{MSlist}\qquad\begin{MSlist}
    \MSForAllSpeedI{\SigRuleBoundi}\MSForAllSpeedIL{\SigThenil}\MSForAllSpeedI{\SigRuleMiddlei}\MSForAllSpeedIL{\IdIf[\SpeedIndexI]{\SpeedIndexL}}\end{MSlist}
  \caption{Meta-signals for encoding.}
  \label{fig:ms:encod}
\end{figure}

Let $\AGCsimMaxSignalsInMacroSignal$ be the maximum number of signals in $\AGCsimRepr(\GenericMSSpeedNbrik)$ for $\GenericMSSpeedNbrik$ a signal of \AGCmachine.
The function $\AGCsimGroup$ looks at most $\SpeedSetNumber . (\AGCsimMaxSignalsInMacroSignal + 2)$ non-\AGCextendedValueVoid values on each side.
Except for the symbol at the center of the configuration (i.e. \AGCconfiguration(0)), whenever a collision rule is encountered, \AGCsimGroup virtually replaces by its input signals, in the reverse order of their speeds: it effectively simulates the configuration at time $t - \epsilon$ for small enough $\epsilon$.
Thus, from now on, we define \AGCsimGroup according to signals only (except for $\AGCconfiguration(0)$).

We now define its value on some configurations, according to this section.
This definition will be completed in later sections, as more meta-signals and collisions of \UniversalMSSpeed are defined.
First, $\AGCsimGroup(\AGCconfiguration)$ is defined to be $\GenericMSSpeedNbrik$ if $\AGCconfiguration(0)$ is \SigMaini, and the closest signals to $0$ are in a configuration that is compatible with the neighbourhood of $main$ in $\AGCsimRepr(\GenericMSSpeedNbrik)$.
This only depends on the $\AGCsimMaxSignalsInMacroSignal$ signals closest to $0$ on each side.
When this is the case, we say that the configuration is \emph{clean} at $0$.
A configuration $\AGCconfiguration$ is clean at position $x$ if its translation by $-x$ is clean at $0$.
For a configuration \AGCconfiguration which is clean at $x$, we define $\AGCsimMSwidth(\AGCconfiguration, x)$ to be width of the macro-signal at $x$, that is the distance between the signals \SigBorderLeft \SigBorderRight closest to $0$.

\section{Macro-Collision Resolution}
\label{sec:collisions}

In the rest of the paper, we define a set of meta-signals and rules to deal with collisions in $\AGCmachine$. The simulation of a collision has two phases: a check phase, which is presented in \RefSec{sec:preparing}, and a resolution phase which is presented in this section.
The simplest space-time diagrams with at least one collision are the ones with exactly one collision, and all signals in the starting configuration are inputs of that collision.
This section presents the sufficient machinery for dealing with this case, and \RefSec{sec:preparing} completes it in order to be able to deal with a fully general diagram.

Let \AGCrule be a collision rule of $\AGCmachine$, let $\SpeedIndexJ=i_0  < i_1 < \ldots < i_{|\AGCruleIn|-1} = \SpeedIndexI$ be the integers such that theses are exactly the indices of the speeds in \AGCruleIn.
Let $\AGCconfigurationSMSymbol$ be a configuration of $\AGCmachine$ whose signals are exactly one of each meta-signal in $\AGCruleIn$ and the positions $x_{i_0} < x_{i_1}, \ldots < x_{i_{|\AGCruleIn|-1}}$ of these signals are such that they all meet at some point $(\AGCspacialPosition, \AGCtemporalPosition)$.

Let $\AGCconfigurationSMotherSymbol'$ be a configuration which is clean at every $x_{k}$, and $\AGCsimStartingMSwidth = max(\{\AGCsimMSwidth(\AGCconfigurationSMSymbol, x_k) | k \in \{ x_i, \ldots, x_j \} \}))$.
Define $\AGCconfigurationSMotherSymbol$ to be $\AGCconfigurationSMotherSymbol'$, with an additional signal \SigCheckOKij at position $x_c = (x_{i} + x_{i_1}) / 2$.
We say that $\AGCconfigurationSMotherSymbol$ is a \emph{\AGCsimStartingMSwidth-width checked configuration} for $\AGCconfigurationSMSymbol$.
The $\SigCheckOKij$ signal acts as a witness that the configuration is locally good.
This configuration must coincide with $\AGCconfigurationSMotherSymbol$ on a wide enough region around $x$ --and thus, also, for a long enough time.
The parameter \AGCsimStartingMSwidth must additionally be small enough with respect to $t$, as will be defined in \RefSec{sec:preparing}.

In the rest of this section, we give subsets $M_{\UniversalMSSpeedCheckedColl}$ and $R_{\UniversalMSSpeedCheckedColl}$ of signals and collision rules of $\UniversalMSSpeed$ (depending only on \SpeedSet).
The machine \UniversalMSSpeedCheckedColl ensures that, if \AGCsimStartingMSwidth is small enough, there is
\AGCsimOutputDelay with $ \AGCsimOutputDelay < K.\AGCsimStartingMSwidth$ for a fixed $0<K$ and
$t + \AGCsimOutputDelay < t'$, the configuration
$\AGCspaceTimeDiagramSMother(t')$ is such that
$\AGCsimGroupConf(\AGCspaceTimeDiagramSMother(t')) = \AGCspaceTimeDiagramSM(t')$, and for any position $x$ such that
$\AGCspaceTimeDiagramSMother(t')(x) \neq \AGCextendedValueVoid$,
$\AGCspaceTimeDiagramSMother(t')$ is clean at $x$ (and has width
$\AGCsimMSwidth(x_i)$).

\subsection{Useless Information Disposal and Id Gathering}
\label{sub:disposal+gathering}

We describe the signals and collision rules of \UniversalMSSpeedCheckedColl through its behaviour on \AGCsimStartingMSwidth-width checked configurations. The signals and rules are then listed in full.

The list of rules of the leftmost macro-signal is used to find the corresponding rule to apply.
All \SigID's are sent onto this list to operate the rule selection.
The rule lists in other macro-signals are just discarded.

\begin{figure}[hbt]
  \centering
  \SubFigure[schema\label{fig:removing-useless:schema}]{\footnotesize\scriptsize\SetUnitlength{2.0em}\newcommand{\TOP}{-5.6}\newcommand{\HalfWidth}{3}\newcommand{\ZigSpeed}{7}\scalebox{.9}{\begin{tikzpicture}[y=\unitlength]
        \ifDebugPicture
        \clip (-10.4-\HalfWidth,-10.2) rectangle (10+\HalfWidth,\TOP+1);
        \else
        \clip (-10.4-\HalfWidth+.5,-10.2) rectangle (10+\HalfWidth,\TOP+.5);
        \fi
        \path (\TOP,\TOP) \CoorNode{Oi};
        \path (0.2,\TOP) \CoorNode{Ok};
        \path (-\TOP,\TOP) \CoorNode{Oj};
        \path (-10,-10) \CoorNode{Mi};
        \path (0,-10) \CoorNode{Mk};
        \path (10,-10) \CoorNode{Mj};
        \path (Mi) ++(-\HalfWidth,0) \CoorNode{MLi} ;
        \path (Mi) ++(\HalfWidth,0) \CoorNode{MRi} ;
        \path (Oi) ++(-\HalfWidth,0) \CoorNode{OLi} ;
        \path (Oi) ++(\HalfWidth,0) \CoorNode{ORi} ;
        \path (Mk) ++(-\HalfWidth,0) \CoorNode{MLk} ;
        \path (Mk) ++(\HalfWidth,0) \CoorNode{MRk} ;
        \path (Ok) ++(-\HalfWidth,0) \CoorNode{OLk} ;
        \path (Ok) ++(\HalfWidth,0) \CoorNode{ORk} ;
        \path (Mj) ++(-\HalfWidth,0) \CoorNode{MLj} ;
        \path (Mj) ++(\HalfWidth,0) \CoorNode{MRj} ;
        \path (Oj) ++(-\HalfWidth,0) \CoorNode{OLj} ;
        \path (Oj) ++(\HalfWidth,0) \CoorNode{ORj} ;
        \path (-3.8,-10) \CoorNode{z0};
        \path[name path=zig] (z0) +(-3*\ZigSpeed,3) -- (z0) ;
        \path[name path=MLiS] (MLi) -- (OLi) ;
        \path[name intersections={of=zig and MLiS}]  (intersection-1) coordinate (z1) ;
        \path[name path=zag] (z1) +(3*\ZigSpeed,3) -- (z1) ;
        \path[name path=MRjS] (MRj) -- (ORj) ;
        \path[name intersections={of=zag and MRjS}]  (intersection-1) coordinate (z2) ;
        \path[name path=MLkS] (MLk) -- (OLk) ;
        \path[name intersections={of=zag and MLkS}]  (intersection-1) coordinate (Lk) ;
        \path[name path=MkS] (Mk) -- (Ok) ;
        \path[name intersections={of=zag and MkS}]  (intersection-1) coordinate (k) ;
        \path[name path=MRkS] (MRk) -- (ORk) ;
        \path[name intersections={of=zag and MRkS}]  (intersection-1) coordinate (Rk) ;
        \path[name path=MLjS] (MLj) -- (OLj) ;
        \path[name intersections={of=zag and MLjS}]  (intersection-1) coordinate (Lj) ;
        \path[name path=zag] (z1) +(3*\ZigSpeed,3) -- (z1) ;
        \path[name path=MRjS] (MRj) -- (ORj) ;
        \path[name intersections={of=zag and MRjS}]  (intersection-1) coordinate (z2) ;
        \path[name path=zig] (z2) +(-3*\ZigSpeed,3) -- (z2) ;
        \path[name path=top] (-10,\TOP) -- (10,\TOP) ;
        \path[name intersections={of=zig and top}]  (intersection-1) coordinate (z3) ;
        \DrawSigBorderLeftiLUAboveRight(MLi)(OLi)
        \DrawSigMainiLUAboveRight(Mi)(Oi)
        \DrawSigBorderRightiLUAboveRight(MRi)(ORi)
        \DrawSigBorderLeftkLUBelowLeft(MLk)(Lk)
        \DrawSigMainkLUAboveLeft(Mk)(Ok)
        \DrawSigBorderRightkLUAboveLeft(MRk)(Rk)
        \DrawSigBorderLeftjLUAboveRight(Lj)(MLj)
        \DrawSigMainjLUAboveRight(Oj)(Mj)
        \DrawSigBorderRightjLUAboveRight(z2)(MRj)
        \DrawSigCollectijLUAbove(z1)(z2)
        \begin{scope}[inner sep=0,outer sep=0]  
          \DrawSigCheckOKijLUAboveRight(z1)(z0)      
          \DrawSigReadyiLUAboveRight(z3)(z2)
        \end{scope}
      \end{tikzpicture}}}

  \SubFigure[example\label{fig:removing-useless:example}]{\includegraphics[width=\textwidth]{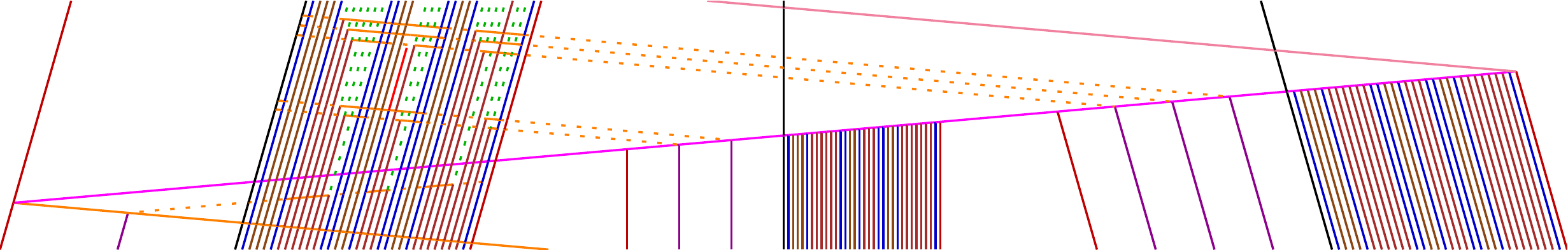}
  }

  \caption{Removing unused lists and sending ids to the decision making area.}
  \label{fig:removing-useless}
\end{figure}

This is done as in \RefFig{fig:removing-useless}.
\RefFigure{fig:removing-useless:schema} depicts the signals that drive the dynamics (\SigCheckOKij then \SigCollectij then \SigReadyi) while there is an actual diagram in \RefFig{fig:removing-useless:example}.

Signal \SigCheckOKij initiates the process.
It first goes on the left to make the id of the leftmost macro-signal act on the rules.
It bounces on \SigBorderLefti to become \SigCollectij.
Signal \SigCollectij crosses the whole configuration and bounces (and erases) the \SigBorderRightj (rightmost) to become \SigReadyi.
The latter will select and apply the rule.

Before turning to \SigCollectij, signal \SigCheckOKij turns each \SigIDi (for \SpeedIndexI only) into \SigCrossBackOKi which heads right for the rule list.
Together, these \SigCrossBackOKi signals encode the id of the macro-signal of speed index \SpeedIndexI.
In \RefFig{fig:removing-useless}, this corresponds to the signals on the bottom left that are changed to fast right-bounds signals.

While crossing the configuration, \SigCollectij erases all the surounding signals of collaborating macro-signals. 
It turns each \SigIDk (for $\SpeedIndexJ\leq\SpeedIndexK<\SpeedIndexI$) into \SigCrossOKk which heads left for the rule list.
Together, the \SigCrossOKk signals encode the id of the macro-signal of speed index \SpeedIndexK.
In \RefFig{fig:removing-useless}, this corresponds to the remaining signals of each macro-signals, which are changed to fast left-bounds signals.

All the needed meta-signals and collision rules are defined in \RefFig{fig:ms+rule:disposal}.
The constant 40 is arbitrary.
It ensures that the delays in \RefFig{fig:phases} are respected.

\begin{figure}[hbt]
  \centerline{
    \begin{tabular}{@{}cc@{}}    
      \begin{tabular}{@{}c@{}}
        \begin{ParameterList}
          \SpeedMaxAbs & \max_{\SpeedIndexI\in\IntegerInterval{1}{\SpeedSetNumber} }|\BaseSpeed{\SpeedIndexI}| \\[.6em]
          \SpeedRapid & 40.\SpeedMaxAbs \\
        \end{ParameterList}
        \\[2em]
        \begin{MSlist}
          \MSForAllSpeedJltI[-\SpeedRapid]{\SigCheckOKij}\MSForAllSpeedJltI{\SigCollectij}\MSForAllSpeedI[\SpeedRapid]{\SigCrossBackOKi}\MSForAllSpeedI[-\SpeedRapid]{\SigCrossOKi}\end{MSlist}
      \end{tabular}&
                         \begin{CRlist}
                           \CRForAllSpeedJltI{\SigIDi, \SigCheckOKij}{\SigCheckOKij, \SigCrossBackOKi}\CRForAllSpeedJltI{\SigBorderLefti, \SigCheckOKij}{\SigBorderLefti, \SigCollectij}\CRForAllSpeedJleKltI{\SigCollectij, \SigBorderLeftk}{\SigCollectij}\CRForAllSpeedJleKltI{\SigCollectij, \SigIDk}{\SigCrossOKk, \SigCollectij}\CRForAllSpeedJltKltI{\SigCollectij, \SigBorderRightk}{\SigCollectij}\CRForAllSpeedJleKltI{\SigCollectij, \SigRuleBoundk}{\SigCollectij}\CRForAllSpeedJleKltI{\SigCollectij, \SigRuleMiddlek}{\SigCollectij}\CRForAllSpeedJleKltIL{\SigCollectij, \SigIfkl}{\SigCollectij}\CRForAllSpeedJleKltIL{\SigCollectij, \SigThenkl}{\SigCollectij}\CRForAllSpeedJltI{\SigCollectij, \SigBorderRighti}{\SigReadyi}\end{CRlist}
    \end{tabular}}
  \caption{Meta-signals and collisions rules for disposal.}
  \label{fig:ms+rule:disposal}
\end{figure}

The function \AGCsimGroup is refined to take account of these new signals and rules.
As before, the value of $\AGCsimGroup(\AGCconfigurationSMotherSymbol)$ is \AGCextendedValueVoid if $\AGCconfigurationSMotherSymbol(0)$ is not some \SigMaink.
It always ignores \SigCollectij and \SigCheckOKij
If there is a \SigCollectij before the first \SigBorderLeft to the left, it looks to the left until the first \SigMaini (for the same value of $i$).
If there is a \SigBorderRighti before the \SigBorderLeftk, then it counts any \SigCrossOKk before that \SigBorderRighti as a \SigIDk.
Likewise, it also counts any \SigCrossBackOKi to the right of \SigMaini as a \SigIDi.

\subsection{Applying Id's onto Rules}
\label{sub:apply-in-rules}

The beam of \SigCrossBackOKi signals acts on every if-part of the rules and tries to cross-out the same number of \SigIfi.
Travelling rightward, each meet \SigRuleMiddlei before the if-part of the rule.
It gets \emph{activated} as \SigCrossBacki.
On meeting \SigIfi, they are both \emph{deactivated} and becomes \SigCrossBackOKi and \SigIfOKi.
This is illustrated on \RefFig{fig:comparison-of-ids:back} where dotted lines indicate deactivation and dashed ones indicate failure.

If the numbers do not match, a mark is left on the rule.
If the \SigCrossBackOKi are too few, then at least one (activated) \SigIfi remains as in \RefFig{fig:comparison-of-ids:back:too-few}.
If the \SigCrossBackOKi are in excess, then  at least one (activated) \SigCrossBacki reaches the \SigRuleBoundi on the right.
It turns it into \SigRuleBoundFaili to indicate failure of the rule as in \RefFig{fig:comparison-of-ids:back:too-much}.
Signal \SigCrossBacki is always deactivated on leaving the rule (on \SigRuleBoundi or \SigRuleBoundFaili).
Signals \SigCrossBackOKi are destroyed after the last rule (on \SigBorderRighti).
The left of \RefFig{fig:removing-useless:example} displays a real application of \SigCrossBackOKi on the rules.

\begin{figure}[hbt]\centering
  \scriptsize\SetUnitlength{1.7em}\newcommand{\SettingCrossBack}[1]{\scalebox{.9}{\hspace*{-3ex}\begin{tikzpicture}[inner sep=0,outer sep=.15\unitlength]\newcommand{\TOP}{9}\newcommand{\SHIFT}{2}\path[name path=l] (0.5,0) \CoorNode{l0} -- + (\SHIFT,\TOP) \CoorNode{lt} ;
        \path[name path=a] (3,0) \CoorNode{a0} -- + (\SHIFT,\TOP) \CoorNode{at} ;
        \path[name path=b] (4,0) \CoorNode{b0} -- + (\SHIFT,\TOP) \CoorNode{bt} ;
        \path[name path=r] (5,0) \CoorNode{r0} -- + (\SHIFT,\TOP) \CoorNode{rt} ;
        \path[name path=cu] (-2.25,2.6) \CoorNode{cu0} -- + (9.5+\SHIFT,\TOP-4) \CoorNode{cut} ;
        \path[name path=cm] (-2.25,1.3) \CoorNode{cm0} -- + (9.5+\SHIFT,\TOP-4) \CoorNode{cmt} ;
        \path[name path=cl] (-2.25,0) \CoorNode{cl0} -- + (9.5+\SHIFT,\TOP-4) \CoorNode{clt} ;
        \SetIntersect{l}{cu}{l3}
        \SetIntersect{l}{cm}{l2}
        \SetIntersect{l}{cl}{l1}
        \SetIntersect{a}{cu}{a3}
        \SetIntersect{a}{cm}{a2}
        \SetIntersect{a}{cl}{a1}
        \SetIntersect{b}{cm}{b2}
        \SetIntersect{r}{cu}{r3}
        \SetIntersect{r}{cm}{r2}
        \SetIntersect{r}{cl}{r1}
        \DrawSigRuleMiddleiLUAboveRight(l0)(lt)
        \DrawSigCrossBackOKiLUAbove(cl0)(l1)
        \DrawSigCrossBackiLUAboveParam[pos=.575](l1)(a1)
        \DrawSigCrossBackOKi(a1)(r1)
        \DrawSigCrossBackOKiLUBelow(r1)(clt)
        \DrawSigIfiLUAboveParam[pos=.15](a0)(a1)
        \DrawSigIfOKiLUAboveRight(a1)(at)
        #1
      \end{tikzpicture}\hspace*{-1ex}}}
  \hspace*{1ex}\SubFigure[too few\label{fig:comparison-of-ids:back:too-few}]{\SettingCrossBack{\DrawSigIfiLUAboveLeft(b0)(bt)
      \DrawSigRuleBoundiLUAboveLeft(r0)(rt)
    }}\hspace*{-1ex}\SubFigure[equal\label{fig:comparison-of-ids:back:equal}]{\hspace*{-3ex}\SettingCrossBack{\DrawSigCrossBackOKiLUAbove(cm0)(l2)
      \DrawSigCrossBackiLUAboveParam[pos=.575](l2)(a2)
      \DrawSigCrossBacki(a2)(b2)
      \DrawSigCrossBackOKi(b2)(r2)
      \DrawSigCrossBackOKiLUBelow(r2)(cmt)
      \DrawSigIfiLUAboveLeft(b0)(b2)
      \DrawSigIfOKiLUAboveRight(b2)(bt)
      \DrawSigRuleBoundiLUAboveLeft(r0)(rt)
    }}\hspace*{-3ex}\SubFigure[too much\label{fig:comparison-of-ids:back:too-much}]{\hspace*{-3ex}\SettingCrossBack{\DrawSigCrossBackOKiLUAbove(cm0)(l2)
      \DrawSigCrossBackiLUAboveParam[pos=.575](l2)(a2)
      \DrawSigCrossBacki(a2)(b2)
      \DrawSigCrossBackOKi(b2)(r2)
      \DrawSigCrossBackOKiLUBelow(r2)(cmt)
      \DrawSigIfiLUAboveLeft(b0)(b2)
      \DrawSigIfOKiLUAboveRight(b2)(bt)
      \DrawSigCrossBackOKiLUAbove(cu0)(l3)
      \DrawSigCrossBackiLUAboveParam[pos=.575](l3)(a3)
      \DrawSigCrossBacki(a3)(r3)
      \DrawSigCrossBackOKiLUBelow(r3)(cut)
      \DrawSigRuleBoundiLUAboveLeft(r0)(r3)
      \DrawSigRuleBoundFailiLUAboveRight(r3)(rt)
    }}
  \caption{Comparison of id's in the if-part of a rule for \SigCrossBacki{}.}
  \label{fig:comparison-of-ids:back}
\end{figure}

For every other present speed \BaseSpeed{\SpeedIndexK} ($\SpeedIndexJ\leq\SpeedIndexK<\SpeedIndexI$), the beam of \SigCrossOKk signals acts on every if-part of the rules similarly and tries to cross-out the same number of \SigIfik.
The difference is that they enter each rule from the right: \SigCrossOKk are activated (into \SigCrossk) by \SigRuleBoundi and mark the excess (of  \SigCrossk) on \SigRuleMiddlei (as \SigRuleMiddleFaili).
Signals \SigCrossOKk/\SigCrossk are destroyed after the last rule (on \SigMaini after activation by the last closing \SigRuleBoundi).

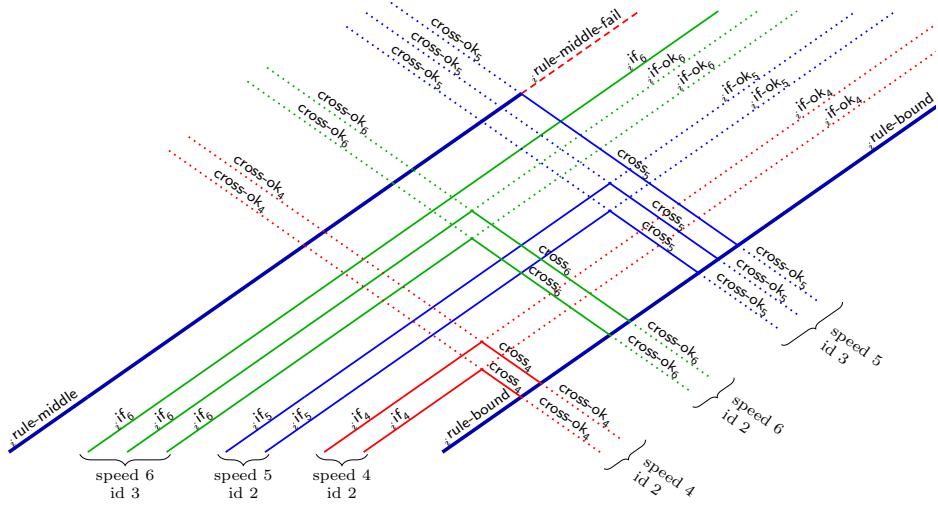
\begin{figure}[hbt]
  \scriptsize\SetUnitlength{1.1em}\centerline{\scalebox{.85}{\begin{tikzpicture}[y=.7\unitlength,inner sep=.2em]\newcommand\A{2}\newcommand\B{4}\path[clip,use as bounding box] (0,-4) rectangle (47,32.6) ;
  \DrawSigRuleMiddleiLUAbove(0,0)(4,4) ;
  \DrawSigRuleMiddlei(4,4)(22+\B,22+\B);
  \DrawSigRuleMiddleFailiLUAbove(22+\B,22+\B)(32,32);
  \DrawSigIfiSixLUAbove(4,0)(8,4) ;
  \DrawSigIfiSix(8,4)(24+\B,20+\B)
  \DrawSigIfiSixLUAbove(24+\B,20+\B)(36,32);
  \DrawSigIfiSixLUAbove(6,0)(10,4) ;
  \DrawSigIfiSix(10,4)(21.5+\A,15.5+\A);
  \DrawSigIfOKiSix(21.5+\A,15.5+\A)(25+\B,19+\B)
  \DrawSigIfOKiSixLUAbove(25+\B,19+\B)(38,32);
  \DrawSigIfiSixLUAbove(8,0)(12,4) ;
  \DrawSigIfiSix(12,4)(21.5+\A,13.5+\A);
  \DrawSigIfOKiSix(21.5+\A,13.5+\A)(26+\B,18+\B);
  \DrawSigIfOKiSixLUAbove(26+\B,18+\B)(40,32);
  \nameunder{4}{8}{speed 6\\id 3}
  \DrawSigIfiFiveLUAbove(11,0)(15,4) ;
  \DrawSigIfiFive(15,4)(26.5+\B,15.5+\B);
  \DrawSigIfOKiFive(26.5+\B,15.5+\B)(27.5+\B,16.5+\B);
  \DrawSigIfOKiFiveLUAbove(27.5+\B,16.5+\B)(43,32);
  \DrawSigIfiFiveLUAbove(13,0)(17,4) ;
  \DrawSigIfiFive(17,4)(26.5+\B,13.5+\B);
  \DrawSigIfOKiFive(26.5+\B,13.5+\B)(28.5+\B,15.5+\B)
  \DrawSigIfOKiFiveLUAbove(28.5+\B,15.5+\B)(45,32);
  \nameunder{11}{13}{speed 5\\id 2}
  \DrawSigIfiFourLUAbove(16,0)(20,4) ;
  \DrawSigIfiFour(20,4)(24,8);
  \DrawSigIfOKiFour(24,8)(30+\B,14+\B);
  \DrawSigIfOKiFourLUAbove(30+\B,14+\B)(48,32);
  \DrawSigIfiFourLUAbove(18,0)(22,4) ;
  \DrawSigIfiFour(22,4)(24,6);
  \DrawSigIfOKiFour(24,6)(31+\B,13+\B);
  \DrawSigIfOKiFourLUAbove(31+\B,13+\B)(50,32);
  \nameunder{16}{18}{speed 4\\id 2}
  \DrawSigRuleBoundiLUAbove(22,0)(26,4);
  \DrawSigRuleBoundi(26,4)(33+\B,11+\B);
  \DrawSigRuleBoundiLUAbove(33+\B,11+\B)(54,32);
  \DrawSigCrossOKFourLUAbove(30,0)(26,4);
  \DrawSigCrossFourLUAbove(26,4)(24,6);
  \DrawSigCrossOKFour(24,6)(15,15);
  \DrawSigCrossOKFourLUAbove(15,15)(8,22);
  \DrawSigCrossOKFourLUAbove(31,1)(27,5);
  \DrawSigCrossFourLUAbove(27,5)(24,8);
  \DrawSigCrossOKFour(24,8)(16,16);
  \DrawSigCrossOKFourLUAbove(16,16)(9,23);
  \nameright{30}{0}{31}{1}{speed 4\\id 2}
  \DrawSigCrossOKSixLUAbove(32.5+\A,2.5+\A)(28.5+\A,6.5+\A);
  \DrawSigCrossSixLUAbove(28.5+\A,6.5+\A)(21.5+\A,13.5+\A);
  \DrawSigCrossOKSix(21.5+\A,13.5+\A)(17.5+\A,17.5+\A);
  \DrawSigCrossOKSixLUAbove(17.5+\A,17.5+\A)(10+\A,25+\A);
  \DrawSigCrossOKSixLUAbove(33.5+\A,3.5+\A)(29.5+\A,7.5+\A);
  \DrawSigCrossSixLUAbove(29.5+\A,7.5+\A)(21.5+\A,15.5+\A);
  \DrawSigCrossOKSix(21.5+\A,15.5+\A)(18.5+\A,18.5+\A);
  \DrawSigCrossOKSixLUAbove(18.5+\A,18.5+\A)(11+\A,26+\A);
  \nameright{32.5+\A}{2.5+\A}{33.5+\A}{3.5+\A}{speed 6\\id 2}
  \DrawSigCrossOKFiveLUAbove(35.0+\B,5.0+\B)(31+\B,9+\B);
  \DrawSigCrossFiveLUAbove(31+\B,9+\B)(26.5+\B,13.5+\B);
  \DrawSigCrossOKFive(26.5+\B,13.5+\B)(20+\B,20+\B);
  \DrawSigCrossOKFiveLUAbove(20+\B,20+\B)(13+\B,27+\B);
  \DrawSigCrossOKFiveLUAbove(36+\B,6+\B)(32+\B,10+\B);
  \DrawSigCrossFiveLUAbove(32+\B,10+\B)(26.5+\B,15.5+\B);
  \DrawSigCrossOKFive(26.5+\B,15.5+\B)(21+\B,21+\B);
  \DrawSigCrossOKFiveLUAbove(21+\B,21+\B)(14+\B,28+\B);
  \DrawSigCrossOKFiveLUAbove(37+\B,7+\B)(33+\B,11+\B);
  \DrawSigCrossFiveLUAbove(33+\B,11+\B)(22+\B,22+\B);
  \DrawSigCrossOKFiveLUAbove(22+\B,22+\B)(15+\B,29+\B);
  \nameright{35.0+\B}{5.0+\B}{37+\B}{7+\B}{speed 5\\id 3}
\end{tikzpicture}
}}
  \caption{Comparison of id's in the if-part of a rule.}
  \label{fig:comparison-of-ids}
\end{figure}

\RefFigure{fig:comparison-of-ids} depicts the process with equality on speed number $4$, too few on speed $6$ and too much on speed $5$.
\RefFigure{fig:removing-useless:example} displays a real application of id's onto rules.

For every speed index \SpeedIndexM not involved in the macro-collision, \SigIfim are unaffected and thus remain active.
Altogether, if the if-part of a rule does not matched the incoming macro-signals, then at least one \SigIfil remains or \SigRuleMiddlei is replaced by \SigRuleMiddleFaili or the left \SigRuleBoundi is replaced by \SigRuleBoundFaili.

All the meta-signals and collision rules needed for the application are detailed in \RefFig{fig:ms+rule:apply}.

\begin{figure}[hbt]
  \centerline{
    \begin{tabular}{@{}cc@{}}    
      \begin{MSlist}
        \MSForAllSpeedI[\SpeedRapid]{\SigCrossBacki} \MSForAllSpeedI[-\SpeedRapid]{\SigCrossi}\MSForAllSpeedIL{\IdIfOK[\SpeedIndexI]{\SpeedIndexL}}
        \MSForAllSpeedI{\SigRuleMiddleFaili}\MSForAllSpeedI{\SigRuleBoundFaili}\end{MSlist}
      &
        \begin{CRlist}
          \CRForAllSpeedI{\SigCrossBackOKi, \SigRuleMiddlei}{\SigRuleMiddlei, \SigCrossBacki}\CRForAllSpeedI{\SigCrossBacki, \SigIfi}{\SigIfOKi, \SigCrossBackOKi}\CRForAllSpeedI{\SigCrossBacki, \SigRuleBoundi}{\SigCrossBackOKi, \SigRuleBoundFaili}\CRForAllSpeedI{\SigCrossBacki, \SigRuleBoundFaili}{\SigCrossBackOKi, \SigRuleBoundFaili}\CRForAllSpeedI{\SigCrossBackOKi, \SigBorderRighti}{\SigBorderRighti}\CRForAllSpeedI{\SigRuleBoundi, \SigCrossOKi}{\SigCrossi, \SigRuleBoundi}\CRForAllSpeedKltI{\SigIfik, \SigCrossk}{\SigCrossOKk, \SigIfOKik}\CRForAllSpeedKltI{\SigRuleMiddlei, \SigCrossk}{\SigCrossOKk, \SigRuleMiddleFaili}\CRForAllSpeedKltI{\SigRuleMiddleFaili, \SigCrossk}{\SigCrossOKk, \SigRuleMiddleFaili}\CRForAllSpeedKltI{\SigMaini, \SigCrossk}{\SigMaini}\end{CRlist}
    \end{tabular}}
  \caption{Meta-signals and collisions rules for applying id's to rules.}
  \label{fig:ms+rule:apply}
\end{figure}

As above, the function \AGCsimGroup has to be refined to take these new meta-signal and rules into account.
It still yields \AGCextendedValueVoid if the configuration is not centred on a \SigMain.
If it is centred on \SigMaink, then it needs to recover the identity corresponding to that main signal.
This is done by counting any \SigCrossOKk and \SigCrossk as a \SigIDk (or \SigCrossBackOKi and \SigCrossBacki when \SpeedIndexK is \SpeedIndexI).
This count is completed by counting the \SigIfOKik in the section encoding \AGCruleMax.
This yields the correct value since the ids of the input signals in \AGCruleMax are maximal.

\subsection{Selecting the Rule}
\label{sub:select-rule}

After all of the \SigCrossBackOKi and \SigCrossOKk have operated on the list, since the simulated machine is deterministic at most one of the rules has no \SigIfil left and no \SigRuleBoundFaili nor \SigRuleMiddleFaili.
This rule is the one corresponding to the collision that is being simulated.
(If there is no rule, then the output is empty: macro-signals just annihilate together.)

When the rule is found two things have to be done:
(a) extracting a copy of the then part of the rule, and 
(b) recording the output speeds.

This is carried out by the \SigReadyi coming from the right.
When it meets some \SigIfil or \SigRuleBoundFaili or \SigRuleMiddleFaili, it becomes \SigReadyNoi.
It is reactivated (i.e. turned back to \SigReadyi) on meeting \SigRuleBoundi.
It is still active only after crossing the correct if-part.

Activated in the correct then-part, \SigReadyi makes a slower copy to be sent on the left and stores the index of the \SigThenil.
The output indices are collected in a set in the exponent part of \SigReadyi, becoming \SigReadyiE where \SpeedSubset is the subset of \IntegerInterval{1}{\SpeedSetNumber} collecting the indices of all out-speed.
This subset is updated each time a \SigThenil is met when active.
It is preserved by activation/deactivation and transmitted to \SigMaini (that becomes \SigMainiE).

\begin{figure}[hbt]
  \centering\scriptsize\SubFigure[scheme\label{fig:rule-selection:scheme}]{\SetUnitlength{2em}
    \scalebox{.88}{\newcommand{\TOP}{7}\begin{tikzpicture}[inner xsep=0,outer xsep=0,y=.68\unitlength]
        \DrawSigBorderLeftiLUAbove(-5,-3)(-5,\TOP)
        \DrawSigMainiLUAbove(0,-3)(0,0)
        \DrawSigMainiELUAboveParam[pos=.8](0,0)(0,\TOP)
        \DrawSigBorderRightiLUAbove(5,-3)(5,\TOP)
        \DrawSigReadyiELUBelowParam[pos=.4](0,0)(2.5,-1)
        \DrawSigReadyi(2.5,-1)(5,-2)
        \DrawSigReadyNoi(2.5,-1)(5,-2)
        \DrawSigReadyiLUAbove(5,-2)(7.5,-3)
        \DrawSigReadyULUAbove(5,-2)(0,3)
        \DrawSigReadyUULUAboveParam[pos=.2](0,3)(-5,5)
        \DrawSigTheniOneLUAboveLeft(2,-3)(2,-.8)
        \DrawSigTheniOne(2,-.8)(2,\TOP)
        \DrawSigIdCopyOneLUBelow(2,-.8)(-3,4.2)
        \DrawSigIdSelectedOneLUAboveRight(-3,4.2)(-3,\TOP)
        \DrawSigTheniTwoLUAboveLeft(3,-3)(3,-1.2)
        \DrawSigTheniTwo(3,-1.2)(3,\TOP)
        \DrawSigIdCopyTwoLUAbove(3,-1.2)(-2,3.8)
        \DrawSigIdSelectedTwoLUAboveRight(-2,3.8)(-2,\TOP)      
      \end{tikzpicture}}}
  \qquad
  \SubFigure[example\label{fig:rule-selection:example}]{\includegraphics[height=3.25cm]{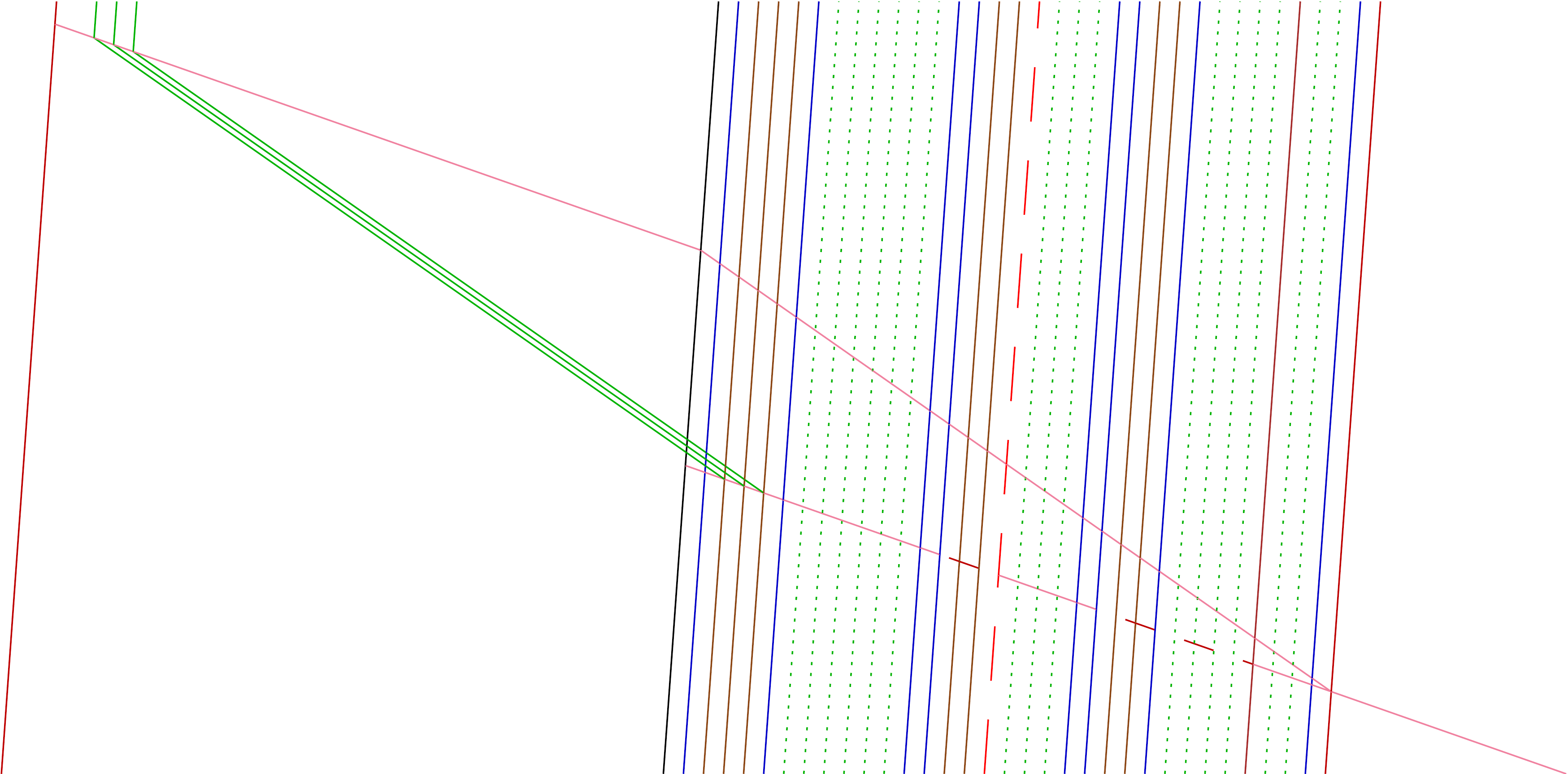}}\caption{Rule selection.}
  \label{fig:rule-selection}
\end{figure}

Extracting a copy of the then-part is done as it is shown in \RefFigure{fig:rule-selection}.
The copy goes to the left to cross \SigMainiE.
They are then made parallel to \SigMainiE by a faster signal \SigReadyUU.
To generate it \SigReadyi emits \SigReadyU on meeting \SigBorderRighti.
On meeting \SigMainiE, \SigReadyU is changed to \SigReadyUU that sets on position all \SigID's and disappear on meeting \SigBorderLefti.

All the needed meta-signals and collision rules are defined in \RefFig{fig:ms+rule:selection}.

\begin{figure}[hbt]
  \centerline{
    \begin{tabular}{@{}cc@{}}    
      \begin{MSlist}
        \MSForAllSpeedIE[-\SpeedRapid]{\SigReadyiE}\MSForAllSpeedIE[-\SpeedRapid]{\SigReadyNoiE}\MS[-\SpeedRapid/2]{\SigReadyU}\MS[-\SpeedRapid]{\SigReadyUU}\MSForAllSpeedIL[-\SpeedRapid]{\IdCopy{\SpeedIndexL}}\MSForAllSpeedIL{\IdSelected{\SpeedIndexL}}\end{MSlist}
      &
        \begin{CRlist}
          \CRForAllSpeedIE{\SigRuleBoundFaili, \SigReadyiE}{\SigReadyNoiE, \SigRuleBoundFaili}\CRForAllSpeedIE{\SigIfil, \SigReadyiE}{\SigReadyNoiE, \SigIfil}\CRForAllSpeedIE{\SigRuleMiddleFaili, \SigReadyiE}{\SigReadyNoiE, \SigRuleMiddleFaili}\CRForAllSpeedIE{\SigRuleBoundi, \SigReadyNoiE}{\SigReadyiE, \SigRuleBoundi}\CRForAllSpeedI{\SigBorderRighti, \SigReadyi}{\SigReadyi, \SigReadyU, \SigBorderRighti}\CRForAllSpeedIE{\SigMaini, \SigReadyiE}{\SigMainiE}\CRForAllSpeedIE{\SigMainiE, \SigReadyU}{\SigReadyUU, \SigMainiE}\CRForAllSpeedI{\SigBorderLefti, \SigReadyUU}{\SigBorderLefti}\CRForAllSpeedILE{\SigThenil, \SigReadyiE}{\SigReadyiEl, \SigIdCopyl, \SigThenil}\CRForAllSpeedILE{\SigIdCopyl, \SigReadyUU}{\SigReadyUU, \SigIdSelectedl}\end{CRlist}
    \end{tabular}}
  \caption{Meta-signals and collisions rules for selecting the collision rule.}
  \label{fig:ms+rule:selection}
\end{figure}

As above, the function \AGCsimGroup has to be refined to take these new meta-signal and rules into account.
It suffices to have \AGCsimGroup ignore the new signals from this section, since they leave the signals used previously by \AGCsimGroup unaffected.

\subsection{Setting the Output Macro-Signals}
\label{sub:output}

\begin{figure}[hbt]
  \centering\scriptsize\SubFigure[scheme\label{fig:output-construction:scheme}]{\SetUnitlength{1.3ex}\newcommand{\BOT}{-10}\newcommand{\Speedi}{.5}\newcommand{\ArmX}{15}\newcommand{\ArmUp}{25}\newcommand{\ArmY}{2}\newcommand{\Dispatch}[2]{\begin{scope}[shift={(#1)}]
        \csname DrawSig#2LUBelowParam\endcsname[pos=.75](0,0)(120:\ArmUp)
        \csname DrawSig#2LUBelowParam\endcsname[pos=.25](60:\ArmUp)(0,0)
        \draw[thick,densely dotted,radius=7\unitlength] (60:7) arc[start angle=60, end angle=120] node[pos=.5,above] {$\SpeedIndexL\in E$};
      \end{scope}
    }
    \begin{tikzpicture}
      \path(0,0) \CoorNode{O} ;
      \path(-\ArmX,\ArmY) \CoorNode{L} +(-.4,0) node[left] {$\left(-1 - \frac{\SpeedMaxAbs}{\SpeedRapid},\frac{1}{\SpeedRapid} \right)$} ;
      \path(\ArmX,\ArmY) \CoorNode{R} +(.4,0) node[right] {$\left(1 - \frac{\SpeedMaxAbs}{\SpeedRapid},\frac{1}{\SpeedRapid} \right)$} ;
      \DrawSigMainiELUAboveLeft(\BOT*\Speedi,\BOT)(O)
      \DrawSigMainjLUBelowRight(O)(-\BOT,\BOT)
      \DrawSigMainkLUBelowRight(O)({-\BOT/3},\BOT)
      \DrawSigFastLeftLUBelow(O)(L)
      \DrawSigFastRightLUBelow(O)(R)
      \DrawSigBorderLeftiLUAboveLeft({(\BOT-\ArmY)*\Speedi-\ArmX},\BOT)(L)
      \DrawSigBorderRightiLUBelowRight({(\BOT-\ArmY)*\Speedi+\ArmX},\BOT)(R)
      \Dispatch{O}{Mainl}
      \Dispatch{L}{BorderLeftl}
      \Dispatch{R}{BorderRightl}
      \draw[dotted] (L) -- (R) ;
      \path (O) +(-.3,-1) node[left] {$\left(0,0 \right)$} ;
      \path ($.4*(L)+.6*(R)$) +(-1,-1.4) node[above right] {$\left(\frac{\BaseSpeed{\SpeedIndexL}}{\SpeedRapid},\frac{1}{\SpeedRapid} \right)$} ;
    \end{tikzpicture}}\hspace*{-2em}\SubFigure[example\label{fig:output-construction:example}]{\includegraphics[height=.32\textwidth]{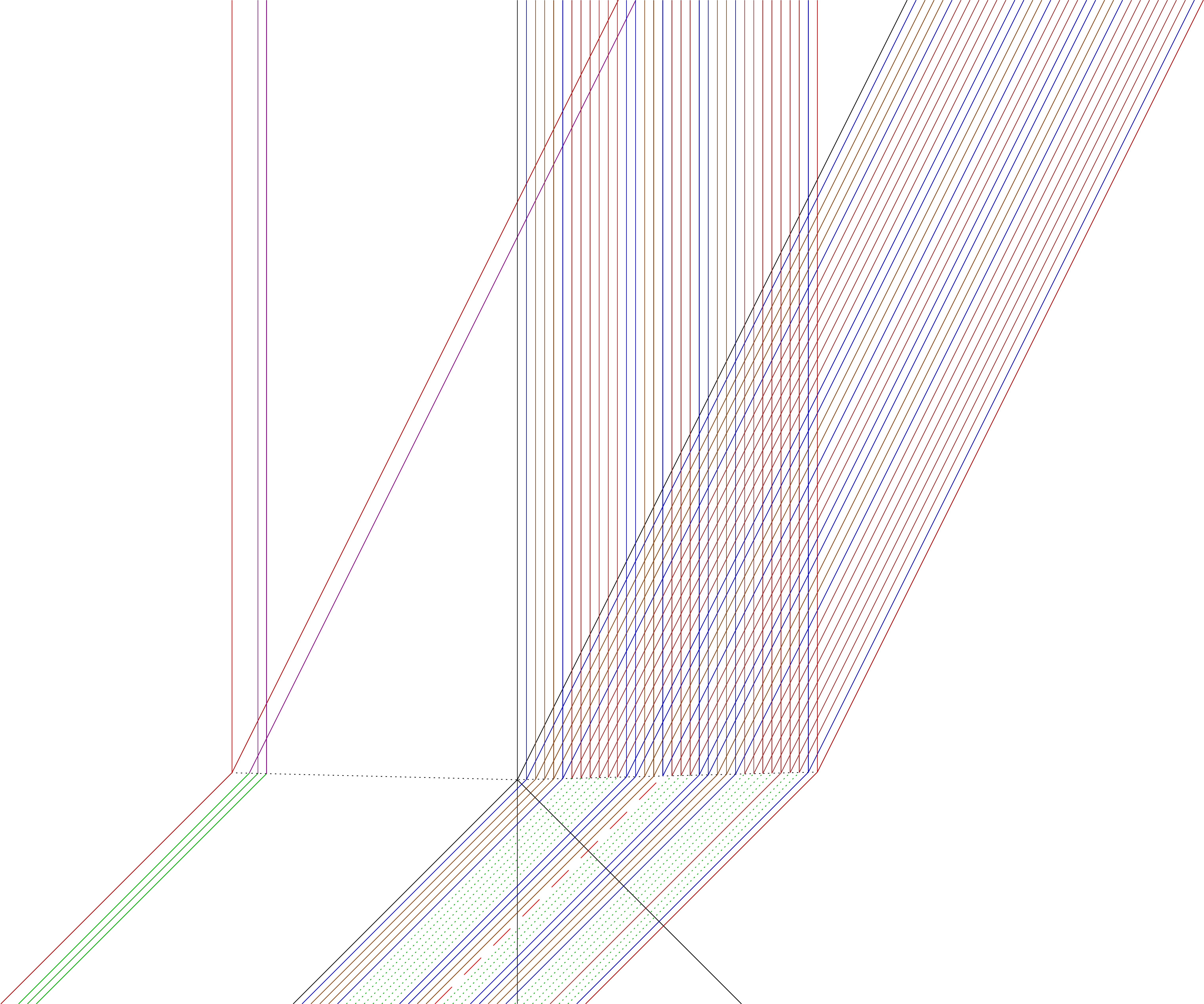}}
  \caption{Generating the output of a macro-collision.}
  \label{fig:output-construction}
\end{figure}

\RefFigure{fig:output-construction} depicts how the output macro-signals are generated.
When \SigMainiE and \SigMainj (and all collaborating main signals) meet, then \SigFastLeft and \SigFastRight are sent and \SigMainl are generated for every \SpeedIndexL of \SpeedSubset. 

On the left, \SigFastLeft sends each \SigIdSelectedl signal on the right direction as \SigIDl.
Then on reaching \SigBorderLefti, it emits one \SigBorderLeftl for each \SpeedIndexL of \SpeedSubset and disappears.
Similarly, on the right, \SigFastRight sends a clean copy of the rules on each speed \SpeedIndexL of \SpeedSubset.
Finally,  on reaching \SigBorderRighti, it emits one \SigBorderRightl for each \SpeedIndexL of \SpeedSubset and disappears.

If the simulated intersection happens at $(0, 0)$ with \SigBorderLefti at $ (-1 - \frac{\BaseSpeed{\SpeedIndexI} + \SpeedMaxAbs}{\SpeedRapid}, 0) $ and \SigBorderRighti at $ (1 - \frac{\BaseSpeed{\SpeedIndexI} + \SpeedMaxAbs}{\SpeedRapid}, 0) $, \SigFastLeft and \SigBorderLefti intersect at $(-1 - \frac{\SpeedMaxAbs}{\SpeedRapid}, \frac{1}{\SpeedRapid})$, while \SigFastRight and \SigBorderRighti intersect at $(1 - \frac{\SpeedMaxAbs}{\SpeedRapid}, \frac{1}{\SpeedRapid})$.
At time $\frac{1}{\SpeedRapid}$, each outgoing \SigMainl will be at position $\frac{\BaseSpeed{\SpeedIndexL}}{\SpeedRapid}$.

This proves that main signals will remain about in the middle of their borders (specifically, at position $\frac{\BaseSpeed{\SpeedIndexI} + \SpeedMaxAbs}{\SpeedRapid}$ if the borders are at $-1$ and $+1$), and that the right part of macro-signals remains no bigger than the left part.
It ensures all \SigIdCopyl cross \SigReadyUU before \SigBorderLefti.

After a while (given explicitly in \RefSec{subsec:safety-zone}), all the initiated macro-signals are separated and ready for macro-collision.

Knowing how the output of a macro-collision is set, we are ready to provide value of $\AGCsimRepr$ on a collision $\AGCrule$, as advertised in \RefSec{sec:encoding_repr}: 
\begin{DisplayRuleInP}
  \SigBorderLefti \quad _{\SpeedIndexI}{<}\text{output id encoding}{>} \quad \AGCrule^{\SpeedSubsetTwo,\SpeedSubset} \quad _{\SpeedIndexI}{<}\text{rules encoding}{>} \quad \SigBorderRighti
  \enspace .
\end{DisplayRuleInP}
The collision $\AGCrule^{\SpeedSubsetTwo,\SpeedSubset}$ is:
$$\{\SigMainl\}_{\SpeedIndexL\in\SpeedSubsetTwo} \cup \{\SigMainiE\} 
\rightarrow \{\SigMainl\}_{\SpeedIndexL\in\SpeedSubset} \cup \{\SigFastLeft, \SigFastRight\}$$
With \SpeedSubsetTwo the set of indices of speeds different from $ \BaseSpeed{\SpeedIndexL} $ of input meta-signals (of \AGCruleIn).
The output ids (of \AGCruleOut) are encoded in unary with \IdSelected{\SpeedIndexL} signals.
The rules are tainted with failures marks and $\SigIfOKik$ signals, as they would be, had the ids of \AGCruleIn been applied to them, so that \AGCsimGroup can recognise \AGCrule.
The signal \SigBorderLefti, \SigBorderRighti and $\AGCrule^{\SpeedSubsetTwo,\SpeedSubset}$ are placed respectively at positions $-1 - \frac{\BaseSpeed{\SpeedIndexI} + \SpeedMaxAbs}{\SpeedRapid}$, $1 - \frac{\BaseSpeed{\SpeedIndexI} + \SpeedMaxAbs}{\SpeedRapid}$, and $0$.

All the needed meta-signals and collision rules are defined in \RefFig{fig:ms+rule:output}.

\begin{figure}[hbt]
  \centering
  \begin{MSlist}
    \MSForAllSpeedE[-\SpeedRapid{-}\SpeedMaxAbs]{\SigFastLeft}\MSForAllSpeedE[\SpeedRapid{-}\SpeedMaxAbs]{\SigFastRight}\end{MSlist}
  \\[.75ex]
  \begin{CRlist}
    \CRForAllSpeedIEFltI{$\SigMainl\}_{\SpeedIndexL\in\SpeedSubsetTwo}\cup$\{\SigMainiE}{{\SigMainl}$\, \}_{\SpeedIndexL\in\SpeedSubset}{\cup}\{$\SigFastLeft, \SigFastRight}\CRForAllSpeedIE[\ensuremath{_{\SpeedIndexL\in\SpeedSubset}}]{\SigBorderLefti, \SigFastLeft}{\SigBorderLeftl}\CRForAllSpeedILinE{\SigIdSelectedl, \SigFastLeft}{\SigFastLeft, \SigIDl}\CRForAllSpeedIE{\SigFastRight, \SigRuleBoundi}{{\SigRuleBoundl}$\, \}_{\SpeedIndexL\in\SpeedSubset}{\cup}\{\, $\SigFastRight}\CRForAllSpeedIE{\SigFastRight, \SigRuleBoundFaili}{{\SigRuleBoundl}$\, \}_{\SpeedIndexL\in\SpeedSubset}{\cup}\{\, $\SigFastRight}\CRForAllSpeedIE{\SigFastRight, \SigRuleMiddlei}{{\SigRuleMiddlel}$\, \}_{\SpeedIndexL\in\SpeedSubset}{\cup}\{\, $\SigFastRight}\CRForAllSpeedIE{\SigFastRight, \SigRuleMiddleFaili}{{\SigRuleMiddlel}$\, \}_{\SpeedIndexL\in\SpeedSubset}{\cup}\{\, $\SigFastRight}\CRForAllSpeedIME{\SigFastRight, \SigIfim}{{\SigIflm}$\, \}_{\SpeedIndexL\in\SpeedSubset}{\cup}\{\, $\SigFastRight}\CRForAllSpeedIME{\SigFastRight, \SigIfOKim}{{\SigIflm}$\, \}_{\SpeedIndexL\in\SpeedSubset}{\cup}\{\, $\SigFastRight}\CRForAllSpeedIME{\SigFastRight, \SigThenim}{{\SigThenlm}$\, \}_{\SpeedIndexL\in\SpeedSubset}{\cup}\{\, $\SigFastRight}\CRForAllSpeedIE[\ensuremath{_{\SpeedIndexL\in\SpeedSubset}}]{\SigFastRight, \SigBorderRighti}{\SigBorderRightl}\end{CRlist}
  \caption{Meta-signals and collisions rules for the output.}
  \label{fig:ms+rule:output}
\end{figure}

\begin{figure}
  \centering
  \mbox{}\hfill\includegraphics[width=.8\textwidth]{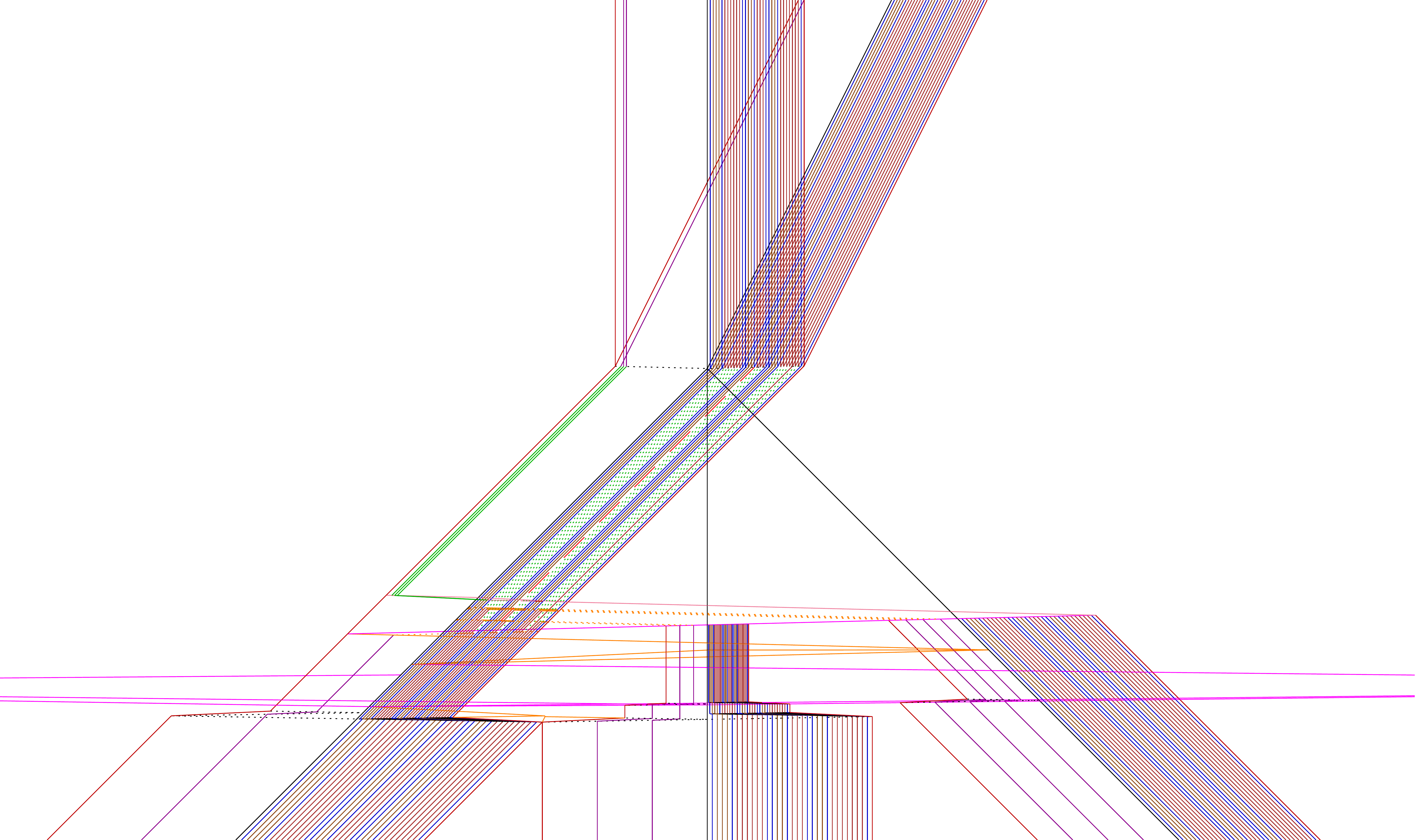}\hfill\mbox{}\caption{Exact 3-signal collision with whole preparation.}
  \label{fig:3-signal collision}
\end{figure}

An exact 3-signal collision simulation is depicted in \RefFigure{fig:3-signal collision}.

Let $\UniversalMSSpeedCheckedColl$ be the signal machine defined by the above signals and collision rules, instantiated for all possible values of \SpeedIndexI and \SpeedIndexJ in \SpeedSet.
The above arguments constitute the proof of the following lemma.

\begin{lemma}
  \label{lem:checked_collision}
  Let $\AGCrule$ be a collision rule of $\AGCmachine$, and let $\AGCconfigurationSMSymbol$ be a configuration of $\AGCmachine$ whose signals are exactly $\AGCruleIn$ and the positions of these signals are such that they all meet at some point $(\AGCspacialPosition, \AGCtemporalPosition)$.  
  Then for small enough \AGCsimStartingMSwidth, let $\AGCconfigurationSMotherSymbol$ be a $\AGCsimStartingMSwidth$-width checked configuration for $\AGCconfigurationSMSymbol$.
  Let \AGCspaceTimeDiagramSM and \AGCspaceTimeDiagramSMother be the respective space-time diagrams of $(\AGCconfigurationSMSymbol, \AGCmachine)$ and $(\AGCconfigurationSMotherSymbol, \UniversalMSSpeedCheckedColl)$.
  There is $\AGCsimOutputDelay = O(\AGCsimStartingMSwidth)$ such that for $t' > t + \AGCsimOutputDelay$, the configuration $\AGCspaceTimeDiagramSMother(t')$ satisfies $\AGCsimGroup(\AGCspaceTimeDiagramSMother(t')) = \AGCspaceTimeDiagramSM(t')$, and for any position $x$ such that $\AGCspaceTimeDiagramSMother(t')(x) \neq \AGCextendedValueVoid$, $\AGCspaceTimeDiagramSMother(t')$ is clean at $x$.
\end{lemma}

We now refine the \AGCsimGroup function in accordance with this section.
If there is no \SigFastLeft or \SigFastRight signal, then no change is needed.
Otherwise, if the centre of the configuration is a collision between \SigMain signals, both the input and the outputs of that collision have to be determined.
This can be done by looking at the signals between \SigMaini and \SigBorderRighti, where one of the rules has been selected.
After the collision, the identity of an outgoing \SigMainl signal can be gathered from the \SigIDl right of \SigFastLeft, and the \SigIdSelectedl left of \SigFastLeft.

With this refinement of \AGCsimGroupConf, the above construction gives a ``conditional simulation'' for configurations with one exact collision, as stated in the following lemma.

\begin{lemma}
  Let $\AGCrule$ be a collision rule of $\AGCmachine$, and let $\AGCconfigurationSMSymbol$ be a configuration of $\AGCmachine$ whose signals are exactly $\AGCruleIn$ and the positions of these signals are such that they all meet at some point $(\AGCspacialPosition, \AGCtemporalPosition)$.
  Then for small enough \AGCsimStartingMSwidth, let $\AGCconfigurationSMotherSymbol$ be a $\AGCsimStartingMSwidth$-width checked configuration for $\AGCconfigurationSMSymbol$.
  Let \AGCspaceTimeDiagramSM and \AGCspaceTimeDiagramSMother be the respective space-time diagrams of \AGCconfigurationSMSymbol and \AGCconfigurationSMotherSymbol.
  We have that for any $t' \geq 0$,
  \begin{equation*}
    \AGCsimGroupConf(\AGCspaceTimeDiagramSMother(t')) = \AGCspaceTimeDiagramSM(t')
    \enspace.
  \end{equation*}
\end{lemma}

Let $\AGCrule$ be a collision rule of $\AGCmachine$, and let $\AGCconfigurationSMSymbol$ be a configuration of $\AGCmachine$ whose signals are exactly $\AGCruleIn$ and the positions of these signals are such that they all meet at the point $(0, 1)$.
Then for small enough \AGCsimStartingMSwidth, let $\AGCconfigurationSMotherSymbol$ be a $\AGCsimStartingMSwidth$-width checked configuration for $\AGCconfigurationSMSymbol$, and $\AGCspaceTimeDiagramSMother$ the associated space-time diagram.
Define $\AGCsimRepr(\AGCrule)$ to be the configuration of \AGCspaceTimeDiagramSMother at time 1, rescaled and translated so that \SigBorderLefti is at $-1 - \frac{\BaseSpeed{\SpeedIndexI} + \SpeedMaxAbs}{\SpeedRapid}$, \SigBorderRighti is at $1 - \frac{\BaseSpeed{\SpeedIndexI} + \SpeedMaxAbs}{\SpeedRapid}$, and thus the collision of the \SigMain is at $0$.

\subsection{Towards Simulating a Collision in a Larger Diagram}

More generally, this construction works for the simulation of a collision when all the participating signals are identified and no other disturbing macro-signal is near.

\begin{lemma}
  \label{lem:collision}
  Let $\OtherUniversalMSSpeed = (M_{\OtherUniversalMSSpeed}, S_{\OtherUniversalMSSpeed} , R_{\OtherUniversalMSSpeed} )$ be a signal machine which contains the meta-signals and rules of \UniversalMSSpeedCheckedColl, where for every rule $\AGCrule_{\OtherUniversalMSSpeed} \in R_{\OtherUniversalMSSpeed}$ with an input signal of speed larger than $\SpeedMaxAbs$, any input signal belonging to \UniversalMSSpeedCheckedColl is also in its output.

  Let $\AGCrule$ be a collision rule of $\AGCmachine$, and let $\AGCconfigurationSMSymbol$ be a configuration of $\AGCmachine$ whose signals are exactly $\AGCruleIn$ and the positions of these signals are such that they all meet at some point $(\AGCspacialPosition, \AGCtemporalPosition)$. Let $\AGCspaceTimeDiagramSM$ be the associated space-time diagram.
  
  Then for small enough \AGCsimStartingMSwidth, let $\AGCconfigurationSMotherSymbol$ be a $\AGCsimStartingMSwidth$-width checked configuration for \AGCconfigurationSMSymbol. There are $W$, $\AGCsimOutputDelay$ and $W'$ such that for any initial configuration $\AGCconfigurationSMotherSymbol_{\OtherUniversalMSSpeed}$ of \OtherUniversalMSSpeed which coincides with $\AGCconfigurationSMotherSymbol$ on a width $W$ around $x$, after a time $t + \AGCsimOutputDelay$, $\AGCconfigurationSMotherSymbol_{\OtherUniversalMSSpeed}$ coincides on a width $W'$ with a configuration $\AGCconfigurationSMotherSymbol_{\operatorname{after}}$ such that:
  \begin{equation*}
\AGCsimGroupConf(\AGCconfigurationSMotherSymbol_{\operatorname{after}}) = \AGCspaceTimeDiagramSM(t + \AGCsimOutputDelay)
\end{equation*}
  $\AGCconfigurationSMotherSymbol_{\operatorname{after}}$ is clean at every position of a signal in $\AGCspaceTimeDiagramSM(t + \AGCsimOutputDelay)$, and any such position is at distance less than $W'$ from $\AGCspacialPosition$.

\end{lemma}
\begin{proof}
  This follows from \RefLem{lem:checked_collision}.
  For any value of \AGCsimStartingMSwidth, note $W = \SpeedMaxAbs . (t + \AGCsimOutputDelay) + 2 \AGCsimStartingMSwidth$.
  Suppose \AGCconfigurationSMotherSymbol coincides with a \AGCsimStartingMSwidth-width checked configuration $\AGCconfigurationSMotherSymbol_{(x,t)}$, on a width $W$ around $x$ at time $0$, then at time $t + \AGCsimOutputDelay$, it coincides with \AGCconfigurationSMotherSymbol on a width $\AGCsimStartingMSwidth$ around $x$ at time $t + \AGCsimOutputDelay$.
\end{proof}

\RefSection{sec:preparing} will deal with ensuring a locally \AGCsimStartingMSwidth-width checked configuration before each macro-collision, with \AGCsimStartingMSwidth small enough with respect to the delay before the collision. This means the following must be ensured:

\begin{enumerate}
\item the width of each macro-signal is small enough with respect to the time remaining before the support zones meet,
  \label{cnd:shrunk}\item any macro-signal that is not part of the collision is sufficiently away to not interfere,
  \label{cnd:alone}\item all \SigMain signals intersect at the same location, where the simulated collision takes place,\label{cnd:mark}\item a signal \SigCheckOKij is arriving on the leftmost macro-signal (index \SpeedIndexI) and \SpeedIndexJ is the speed index of the rightmost one.
  It witnesses that the resolution presented in this section is ready to start.
  \label{cnd:extreme}\end{enumerate}

\section{Preparing for Macro-Collision}
\label{sec:preparing}

We now define the rest of the signals and collisions of \UniversalMSSpeed.
Again, we explain first how correct diagrams work, then list explicitly the meta-signals and collision rules.

What is needed from \UniversalMSSpeed is to make sure that before every collision of \AGCmachine, there is a $\AGCsimStartingMSwidth$-width checked configuration for the inputs of that configuration, for small enough $\AGCsimStartingMSwidth$. 
This is done through the following phases:
\begin{description}
\item[detection] of an imminence macro-collision when some \SigBorderRighti and \SigBorderLeftk meet,
\item[shrinking] as described in \RefSec{subsec:shrinking} which is done by an elementary shrinking gadget, and checks appropriate sizing of width of macro-signals of both sides,
\item[testing around] described in \RefSec{subsec:safety-zone}, which ensures a \emph{safety zone} around the macro-collision. 
  This is done by checking wether any unexpected disturbing signal enters the zone, and
\item[check participating macro-signals] described in \RefSec{subsec:check}, 
  which through checking actual position of signals around, acquires the list of actual participating signals in the ongoing macro-collision.
\end{description}

The bottom half of \RefFigure{fig:3-signal collision} shows the full testing and information gathering before resolving a macro-collision.
The last two phases may fail.
In such a case, the whole process is cancelled to be eventually restarted later.
In the rest of this section, only the positive cases are presented.
Failure cases are not fully detailed.

\paragraph{Detection Phase} The detection phase is only one collision: any collision between a \SigBorderRighti and \SigBorderLeftk sends a collection of signals: \SigShrinkBottomBothRi, \SigShrinkTopTestRi, \SigShrinkTestik,
\SigShrinkTopTestLk, and \SigShrinkBottomBothLk, which initiate the shrinking phase.

\paragraph{Shrinking and Width Checking} 
Shrinking is done by a elementary and widely used gadget in signal machines. 
To ensure every controlling signal we will send for future uses passes through all the participating meta-signals and come back in a reasonable time, we ensure that width of left-most participating meta-signal is the larger one, among all participating macro-signals. 
Comparing width of two meta-signals is done by sending a signal from middle and waiting for the echoes. 
The echo that arrives first indicates the thiner macro-signal. 

\paragraph{Testing for Safety Zone} 
A zone around any (potential) macro-collision is considered to not contain any disrupting signal. This property helps to ensure that no other signal may have collision with the detected meta-signals during the process of handling ongoing macro-collision. 
The zone is surrounded by four boarders. 
The existence of disturbing signals in the safety-zone is checked by sending some signals at two bottom boarders and test for any unexpected collision.
Any unexpected collision cancels the process. 
The two other boarders have a high absolute slope, thus, no other signal may cross those. 

\paragraph{Checking Participating Meta-signals} 
Once a potential macro-collision is detected, since all speeds of macro-signals are of a fixed set (\SpeedSet), for the ongoing macro-collision, all other potentially present macro-signals could be detected.
In order to check their existence, some welcoming signals (\SigCheckInterceptikm) are sent to the positions we expect for other meta-signals's \SigMainl to be present at.
Any encounter of \SigMainl without the corresponding welcoming signal violates the condition of safety-zone. 
In this case, the process is aborted.
Note that, the macro-signals are already shrunk, and next collision would be processed with thiner macro-signals and, thus with a smaller safety-zone.

\subsection{Shrinking for Delay and Separating Macro-Signals}
\label{subsec:shrinking}

In order to gain some delay macro-signals are shrunk so that either the macro-collision as presented above is processed or aborted.
Aborting is just not to go to the main step; the shrunk macro-signals are operational and can restart a macro-collision.

\begin{figure}[hbt]
  \centering
  \SubFigure[elementary shrinking\label{fig:elementary:shrinking}]{
    \newcommand{\NodeA}[1]{\node at ([xshift=-.5\unitlength]A#1) {$a_{#1}$} ;}
    \newcommand{\NodeB}[1]{\node at ([xshift=.5\unitlength]B#1) {$b_{#1}$} ;}
    \scriptsize\SetUnitlength{1.25em}\begin{tikzpicture}
      \coordinate (A0) at (0,1) ;
      \coordinate (A1) at (0,2) ;
      \coordinate (A2) at (6,7) ;
      \coordinate (A3) at (6,8) ;
      \coordinate (B0) at (9,1) ;
      \coordinate (B1) at (9,5) ;
      \coordinate (B2) at (9,8) ;
      \NodeA{0}
      \NodeA{1}
      \NodeA{2}
      \NodeA{3}
      \NodeB{0}
      \NodeB{1}
      \NodeB{2}
      \draw[thick] (A1) -- (B1) --(A2) ; 
      \draw[thick] (A0) -- (A1) -- (A2) --(A3) ; 
      \draw[thick] (B0) -- (B1) -- (B2) ;
      \foreach \w/\lab in {1/b,.33/c,3/a} {
        \draw[thin,densely dotted] 
        (barycentric cs:A0=\w,B0=1) -- node[sloped,above=-.1\unitlength] {\AGCsigPolice{\lab}}
        (barycentric cs:A1=0\w,B1=1) -- node[sloped,above=-.1\unitlength] {\AGCsigPolice{\lab'}}
        (barycentric cs:A2=0\w,B1=1) -- node[sloped,above=-.1\unitlength] {\AGCsigPolice{\lab}}
        (barycentric cs:A3=0\w,B2=1) ;
      } ;
    \end{tikzpicture}
  }
  \qquad
  \SubFigure[shrink and check relative widths\label{fig:real-shrink-relative-width:example}]{
    \includegraphics[height=2.5cm]{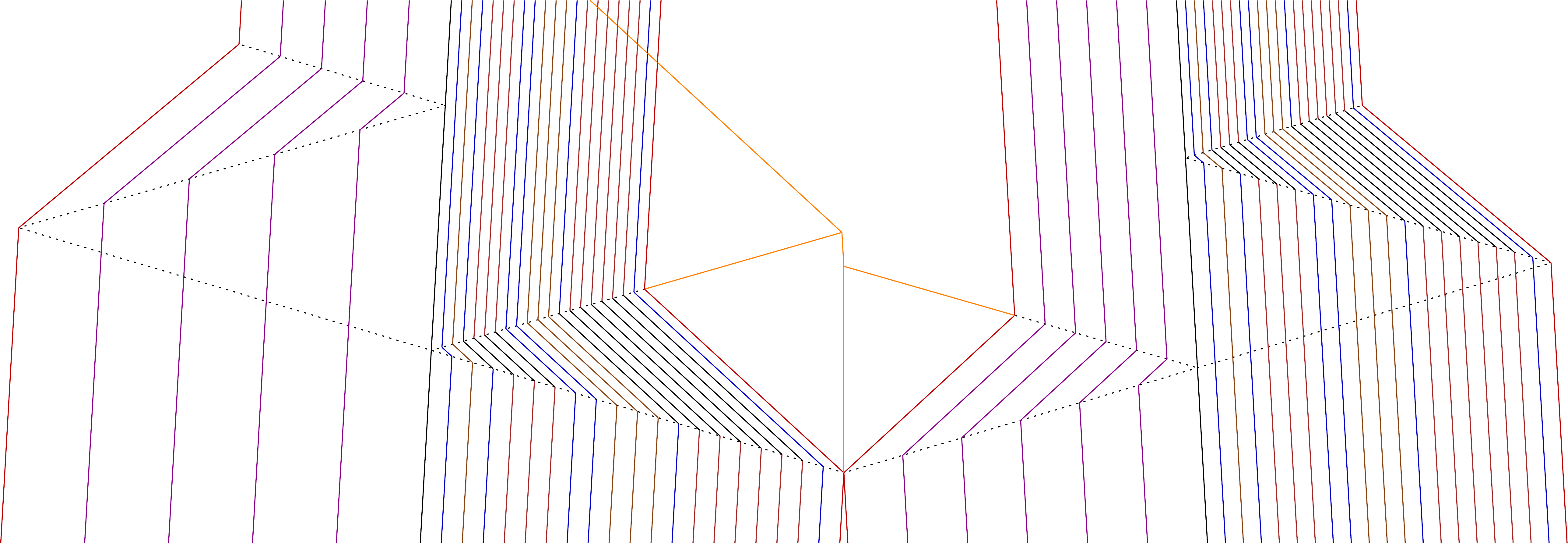}
  }
  \caption{Shrinking.}
\end{figure}

\RefFigure{fig:real-shrink-relative-width:example} provides an example of the shrinking process.
It is started in the middle where the support zones meet.
Shrinking processes are always initiated from the frontier and each half of a macro-signal is shrunk independently; this is done in order to handle concurrent shrinking on the same macro-signal.

Shrinking parallel signals is an application of proportion as illustrated in \RefFig{fig:elementary:shrinking}.
The thick signals control the shrinking while the dotted ones undergo it.
Basic geometry shows that the relative position of the intersection of the dotted signals on the segments 
$[a_0,b_0]$ then
$[a_1,b_1]$ then
$[a_2,b_1]$ then
$[a_3,b_2]$ are identical.
Thus they are shrink with the same relative positions and order.

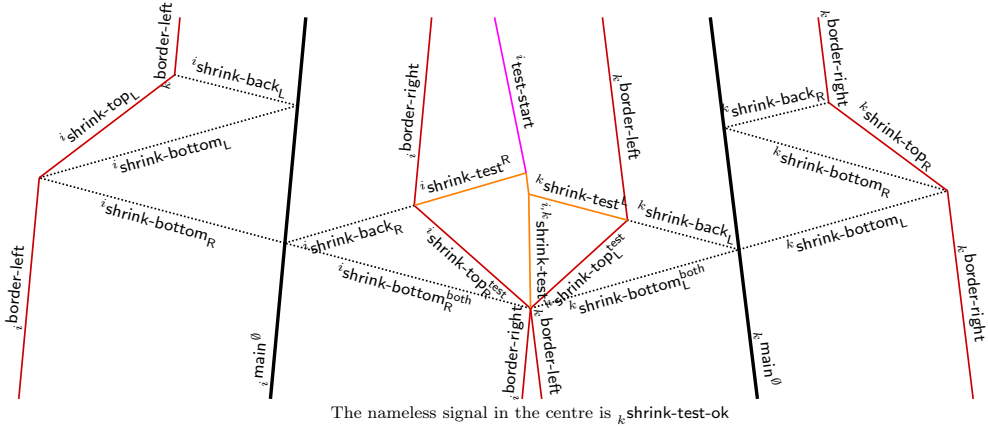
\begin{figure}[hbt]
  \centering\tiny\footnotesize\SetUnitlength{2em}\newcommand{\BOT}{-2.8}\newcommand{\TOP}{9}\newcommand{\WidthI}{-7}\newcommand{\WidthK}{6}\newcommand{\SpeedIval}{1/12}\newcommand{\SpeedKval}{-1/9}\newcommand{\SpeedSch}{1.1}\newcommand{\SpeedBotKR}{(\SpeedKval-3*\SpeedSch)}\newcommand{\SpeedBotKL}{(\SpeedKval+3*\SpeedSch)}\newcommand{\SpeedTopKR}{(\SpeedKval-\SpeedSch)}\newcommand{\SpeedTopKL}{(\SpeedKval+\SpeedSch)}\newcommand{\SpeedBotIR}{(\SpeedIval-3*\SpeedSch)}\newcommand{\SpeedBotIL}{(\SpeedIval+3*\SpeedSch)}\newcommand{\SpeedTopIR}{(\SpeedIval-\SpeedSch)}\newcommand{\SpeedTopIL}{(\SpeedIval+\SpeedSch)}\newcommand{\SpeedTestIK}{((\SpeedIval+\SpeedKval)/2)}\scalebox{.8}{\begin{tikzpicture}[y=.9\unitlength]\path (0,0) \CoorNode{O};
    \path (\BOT*\SpeedKval,\BOT) \CoorNode{KL0} ++ (\WidthK,0) \CoorNode{K0} + (\WidthK,0) \CoorNode{KR0} ;
    \path (\BOT*\SpeedIval,\BOT) \CoorNode{IR0} ++ (\WidthI,0) \CoorNode{I0} + (\WidthI,0) \CoorNode{IL0} ;
    \path[name path=left-i] (IL0) -- + ({(\TOP - ( \BOT ) )*\SpeedIval},{\TOP - ( \BOT )}) \CoorNode{ILt} ;
    \path[name path=main-i] (I0) -- + ({(\TOP - ( \BOT ) )*\SpeedIval},{\TOP - ( \BOT )}) \CoorNode{It} ;
    \path[name path=bot-i] (O) -- + ({2*\WidthI},{-2*\WidthI/\SpeedBotIL}) \CoorNode{z1} ; 
    \path[name path=new-i-right]  (\BOT*\SpeedIval,\BOT) ++ (\WidthI/2,0) \CoorNode{nIR0} -- + ({(\TOP - ( \BOT ) )*\SpeedIval},{\TOP - ( \BOT )}) \CoorNode{nIRt} ; 
    \path[name path=new-i-left]  (\BOT*\SpeedIval,\BOT)  ++ (3*\WidthI/2,0) \CoorNode{nIL0} -- + ({(\TOP - ( \BOT ) )*\SpeedIval},{\TOP - ( \BOT )}) \CoorNode{nILt} ; 
    \path[name path=top-i-r] (O) -- + (\WidthI,{\WidthI/\SpeedTopIR}) \CoorNode{z2} ;
    \path[name path=main-k] (K0) -- + ({(\TOP - ( \BOT ) )*\SpeedKval},{\TOP - ( \BOT )}) \CoorNode{Kt} ;
    \path[name path=right-k] (KR0) -- + ({(\TOP - ( \BOT ) )*\SpeedKval},{\TOP - ( \BOT )}) \CoorNode{KRt} ; 
    \path[name path=bot-k] (O) -- + ({2*\WidthK},{2*\WidthK/\SpeedBotKL})  \CoorNode{z3} ;
    \path[name path=new-k-left]  (\BOT*\SpeedKval,\BOT) ++ (\WidthK/2,0) \CoorNode{nKL0} -- + ({(\TOP - ( \BOT ) )*\SpeedKval},{\TOP - ( \BOT )}) \CoorNode{nKLt} ; 
    \path[name path=new-k-right]  (\BOT*\SpeedKval,\BOT)  ++ (3*\WidthK/2,0) \CoorNode{nKR0} -- + ({(\TOP - ( \BOT ) )*\SpeedKval},{\TOP - ( \BOT )}) \CoorNode{nKRt} ; 
    \path[name path=top-k-l] (O) -- + (\WidthK,{\WidthK/\SpeedTopKL})  \CoorNode{z4} ; 
    \SetIntersect{bot-i}{left-i}{IL1}
    \SetIntersect{main-k}{bot-k}{K1}
    \SetIntersect{main-i}{bot-i}{I1}
    \SetIntersect{bot-k}{right-k}{KR1}
    \SetIntersect{top-k-l}{new-k-left}{nKL1}
    \SetIntersect{top-i-r}{new-i-right}{nIR1} \path[name path=top-k-r] (KR1) -- + (-\WidthK,{-\WidthK/\SpeedTopKR})  \CoorNode{z5} ;
    \path[name path=bot-k-r] (KR1) -- + (-2*\WidthK,{2*\WidthK/\SpeedBotKL})  \CoorNode{z6} ;
    \path[name path=top-i-l] (IL1) -- + (-\WidthI,{-\WidthI/\SpeedTopIL})  \CoorNode{z7} ; 
    \path[name path=bot-i-l] (IL1) -- + (-2*\WidthI,{2*\WidthI/\SpeedBotIR})  \CoorNode{z8} ;
    \SetIntersect{main-k}{bot-k-r}{K2}
    \SetIntersect{main-i}{bot-i-l}{I2}
    \SetIntersect{new-k-right}{top-k-r}{nKR1}
    \SetIntersect{new-i-left}{top-i-l}{nIL1}
    \DrawSigBorderLeftiLUAbove(IL0)(IL1)
    \DrawSigMainiLUAboveParam[pos=.1](I0)(It)
    \DrawSigBorderRightiLUAbove(IR0)(O)
    \DrawSigBorderRightiLUAbove(nIR1)(nIRt)
    \DrawSigBorderLeftkLUAbove(nKL1)(nKLt)
    \DrawSigBorderLeftkLUAbove(KL0)(O)
    \DrawSigMainkLUAboveParam[pos=.1](K0)(Kt)
    \DrawSigBorderRightkLUAbove(KR0)(KR1)
    \DrawSigBorderRightkLUAbove(nKR1)(nKRt)
    \DrawSigBorderLeftkLUAbove(nIL1)(nILt)
    \DrawSigShrinkBottomBothLkLUBelow(O)(K1)
    \DrawSigShrinkBottomLkLUBelow(K1)(KR1)
    \DrawSigShrinkBottomRkLUBelow(KR1)(K2)
    \DrawSigShrinkBottomLiLUBelow(IL1)(I2)
    \DrawSigShrinkTopRkLUAbove(KR1)(nKR1)
    \DrawSigShrinkTopLiLUAbove(IL1)(nIL1)
    \DrawSigShrinkBackRkLUAbove(K2)(nKR1)
    \DrawSigShrinkBackLiLUAbove(I2)(nIL1)
    \DrawSigShrinkBottomBothRiLUBelow(O)(I1)
    \DrawSigShrinkBottomRiLUBelow(I1)(IL1)
    \DrawSigShrinkTopTestLkLUBelow(O)(nKL1)
    \DrawSigShrinkBackLkLUAbove(K1)(nKL1)
    \DrawSigShrinkTopTestRiLUBelow(O)(nIR1)
    \DrawSigShrinkBackRiLUBelow(I1)(nIR1)
    \path[name path=i-test] (I1) -- + ($2*(nIR1) - 2*(I1)$) ;
    \path[name path=k-test] (K1) -- + ($3*(nKL1) - 3*(K1)$) ;
    \path[name path=test] (O) -- + ({4*\SpeedTestIK},4) ;
    \SetIntersect{test}{k-test}{T2}
    \path[name path=test-ok] (T2) -- + ({4*\SpeedKval},4) ;
    \SetIntersect{test-ok}{i-test}{T3}
    \DrawSigShrinkTestikLUAbove(O)(T2)
    \DrawSigShrinkTestiRLUAbove(nIR1)(T3)
    \DrawSigShrinkTestkLLUAbove(nKL1)(T2)
    \DrawSigTestStartiLUAbove(T3)(-1,\TOP)
    \DrawSigShrinkTestOKk(T2)(T3)
    \path (0,\BOT-.5) node{The nameless signal in the centre is \SigShrinkTestOKk};
  \end{tikzpicture}}
  \caption{Scheme to shrink and check relative widths.}
  \label{fig:real-shrink-relative-width:scheme}
\end{figure}

The elementary shrinking process is quite an usual primitive of signal machines \cite{durand-lose06fi-mcu,durand-lose09nc}, it is not developed more in this article.
\RefFigure{fig:real-shrink-relative-width:scheme} details the signal scheme to handle the multiple shrinkings.

Process of handling macro-collision requires sending some controlling signals through all the macro-signals and gets back in a reasonable time. 
By ensuring that the width of the left-most macro-signal is the largest among all participating macro-signals, the width of all potential participating macro-signals could be limited. 
A gadget in the middle of Figs.~\ref{fig:real-shrink-relative-width:example} and \ref{fig:real-shrink-relative-width:scheme} ensures that left macro-signal is wider than the right one.
This is done by sending some \SigShrinkTestik signal when borders meet.
This signal has speed average of  \SigMaini and \SigMaink.
The \SigShrinkBottomBothRi and \SigShrinkBottomBothLk bounce on \SigMaini and \SigMaink.
If the two macro-collision would have the same width, then the \emph{echos} \SigShrinkTestiR and \SigShrinkTestkL  would arrive simultaneously.
If the left macro is larger, then \SigShrinkTestkL arrives first.

\begin{figure}[hbt]
  \centering\SubFigure[scheme\label{fig:real-shrink-relative-width:fail:scheme}]{\tiny\SetUnitlength{3em}\newcommand{\BOT}{-2}\newcommand{\TOP}{12}\newcommand{\WidthI}{-3}\newcommand{\WidthK}{6.5}\newcommand{\SpeedIval}{1/12}\newcommand{\SpeedKval}{-1/9}\newcommand{\SpeedSch}{1.1}\newcommand{\SpeedBotKR}{(\SpeedKval-3*\SpeedSch)}\newcommand{\SpeedBotKL}{(\SpeedKval+3*\SpeedSch)}\newcommand{\SpeedTopKR}{(\SpeedKval-\SpeedSch)}\newcommand{\SpeedTopKL}{(\SpeedKval+\SpeedSch)}\newcommand{\SpeedBotIR}{(\SpeedIval-3*\SpeedSch)}\newcommand{\SpeedBotIL}{(\SpeedIval+3*\SpeedSch)}\newcommand{\SpeedTopIR}{(\SpeedIval-\SpeedSch)}\newcommand{\SpeedTopIL}{(\SpeedIval+\SpeedSch)}\newcommand{\SpeedTestIK}{((\SpeedIval+\SpeedKval)/2)}\scalebox{.9}{\begin{tikzpicture}\path[clip] (-2.5,-1) rectangle (11,12) ;
    \path (0,0) \CoorNode{O};
    \path (\BOT*\SpeedKval,\BOT) \CoorNode{KL0} ++ (\WidthK,0) \CoorNode{K0} + (\WidthK,0) \CoorNode{KR0} ;
    \path (\BOT*\SpeedIval,\BOT) \CoorNode{IR0} ++ (\WidthI,0) \CoorNode{I0} + (\WidthI,0) \CoorNode{IL0} ;
    \path[name path=left-i] (IL0) -- + ({(\TOP - ( \BOT ) )*\SpeedIval},{\TOP - ( \BOT )}) \CoorNode{ILt} ;
    \path[name path=main-i] (I0) -- + ({(\TOP - ( \BOT ) )*\SpeedIval},{\TOP - ( \BOT )}) \CoorNode{It} ;
    \path[name path=bot-i] (O) -- + ({2*\WidthI},{-2*\WidthI/\SpeedBotIL}) \CoorNode{z1} ; 
    \path[name path=n-i-right]  (\BOT*\SpeedIval,\BOT) ++ (\WidthI/2,0) \CoorNode{nIR0} -- + ({(\TOP - ( \BOT ) )*\SpeedIval},{\TOP - ( \BOT )}) \CoorNode{nIRt} ; 
    \path[name path=n-i-left]  (\BOT*\SpeedIval,\BOT)  ++ (3*\WidthI/2,0) \CoorNode{nIL0} -- + ({(\TOP - ( \BOT ) )*\SpeedIval},{\TOP - ( \BOT )}) \CoorNode{nILt} ; 
    \path[name path=top-i-r] (O) -- + (\WidthI,{\WidthI/\SpeedTopIR}) \CoorNode{z2} ;
    \path[name path=main-k] (K0) -- + ({(\TOP - ( \BOT ) )*\SpeedKval},{\TOP - ( \BOT )}) \CoorNode{Kt} ;
    \path[name path=right-k] (KR0) -- + ({(\TOP - ( \BOT ) )*\SpeedKval},{\TOP - ( \BOT )}) \CoorNode{KRt} ; 
    \path[name path=bot-k] (O) -- + ({2*\WidthK},{2*\WidthK/\SpeedBotKL})  \CoorNode{z3} ;
    \path[name path=n-k-left]  (\BOT*\SpeedKval,\BOT) ++ (\WidthK/2,0) \CoorNode{nKL0} -- + ({(\TOP - ( \BOT ) )*\SpeedKval},{\TOP - ( \BOT )}) \CoorNode{nKLt} ; 
    \path[name path=n-k-right]  (\BOT*\SpeedKval,\BOT)  ++ (3*\WidthK/2,0) \CoorNode{nKR0} -- + ({(\TOP - ( \BOT ) )*\SpeedKval},{\TOP - ( \BOT )}) \CoorNode{nKRt} ; 
    \path[name path=nn-k-left]  (\BOT*\SpeedKval,\BOT) ++ (3*\WidthK/4,0) \CoorNode{nnKL0} -- + ({(\TOP - ( \BOT ) )*\SpeedKval},{\TOP - ( \BOT )}) \CoorNode{nnKLt} ; 
    \path[name path=nn-k-right]  (\BOT*\SpeedKval,\BOT)  ++ (5*\WidthK/4,0) \CoorNode{nnKR0} -- + ({(\TOP - ( \BOT ) )*\SpeedKval},{\TOP - ( \BOT )}) \CoorNode{nnKRt} ; 
    \path[name path=top-k-l] (O) -- + (\WidthK,{\WidthK/\SpeedTopKL})  \CoorNode{z4} ; 
    \SetIntersect{bot-i}{left-i}{IL1}
    \SetIntersect{main-k}{bot-k}{K1}
    \SetIntersect{main-i}{bot-i}{I1}
    \SetIntersect{bot-k}{right-k}{KR1}
    \SetIntersect{top-k-l}{n-k-left}{nKL1}
    \SetIntersect{top-i-r}{n-i-right}{nIR1} \path[name path=top-k-r] (KR1) -- + (-\WidthK,{-\WidthK/\SpeedTopKR})  \CoorNode{z5} ;
    \path[name path=bot-k-r] (KR1) -- + (-2*\WidthK,{2*\WidthK/\SpeedBotKL})  \CoorNode{z6} ;
    \path[name path=top-i-l] (IL1) -- + (-\WidthI,{-\WidthI/\SpeedTopIL})  \CoorNode{z7} ; 
    \path[name path=bot-i-l] (IL1) -- + (-3*\WidthI,{3*\WidthI/\SpeedBotIR})  \CoorNode{z8} ;
    \SetIntersect{main-k}{bot-k-r}{K2}
    \SetIntersect{main-i}{bot-i-l}{I2}
    \SetIntersect{n-k-right}{top-k-r}{nKR1}
    \SetIntersect{n-i-left}{top-i-l}{nIL1}
    \DrawSigBorderLeftiLUAbove(IL0)(IL1)
    \DrawSigMainiLUAboveParam[pos=.1](I0)(It)
    \DrawSigBorderRightiLUAbove(IR0)(O)
    \DrawSigBorderRightiLUAbove(nIR1)(nIRt)
    \DrawSigBorderLeftkLUAbove(KL0)(O)
    \DrawSigMainkLUAboveParam[pos=.1](K0)(Kt)
    \DrawSigBorderRightkLUAbove(KR0)(KR1)
    \DrawSigBorderLeftkLUAbove(nIL1)(nILt)
    \DrawSigShrinkBottomBothLkLUBelow(O)(K1)
    \DrawSigShrinkBottomLkLUBelow(K1)(KR1)
    \DrawSigShrinkBottomRkLUBelow(KR1)(K2)
    \DrawSigShrinkBottomLiLUBelow(IL1)(I2)
    \DrawSigShrinkTopRkLUAbove(KR1)(nKR1)
    \DrawSigShrinkTopLiLUAbove(IL1)(nIL1)
    \DrawSigShrinkBackRkLUAbove(K2)(nKR1)
    \DrawSigShrinkBackLiLUAbove(I2)(nIL1)
    \DrawSigShrinkBottomBothRiLUBelow(O)(I1)
    \DrawSigShrinkBottomRiLUBelow(I1)(IL1)
    \DrawSigShrinkTopTestLkLUBelow(O)(nKL1)
    \DrawSigShrinkBackLkLUAbove(K1)(nKL1)
    \DrawSigShrinkTopTestRiLUBelow(O)(nIR1)
    \DrawSigShrinkBackRiLUAbove(I1)(nIR1)
    \path[name path=i-test] (I1) -- + ($2*(nIR1) - 2*(I1)$) ;
    \path[name path=k-test] (K1) -- + ($3*(nKL1) - 3*(K1)$) ;
    \path[name path=test] (O) -- + ({4*\SpeedTestIK},4) ;
    \SetIntersect{test}{i-test}{T1}
    \path[name path=test-ok] (T1) -- + ({4*\SpeedIval},4) ;
    \SetIntersect{test-ok}{k-test}{T3}
    \path[name path=order] (T3) -- + (1*\WidthK,{1*\WidthK/\SpeedTopKL})  \CoorNode{z8} ;
    \SetIntersect{order}{n-k-left}{nKL2}
    \path[name path=n-bot-l] (nKL2) -- + (2*\WidthK,{2*\WidthK/\SpeedBotKL})  \CoorNode{z9} ;
    \SetIntersect{n-bot-l}{n-k-right}{nKR2}
    \path[name path=n-top-l] (nKL2) -- + (1*\WidthK,{1*\WidthK/\SpeedTopKL})  \CoorNode{z10} ;
    \SetIntersect{n-top-l}{nn-k-left}{nnKL1}
    \SetIntersect{n-bot-l}{main-k}{K3}
    \path[name path=n-bot-r] (nKR2) -- + (-1*\WidthK,-{1*\WidthK/\SpeedBotKR})  \CoorNode{z11} ;
    \path[name path=n-top-r] (nKR2) -- + (-1/2*\WidthK,-{1/2*\WidthK/\SpeedTopKR})  \CoorNode{z12} ;
    \SetIntersect{n-top-r}{nn-k-right}{nnKR1}
    \SetIntersect{n-bot-r}{main-k}{K4}
    \DrawSigShrinkTestikLUAbove(O)(T1)
    \DrawSigShrinkTestiRLUAbove(nIR1)(T1)
    \DrawSigShrinkTestkLLUAbove(nKL1)(T3)
    \DrawSigShrinkTestFailiLUAbove(T1)(T3)
    \DrawSigShrinkOrderkLUAbove(T3)(nKL2)
    \DrawSigBorderLeftkLUAbove(nKL1)(nKL2)
    \DrawSigBorderLeftkLUAbove(nnKL1)(nnKLt)
    \DrawSigBorderRightkLUAbove(nKR1)(nKR2)
    \DrawSigBorderRightkLUAbove(nnKR1)(nnKRt)
    \DrawSigShrinkBottomBothLkLUBelow(nKL2)(K3)
    \DrawSigShrinkBottomLkLUBelow(K3)(nKR2)
    \DrawSigShrinkTopLkLUAbove(nKL2)(nnKL1)
    \DrawSigShrinkBackLkLUAbove(K3)(nnKL1)
    \DrawSigShrinkBottomRkLUBelow(nKR2)(K4)
    \DrawSigShrinkTopRkLUAbove(nKR2)(nnKR1)
    \DrawSigShrinkBackRkLUAbove(K4)(nnKR1)
  \end{tikzpicture}}}
  \SubFigure[example\label{fig:real-shrink-relative-width:fail:example}]{\includegraphics[width=5.5cm]{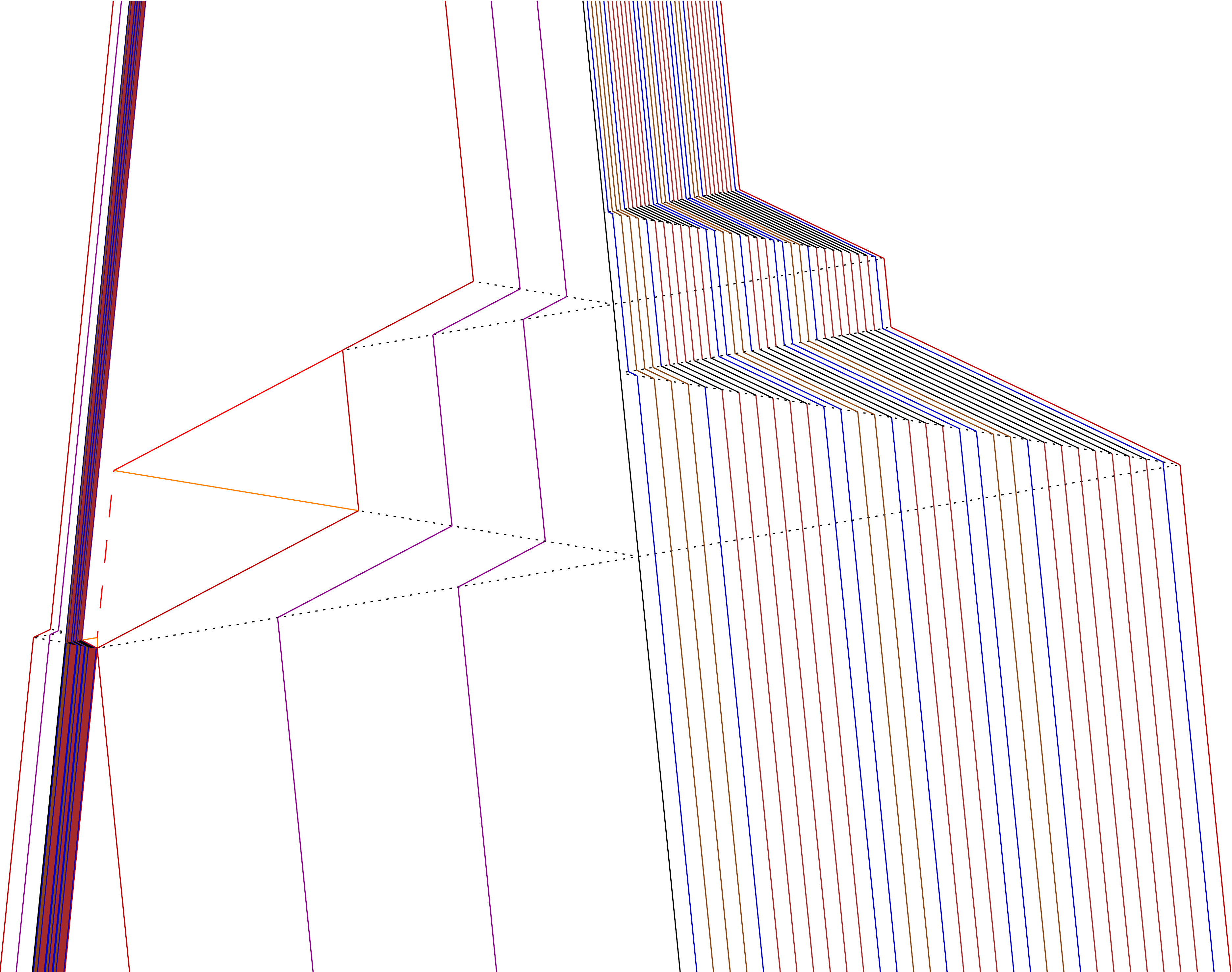}}\caption{Shrink and check relative width of two neighbouring support zones.}
  \label{fig:real-shrink-relative-width:fail}
\end{figure}

Otherwise, if \SigShrinkTestiR arrives first, the right macro-signal is shrunk again and the process is cancelled, to be restarted later when the support zones meet anew, this time with a now-thinner right macro-signal.
\RefFigure{fig:real-shrink-relative-width:fail} depicts the case where the right macro-signal is larger than the left one.

In case a new shrink is started when one is already going on, a new signal \SigShrinkDelayedLk (resp. \SigShrinkDelayedRk) is sent from the collision with \SigShrinkTopLk (resp. \SigShrinkTopRk) so that it will collide with \SigBorderLeftk (resp. \SigBorderRightk) to do the shrink after.
This is not detailed in the paper, meta-signals and collision rules and the special cases later on are omitted.

The used meta-signals and collision rules are defined in Figs. \ref{fig:ms:shrink} and \ref{fig:rule:shrink}.

\begin{figure}[hbt]
  \centerline{
    \begin{tabular}{@{}c@{}}
      \begin{ParameterList}
        \SpeedShrink & \frac{\SpeedRapid}{2} 
      \end{ParameterList}
      \\[1em]
      \begin{MSlist}
        \MSForAllSpeedI[\SpeedI{+}3\SpeedShrink]{\SigShrinkBottomLi}\MSForAllSpeedI[\SpeedI{+}3\SpeedShrink]{\SigShrinkBottomBothLi}\MSForAllSpeedI[\SpeedI{-}3\SpeedShrink]{\SigShrinkBottomRi}\MSForAllSpeedI[\SpeedI{-}3\SpeedShrink]{\SigShrinkBottomBothRi}\MSForAllSpeedI[\SpeedI{+}\SpeedShrink]{\SigShrinkTopLi}\MSForAllSpeedI[\SpeedI{+}\SpeedShrink]{\SigShrinkTopTestLi}\MSForAllSpeedI[\SpeedI{-}\SpeedShrink]{\SigShrinkTopRi}\MSForAllSpeedI[\SpeedI{-}\SpeedShrink]{\SigShrinkTopTestRi}\MSForAllSpeedI[\SpeedI{-}3\SpeedShrink]{\SigShrinkBackLi}\MSForAllSpeedI[\SpeedI{+}3\SpeedShrink]{\SigShrinkBackRi}\end{MSlist}
    \end{tabular}
    \quad
    \begin{MSlist}
      \MSForAllSpeedI[\SpeedI{+}\SpeedShrink]{\SigShrinkIDi}\MSForAllSpeedI[\SpeedI{-}\SpeedShrink]{\SigShrinkRuleBoundi}\MSForAllSpeedI[\SpeedI{-}\SpeedShrink]{\SigShrinkRuleMiddlei}\MSForAllSpeedIL[\SpeedI{-}\SpeedShrink]{\SigShrinkIfil}\MSForAllSpeedIL[\SpeedI{-}\SpeedShrink]{\SigShrinkThenil}\hline\MSForAllSpeedI[\SpeedI{-}3\SpeedShrink]{\SigShrinkTestiL}\MSForAllSpeedI[\SpeedI{+}3\SpeedShrink]{\SigShrinkTestiR}\MSForAllSpeedKltI[\frac{\SpeedI{+}\SpeedK}{2}]{\SigShrinkTestik}\MSForAllSpeedI{\SigShrinkTestOKi}\MSForAllSpeedI{\SigShrinkTestFaili}\MSForAllSpeedI[\SpeedI{+}\SpeedShrink]{\SigShrinkOrderi}\MSForAllSpeedI[\SpeedI{-}\SpeedShrink]{\SigTestStarti}\end{MSlist}}
  \caption{Meta-signals for shrinking.}
  \label{fig:ms:shrink}
\end{figure}

\begin{figure}[hbt]
  \centerline{
    \begin{CRlist}
      \CRForAllSpeedKltI{\SigBorderRighti, \SigBorderLeftk}{\begin{tabular}{l}
          \SigShrinkBottomBothRi, \SigShrinkTopTestRi, \SigShrinkTestik,
          \\
          \SigShrinkTopTestLk, \SigShrinkBottomBothLk
        \end{tabular}}\CRForAllSpeedI{\SigShrinkBottomBothLi, \SigIDi}{\SigShrinkIDi, \SigShrinkBottomBothLi}\CRForAllSpeedI{\SigShrinkBottomLi, \SigIDi}{\SigShrinkIDi, \SigShrinkBottomLi}\CRForAllSpeedI{\SigShrinkIDi, \SigShrinkBackLi}{\SigShrinkBackLi, \SigIDi}\ShiftR{\SigRuleBoundi}{\SigShrinkRuleBoundi}\ShiftR{\SigRuleMiddlei}{\SigShrinkRuleMiddlei}\ShiftR{\SigIfil}{\SigShrinkIfil}\ShiftR{\SigThenil}{\SigShrinkThenil}\hline
      \CRForAllSpeedI{\SigShrinkBottomBothLi, \SigMaini}{\SigShrinkBackLi, \SigMaini, \SigShrinkBottomLi}\CRForAllSpeedI{\SigMaini, \SigShrinkBottomBothRi}{\SigShrinkBottomRi, \SigMaini, \SigShrinkBackRi}\CRForAllSpeedI{\SigShrinkBottomBothLi, \SigMaini, \SigShrinkBottomBothRi}{\SigShrinkBackLi, \SigMaini, \SigShrinkBackRi}\CRForAllSpeedI{\SigShrinkBottomLi, \SigMaini}{\SigShrinkBackLi, \SigMaini}\CRForAllSpeedI{\SigMaini, \SigShrinkBottomRi}{\SigMaini, \SigShrinkBackRi}\CRForAllSpeedI{\SigBorderRighti, \SigShrinkBottomRi}{\SigShrinkTopLi, \SigShrinkBottomLi}\CRForAllSpeedI{\SigShrinkBottomLi, \SigBorderRighti}{\SigShrinkBottomRi, \SigShrinkTopRi}\CRForAllSpeedI{\SigShrinkBottomLi, \SigMaini}{\SigMaini, \SigShrinkBackLi}\CRForAllSpeedI{\SigShrinkTopLi, \SigShrinkBackLi}{\SigBorderLefti}\CRForAllSpeedI{\SigShrinkBackRi, \SigShrinkTopRi}{\SigBorderRighti}\hline
      \CRForAllSpeedI{\SigShrinkTopTestLi, \SigShrinkBackLi}{\SigShrinkTestiL, \SigBorderLefti}\CRForAllSpeedI{\SigShrinkBackRi, \SigShrinkTopTestRi}{\SigBorderRighti, \SigShrinkTestiR}\CRForAllSpeedKltI{\SigShrinkTestik, \SigShrinkTestkL}{\SigShrinkTestOKk}\CRForAllSpeedKltI{\SigShrinkTestiR, \SigShrinkTestOKk}{\SigTestStarti}\CRForAllSpeedKltI{\SigShrinkTestiR, \SigShrinkTestik, \SigShrinkTestkL}{\SigTestStarti}\hline
      \CRForAllSpeedKltI{\SigShrinkTestiR, \SigShrinkTestik}{\SigShrinkTestFaili}\CRForAllSpeedKltI{\SigShrinkTestFaili, \SigShrinkTestkL}{\SigShrinkOrderk}\CRForAllSpeedI{\SigShrinkOrderi, \SigBorderLefti}{\SigShrinkBottomBothLi, \SigShrinkTopLi}\end{CRlist}}
  \caption{Collision rules for shrinking and test relative width.}
  \label{fig:rule:shrink}
\end{figure}

The \AGCsimGroup function has to be refined to take the content of this section in account.
For this, it is enough to count any \SigShrinkIDk as if it were a \SigIDk, and ignore the other meta-signals of this section. Note that the number of additional signals accounted for by this section for one collision is bounded, therefore \AGCsimGroup is indeed defined locally.

\subsection{Testing Isolation on Both Sides}
\label{subsec:safety-zone}

It must be ensured that the macro-collision will happen far away enough from any other macro-collision or macro-signals.
The outside signals to consider are of two kinds: probe signals (the ones used here and in the next sub-section) and the one that delimits macro-signals and macro-collisions.
Probe signals are only testing for the presence of other signals; they are not interacting with any other signals nor collisions, so there is no use to bother with them.
The delimiting ones have their speed in $[-\SpeedMaxAbs,\SpeedMaxAbs]$ (\SpeedMaxAbs is the maximum absolute value of any speed in \SpeedSet
), so that it is enough to consider only extreme speed on both side.

\begin{figure}[hbt]
  \centering\footnotesize\scalebox{.9}{\SetUnitlength{1.5em}\begin{tikzpicture}[inner sep=.2em]\newcommand{\TOP}{15}\newcommand{\TOPhat}{13}\newcommand{\BOT}{-5.5}\newcommand{\HEI}{(\TOP-\BOT)}\newcommand{\SpeedJval}{(-1)}\newcommand{\SpeedIval}{(1)}\newcommand{\SpeedM}{(-1/3)}\newcommand{\SpeedMpo}{(5/12)}\newcommand{\LenI}{3}\newcommand{\LenJ}{2}\newcommand{\EpsilonVal}{2}\path (0,0) \CoorNode{O}
  +({\BOT*\SpeedIval},\BOT) \CoorNode{Oi}
  +({\BOT*\SpeedJval},\BOT) \CoorNode{Oj}
  +({\SpeedMpo*\TOP},\TOP) \CoorNode{Oll}
  +({\SpeedM*\TOP},\TOP) \CoorNode{Ol} ;
  \path (Oi) +(-\LenI,0)  \CoorNode{Li}  +(\LenI,0)  \CoorNode{Ri} ;
  \path (Oj) +(-\LenJ,0)  \CoorNode{Lj}  +(\LenJ,0)  \CoorNode{Rj} ;
  \path[name path=fast-left] (O) -- +(-4,.5) ;
  \path[name path=fast-right] (O) -- +(4,.5) ;
  \path[name path=li] (Li) -- +({\HEI*\SpeedIval},{\HEI}) ;
  \path[name path=ri] (Ri) -- +({\HEI*\SpeedIval},{\HEI}) ;
  \SetIntersect{fast-left}{li}{L}
  \SetIntersect{fast-right}{ri}{R}
  \path[name path=lj] (Lj) -- +({\HEI*\SpeedJval},{\HEI}) ;
  \path[name path=rj] (Rj) -- +({\HEI*\SpeedJval},{\HEI}) ;
  \SetIntersect{ri}{lj}{O'}
  \path[name path=r1] (R) -- +({\SpeedMpo*\TOP},{\TOP}) \CoorNode{Rll} ;
  \path[name path=ln] (L) -- +({\SpeedM*\TOP},{\TOP}) \CoorNode{Ll} ;
  \path[name path=lll] (L) -- +({\SpeedMpo*\TOP},{\TOP}) \CoorNode{Lll} ;
  \path[name path=rl] (R) -- +({\SpeedM*\TOP},{\TOP}) \CoorNode{Rl} ;
  \path (O) -- ++({0},{\TOPhat}) \CoorNode{U} ;
  \SetIntersect{ri}{rj}{Rjj}
  \SetIntersect{r1}{ln}{O''}
  \SetIntersect{rl}{lll}{Cm}
  \path[draw] ({\TOPhat-\BOT-\EpsilonVal},{\BOT+\EpsilonVal}) \CoorNode{Ur} ;
  \path[draw] ({\BOT-\TOPhat+\EpsilonVal},{\BOT+\EpsilonVal}) \CoorNode{Ul} ;
  \draw[HYPOTETIC] (Ul) -- node[sloped,above] {(\SpeedMaxAbs)} (U) -- node[sloped,above] {(-\SpeedMaxAbs)}  (Ur) -- (Oi) -- cycle ; 
  \DrawSigMainiLUAboveLeft(Oi)(O) ;
  \DrawSigMainjLUAboveRight(O)(Oj) ;
  \DrawSigMainmLUBelowLeft(Ol)(O) ;
  \DrawSigMainmpoLUBelowRight(O)(Oll) ;
  \DrawSigFastLeftLUBelowLeft(L)(O) ;
  \DrawSigFastRightLUBelowRight(O)(R) ;
  \DrawSigBorderLeftiLUAboveLeft(Li)(L) ;
  \DrawSigBorderLeftmLUBelowLeft(Ll)(L) ;
  \DrawSigBorderLeftjLUAboveRight(O')(Lj) ;
  \DrawSigBorderLeftmpoLUBelowRight(L)(Lll) ;
  \DrawSigBorderRightmpoLUBelowRight(R)(Rll) ;
  \DrawSigBorderRightmLUBelowLeft(Rl)(R) ;
  \DrawSigBorderRightiLUAboveLeft(Ri)(R) ;
  \DrawSigBorderRightjLUAboveRight(Rjj)(Rj) ;
  \begin{scope}[inner sep=.4em]
    \path (Oi) node[below] {\SafeBot${=}(-\BaseSpeed{i},-1)$} ;
    \path (L) node[left] {\OutputLeft} ;
    \path (R) node[right] {\OutputRight} ;
    \path (O) node[above right=0ex] {$(0,0)$} ;
    \path (Cm) node[right=.1ex] {\SafeCm} ;
    \path (U) node[above right] {\SafeTop} ;
    \path (Ur) node[right] {\SafeRight} ;
    \path (Ul) node[left] {\SafeLeft} ;
  \end{scope}
\end{tikzpicture}}\caption{Safety zone (inside the dotted perimeter).}
  \label{fig:safety-zone}
\end{figure}
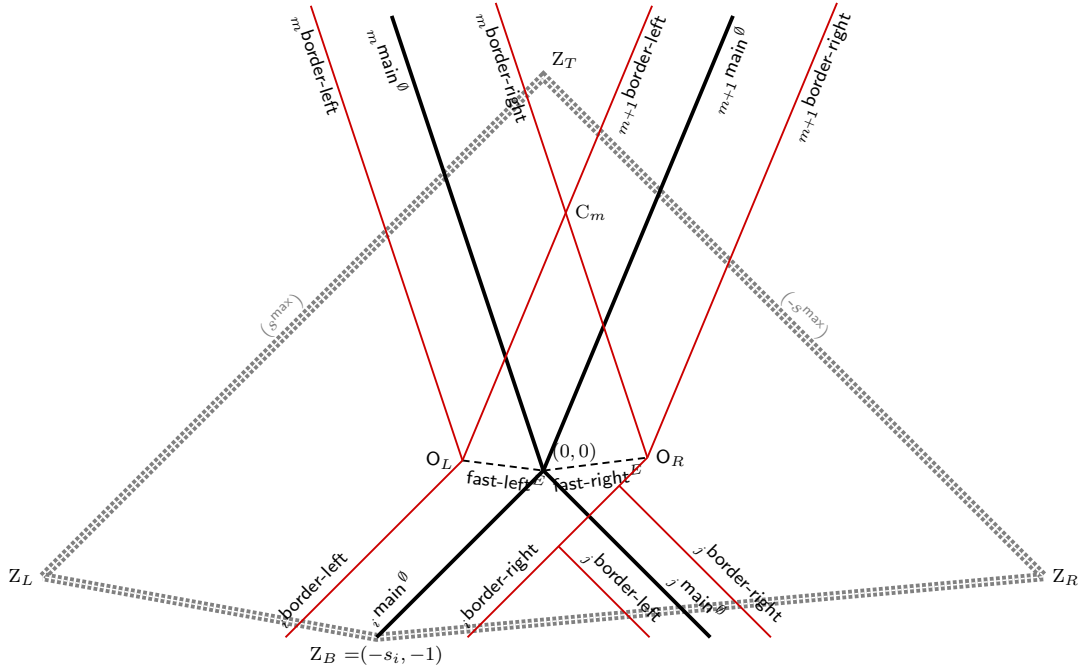

\RefFigure{fig:safety-zone} shows the extent of the safety zone: all the preparation and the resolution is restrained inside it.
This large area can be guaranteed from the positions of \SigMaini and \SigMaink (right next to \SigMaini),  provided the macro-signals have not met yet, so that their width is bounded by the distances between \SigMain.
The extreme points on top of the collisions are when output macro-signals are separating one from the other as the point \SafeCm for macro-signals of speed \BaseSpeed{\SpeedIndexM} and \BaseSpeed{\SpeedIndexM+1} in \RefFig{fig:safety-zone}.
To ensure a large enough safety zone, it will be delimited by four point: \SafeTop, \SafeLeft, \SafeRight and \SafeBot such that:
\begin{enumerate}
\item all \SafeCm ($1\le \SpeedIndexM<\SpeedSetNumber$) are in the zone,
\item the slope of segment from \SafeLeft to \SafeTop correspond to the speed \SpeedMaxAbs,
\item the slope of segment from \SafeRight to \SafeTop correspond to the speed $-\SpeedMaxAbs$, 
\item \SafeLeft and  \SafeRight are low/early enough so that the whole macro-collision is wholly inside the area, and
\item \SafeLeft and  \SafeRight can be reached by signals from  \SafeBot.
\end{enumerate}

First, the position of \SafeBot is the one where \SigMaini and \SigTestStarti meet.
The position of \SafeTop is computed from this position and the position of the simulated collision, that is the intersection of \SigMaini and \SigMaink.
To compute the speed to get to the right positions, a coordinate system is introduced where these points have coordinates $(-\BaseSpeed{i},-1)$ and $(0,0)$ respectively.
We take $4 \SpeedMaxAbs$ as the width of the left macro-signal as it is an upper bound of it and induces upper bounds on the \SafeCm.

Thanks to the choice of relative width of each half of each macro-signal, the point \OutputLeft and \OutputRight have coordinates
$\left(
  x_{\OutputLeft} = -2 \SpeedMaxAbs - 2 \SpeedMaxAbs \frac{\SpeedMaxAbs}{\SpeedRapid}
  ,
  t_{\OutputLeft} = \frac{2 \SpeedMaxAbs}{\SpeedRapid}
\right)$
and
$\left(
  x_{\OutputRight} = 2 \SpeedMaxAbs - 2 \SpeedMaxAbs \frac{\SpeedMaxAbs}{\SpeedRapid}
  ,
  t_{\OutputRight} = \frac{2 \SpeedMaxAbs}{\SpeedRapid}
\right)$.
The point \SafeCm has coordinates
$\left(
  x_{\SpeedIndexM} = x_{\OutputLeft} + 4 \SpeedMaxAbs \frac{\BaseSpeed{\SpeedIndexM+1}}{\BaseSpeed{\SpeedIndexM+1}-\BaseSpeed{\SpeedIndexM}}
  ,
  t_{\SpeedIndexM} = t_{\OutputLeft} + \frac{4 \SpeedMaxAbs}{\BaseSpeed{\SpeedIndexM+1}-\BaseSpeed{\SpeedIndexM}}
\right)$.

We take, as coordinates of \SafeTop: 
\begin{equation*}
  \SafeTop
  \left(
    x_T=0
    ,\ 
    t_T=2 \max\left\{\,t_m \,\middle|\, 1\leq m <\SpeedSetNumber\,\right\}
  \right)
  \enspace .
\end{equation*}
So that, at time $\max\{\,t_m\}$, all the signals are separated and still within the safe zone. 

By setting the point \SafeLeft and \SafeRight to be to be on the line $t=\TestEpsilon-1$ for a sufficiently small positive \TestEpsilon, they have coordinates 
$\left(
  x_{L}=-\SpeedMaxAbs(t_T-\TestEpsilon)
  ,
  t_{L}=\TestEpsilon-1
\right)$
and
$\left(
  x_{R}=\SpeedMaxAbs(t_T-\TestEpsilon)
  ,
  t_{R}=\TestEpsilon-1
\right)$.

It is enough to ensure that no signal (except for probing ones) enters through the bottom of the safety zone since their speed prevent them from entering from the other two sides.
The scheme to send signals to \SafeLeft and \SafeRight is depicted in \RefFig{fig:test-safety-identify}.

After the shrinking, pairs of fast enough signals are issued on both side so that they meet on the extreme points: 
\SigTestLefti and \SigTestLeftUpik on the left and \SigTestRighti and \SigTestRightUpi on the right.
The signals \SigTestLefti, \SigTestRighti and \SigTestRightUpi are issued from \SigMaini while \SigTestLeftUpik and \SigTestRightik (later to become \SigTestRightikj) are issued from the collision between \SigMaink and \SigTestRighti at \Uik.

\begin{figure}[hbt]
  \centering\SetUnitlength{14em}\scalebox{.75}{\SafeZonePictureParameter\begin{tikzpicture}[y=1.5\unitlength]
  \path[use as bounding box,clip] (\SpeedIval+-2+\EpsilonVal-.05,-2.2em) rectangle (\SpeedIval+2-\EpsilonVal+.05,.59) ;
  \fill (0,0) \CoorNode{0-i} node[below left=.4ex]{\SafeBot${=}(\SpeedI,-1)$} circle (.015\unitlength) ;
  \fill (-\SpeedKval+\SpeedIval+.1,0) \CoorNode{0-k} ;
  \fill (-\SpeedJval+\SpeedIval,0) \CoorNode{0-j} ;
  \fill (\SpeedIval,1) \CoorNode{top} node[anchor=south west]{$(\SpeedI,1)$} circle (.015\unitlength) ; 
  \DrawSigMainiLUAbove(0-i)(top)
  \DrawSigMainkLUBelowParam[pos=.25](0-k)(top)
  \DrawSigMainjLUAbove(0-j)(top)
  \fill (\SpeedIval+1.75-\EpsilonVal,\EpsilonVal) \CoorNode{SZ-right} node[anchor=south east,inner xsep=0,outer xsep=0]{\SafeRight}  circle (.015\unitlength) ; 
  \path (\SpeedIval+1.75-\EpsilonVal,\EpsilonVal/2) \CoorNode{SZ-right-half};
  \fill (\SpeedIval-1.75+\EpsilonVal,\EpsilonVal) \CoorNode{SZ-left} node[anchor=south west,inner xsep=0,outer xsep=0]{\SafeLeft} circle (.015\unitlength) ; 
  \fill (1.5*\SpeedIval*\EpsilonVal,1.5*\EpsilonVal) \CoorNode{up-i-3} node[below right=-.2em,inner sep=.1em,outer sep=0]{$\left(
      \begin{array}{@{}c@{}}
        1.5\TestEpsilon\SpeedI{-}\SpeedI\\
        1.5\TestEpsilon{-}1
      \end{array}
    \right)$} circle (.015\unitlength) ; 
  \fill (2*\SpeedIval*\EpsilonVal,2*\EpsilonVal) \CoorNode{up-i-4} node[above left=1ex]{$\left(
      \begin{array}{@{}c@{}}
        2\varepsilon\SpeedI{-}\SpeedI\\
        2\varepsilon{-}1
      \end{array}
    \right)$} circle (.015\unitlength) ;
  \DrawSigTestLeftiLUBelow(0-i)(SZ-left)
  \path[name path=main-j-half] (0-i) -- (SZ-right-half) ;
  \path[name path=main-j] (0-j) -- (top) ;
  \path[name path=main-k] (0-k) -- (top) ;
  \SetIntersect{main-k}{main-j-half}{U}
  \DrawSigTestRightUpiLUAbove(0-i)(SZ-right)
  \path[name path=test-right] (U) -- (SZ-right) ;
  \SetIntersect{main-k}{test-right}{r-k}
  \DrawSigTestRightiLUBelow(0-i)(U)
  \SetIntersect{main-j}{test-right}{r-j}
  \DrawSigTestRightikLUBelow(r-k)(r-j)
  \DrawSigTestRightikjLUBelow(r-j)(SZ-right)
  \DrawSigTestLeftUpikLUAbove(U)(SZ-left)
  \DrawSigTestLeftOKiLUAboveParam[pos=.6](SZ-left)(up-i-3) ;
  \DrawSigMainTestOKiLLeft(up-i-3)(up-i-4)
  \DrawSigTestRightOKijLUAbove(up-i-4)(SZ-right)
  \DrawSigTestStartiLUBelow(.3,-.06)(0,0)
  \begin{scope}[inner sep=0,outer sep=0]        
    \DrawSigCheckMaybeijLUBelowRight(up-i-4)([shift={(.36,.1)}]up-i-4)
    \DrawSigCheckUpijLUAboveRight(up-i-4)([shift={(.36,.2)}]up-i-4)
  \end{scope}
  \fill (U) node[anchor=south west]{\Uik} circle (.015\unitlength) ; 
\end{tikzpicture}}
  
  \scriptsize
  The point where \SigMaini and \SigMaink meet has coordinate $(0,0)$.
  For clarity, border signals are not displays.
  \caption{Testing for the safety zone and identifying the rightmost speed.}
  \label{fig:test-safety-identify}
\end{figure}

On the left, after crossing \SigBorderLefti, if signal \SigTestLefti meets nothing before \SigTestLeftUpik then it returns as \SigTestLeftOKi (and \SigTestLeftUpik is destroyed).
When \SigTestLeftOKi meet \SigMaini, then the later turns into \SigMainTestOKi to record the success on left.
Otherwise, anything met  on the left is either too close or participates in the macro-collision (and \SigMaini is not the left-most involved).
In both cases, the macro-collision should be aborted.
This is done by coming back as \SigTestLeftFaili.
This ones cancels \SigTestLeftUpik and any signal returning from the right side and disappears, the macro-collision is not started.
This is depicted in \RefFig{fig:test-fail-left} where it can be seen that the process is restarted later with success.

On the right, \SigTestRighti tests for obvious non participating macro-signals and collects the index of the rightmost potentially participating macro-signal (next stage checks whether all potentially participating signals are rightly positioned).
It verifies that \SigMainj are in strictly decreasing speed order.
It also verifies that a \SigMainj is reached for any \SigBorderLeftj encountered by turning into \SigTestRightWaitikj  in between (this is not indicated in \RefFig{fig:test-safety-identify}).
The signal \SigTestRightikj also initiate a shrinking process on each macro-signal on the right when it meets a \SigBorderLeftl (to avoid useless macro-collision initialisation).

Signal \SigTestRightikj updates the least speed index encountered when it meets \SigBorderLeftl (becoming \SigTestRightWaitikl) with $\SpeedIndexL<\SpeedIndexJ$ and at crossing \SigMainl becomes \SigTestRightikl.
When \SigTestRightikj and \SigTestRightUpi meet, they comes back as \SigTestRightOKij.
That way, it brings back the index of the rightmost speed.
When \SigTestRightOKij meets \SigMainTestOKi, the next stage of the macro-collision starts.

The signals \SigTestLefti and \SigTestRightUpi head straight to \SafeLeft and \SafeRight, so that their speed are:
$\frac{x_{L} + \BaseSpeed{\SpeedIndexI}}{\TestEpsilon}$
and
$\frac{x_{R} + \BaseSpeed{\SpeedIndexI}}{\TestEpsilon}$ ($=\SpeedTestRightUpi$) respectively.
The speed of \SigTestRighti can be anything greater than the speed of \SigTestRightUpi, double that speed  ($=2\SpeedTestRightUpi$) for example.

To compute the other speed, some coordinates have to be computed (in particular \Uik).
This is straightforward once an appropriate formula is given.
If the speeds (\SpeedA, \SpeedB and \SpeedSS) are given as in \RefFig{fig:computing}, then the coordinates of the intersection point, $M$, are:
\begin{equation}
  \label{eq:intersection}
  \left(\displaystyle
    \XZero,\TZero\right)
  \enspace .
\end{equation}

\begin{figure}[hbt]
  \centering\scriptsize\SetUnitlength{1.3em}\newcommand{\SpeedIval}{.5}
  \newcommand{\SpeedJval}{-.8}
  \newcommand{\SpeedSval}{4}
  \pgfmathsetmacro\Tzeroval{-(\SpeedSval-\SpeedIval)/(\SpeedSval-\SpeedJval)}
  \pgfmathsetmacro\Xzeroval{\SpeedJval*\Tzeroval}
  \SetUnitlength{10em}
  \begin{tikzpicture}
    \fill (-\SpeedIval,-1) coordinate (0-0) node[anchor=north east]{$(-\SpeedA,-1)$} circle (.01) ;
    \fill (0,0) coordinate (top) node[anchor=south west]{$(0,0)$} circle (.01) ; 
    \fill (\Xzeroval,\Tzeroval) coordinate (m0) node[above right=.3ex]{$M$} node[below right=.3ex]{$\left(\displaystyle
        \XZero,\TZero\right)$} circle (.01) ;
    \draw (0-0) -- node[sloped,above] {$x=\SpeedA.y$} (top) ;
    \draw (0-0) -- node[sloped,above] {$x+\SpeedA=\SpeedSS.(y+1)$} (m0) ;
    \draw (top) -- node[sloped,above] {$x=\SpeedB.y$} (m0) ;
    \draw[<->,shift={(-.1,0)}] (- \SpeedIval, -1) -- node[sloped,above] {$1$} (- \SpeedIval,0) ;
    \draw[<->,shift={(0,.1)}] (- \SpeedIval,0) -- node[sloped,above] {\SpeedA} (0,0) ;
  \end{tikzpicture}
  \caption{Computing coordinates.}
  \label{fig:computing}
\end{figure}

Thus, the coordinates of \Uik are: 
\begin{equation*}
  \left(
    x_{U}^{i,k} = -\BaseSpeed{\SpeedIndexK} \frac{2 \SpeedTestRightUpi - \BaseSpeed{\SpeedIndexI}}{2 \SpeedTestRightUpi - \BaseSpeed{\SpeedIndexK}}
    , 
    t_{U}^{i,k} = - \frac{2 \SpeedTestRightUpi - \BaseSpeed{\SpeedIndexI}}{2 \SpeedTestRightUpi - \BaseSpeed{\SpeedIndexK}}
  \right)  
  \enspace.
\end{equation*}

The used meta-signals and (success) collision rules are defined in Figs.\,\ref{fig:ms:test} and \ref{fig:rule:test-ok}.
To be coherent with \RefFig{fig:phases}, the duration of $2\TestEpsilon$ with a start at $8/10$ is assured with the value:
\begin{math}
  \TestEpsilon=1/16 
  \enspace.
\end{math}

\begin{figure}[hbt]
  \centerline{\begin{ParameterList}
      \TestEpsilon & \frac{1}{16} \\[.6em]
      $\ForAllSpeedI,\, \SpeedTestRightUpi$ & \frac{x_{R} + \BaseSpeed{\SpeedIndexI}}{\TestEpsilon}\\[.6em]
      $\ForAllSpeedKltI,\, t_{U}^{i,k}$ & - \BaseSpeed{\SpeedIndexK} \frac{2 \SpeedTestRightUpi - \BaseSpeed{\SpeedIndexI}}{2 \SpeedTestRightUpi - \BaseSpeed{\SpeedIndexK}} \\[.6em]$\ForAllSpeedKltI,\, x_{U}^{i,k}$ & \BaseSpeed{\SpeedIndexK}. t_{U}^{i,k} \\[.6em]
      $\ForAllSpeedI,\, \SpeedTestRightik$ & \frac{x_R - x_{U}^{i,k}} {t_R - t_{U}^{i,k}} \\[.6em]
      $\ForAllSpeedI,\, \SpeedTestBackRighti$ &
      \frac{ 2\TestEpsilon.\SpeedI - \SpeedI - x_R }{ \TestEpsilon } \\[.6em]
      $\ForAllSpeedI,\, \SpeedTestBackLefti$ & \frac{1.5 \TestEpsilon.\SpeedI - \SpeedI - x_L }{ 0.5 \TestEpsilon } \\[.6em]
    \end{ParameterList}
    \quad
    \begin{MSlist}
      \MSForAllSpeedI[\frac{x_L + \SpeedI}{\TestEpsilon}]{\SigTestLefti}\MSForAllSpeedI[\SpeedTestRightUpi]{\SigTestRightUpi}\MSForAllSpeedI[2 \SpeedTestRightUpi]{\SigTestRighti}\MSForAllSpeedJteKltI[\SpeedTestRightik]{\SigTestRightikj}\MSForAllSpeedJltKltI[\SpeedTestRightik]{\SigTestRightWaitikj}\MSForAllSpeedKltI[\frac{x_L - x_{U}^{i,k}}{t_L - t_{U}^{i,k} }]{\SigTestLeftUpik}\MSForAllSpeedKltI[\SpeedTestBackLefti]{\SigTestLeftOKi}\MSForAllSpeedJltI[\SpeedTestBackRighti]{\SigTestRightOKij}\MSForAllSpeedI{\SigMainTestOKi}\hline
      \MSForAllSpeedKltI[\SpeedTestBackLefti]{\SigTestLeftFaili}\MSForAllSpeedJltI[\SpeedTestBackRighti]{\SigTestRightFaili}\MSForAllSpeedI{\SigMainTestFailLi}\MSForAllSpeedI{\SigMainTestFailRi}\end{MSlist}}
  \caption{Meta-signals for testing.}
  \label{fig:ms:test}
\end{figure}

\begin{figure}[hbt]
  \centerline{
    \begin{CRlist}
      \CRForAllSpeedI{\SigMaini, \SigTestStarti}{\SigTestLefti,\SigMaini,\SigTestRightUpi,\SigTestRighti}\CRForAllSpeedKltI{\SigTestRighti, \SigMaink}{\SigTestLeftUpik,\SigMaink,\SigTestRightikk}\CRForAllSpeedLltJleKltI{\SigTestRightikj, \SigBorderLeftl}{\SigShrinkTopLl, \SigTestRightWaitikl, \SigShrinkBottomBothLl}\CRForAllSpeedJltKltI{\SigTestRightWaitikj, \SigMainj}{\SigMainj, \SigTestRightikj}\CRForAllSpeedKltI{\SigTestLefti, \SigTestLeftUpik}{\SigTestLeftOKi}\CRForAllSpeedI{\SigTestLeftOKi, \SigMaini}{\SigMainTestOKi}\CRForAllSpeedJleKltI{\SigTestRightUpi, \SigTestRightikj}{\SigTestRightOKij}\CRForAllSpeedI{\SigTestRightUpi, \SigTestRightWaitij, \SigMainj}{\SigTestRightOKij, \SigMainj}\CRForAllSpeedJltI{\SigMainTestOKi, \SigTestRightOKij}{\SigMaini, \SigCheckUpij, \SigCheckMaybeij}\end{CRlist}}
  \caption{Collision rules for testing, success case.}
  \label{fig:rule:test-ok}
\end{figure}

\subsubsection{Test  Failure}
\label{sub:test-fail}
\begin{figure}[hbt]
  \centering\includegraphics[height=2.75cm]{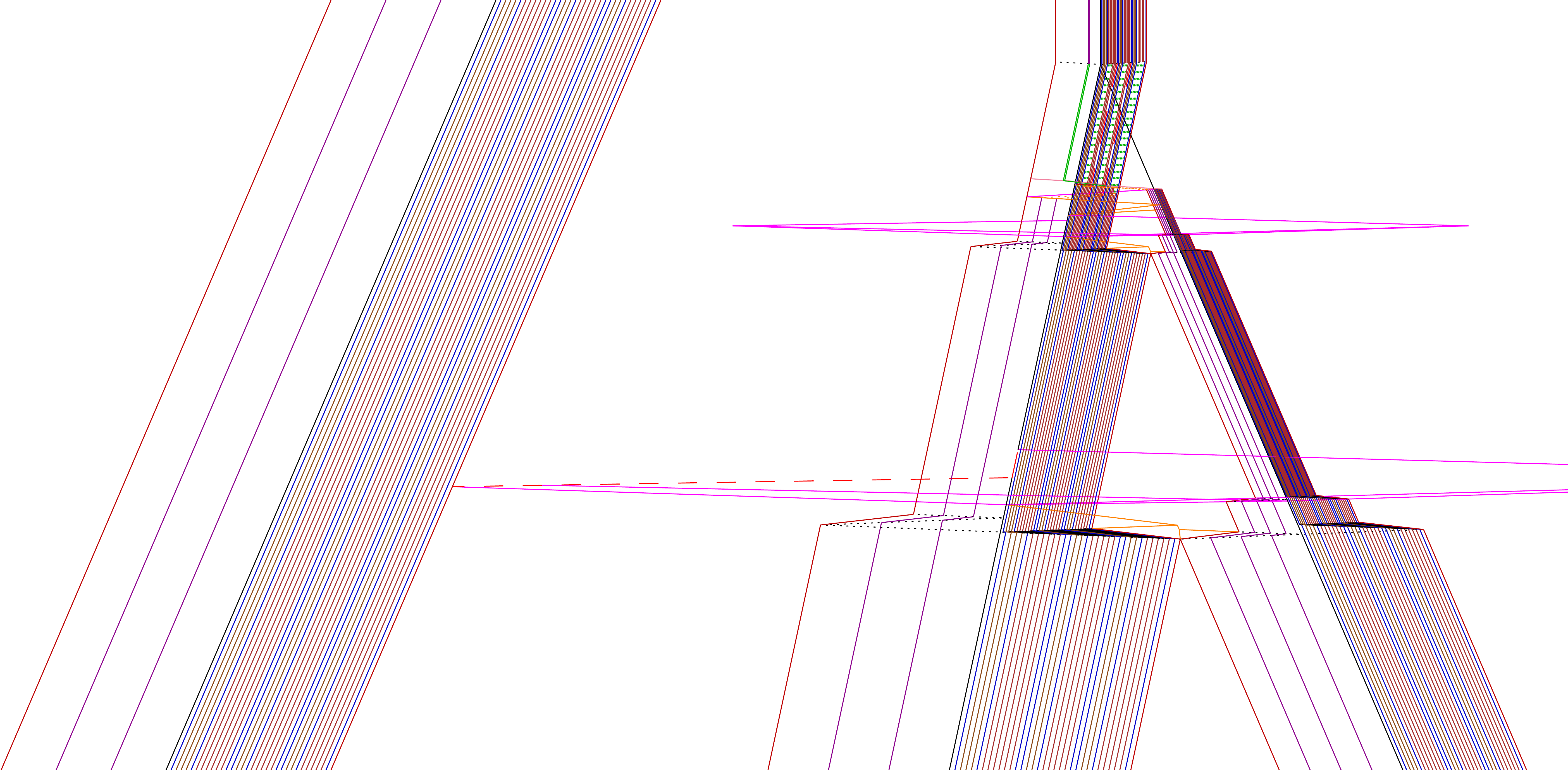}\qquad\includegraphics[height=2.75cm]{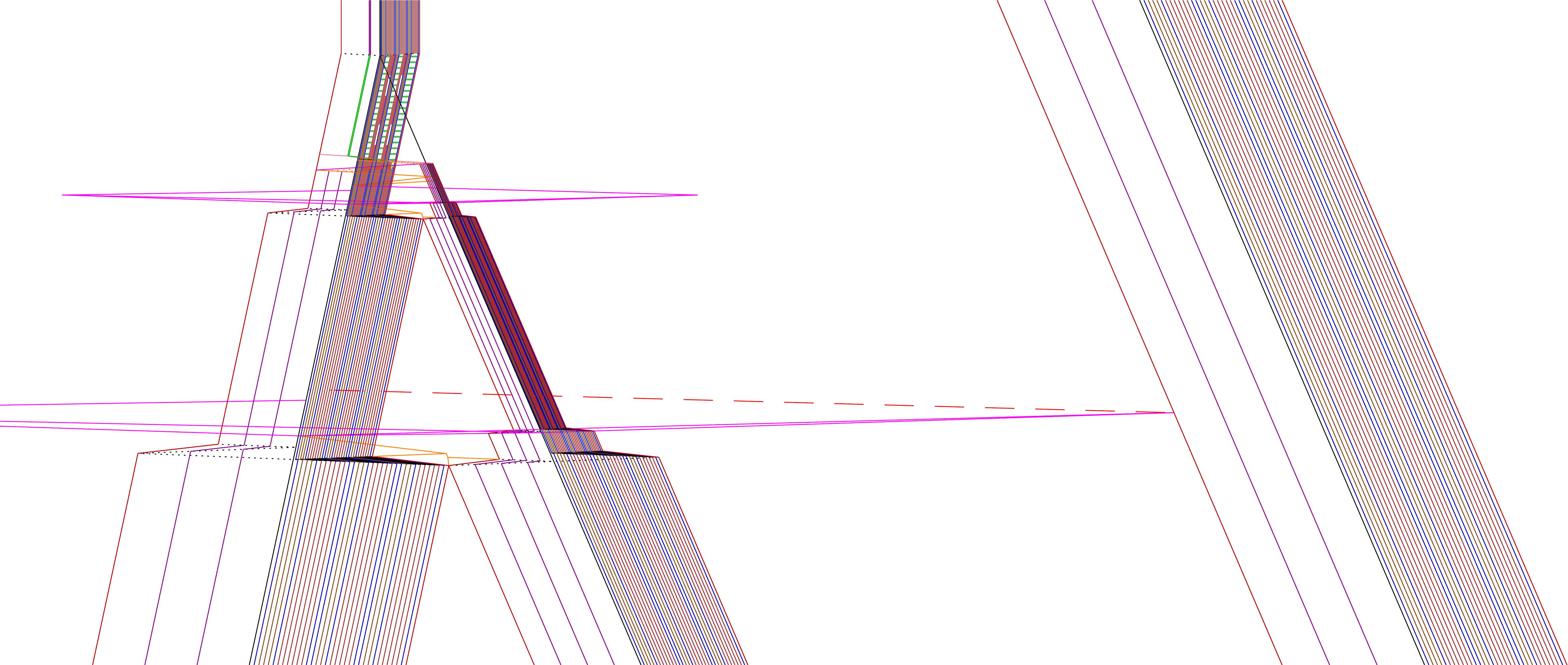}\caption{Detection of a signal not participating on the left and on the right.}
  \label{fig:test-fail-left}
\end{figure}

As depicted on \RefFig{fig:test-fail-left}, the test can fail because of the presence of unwanted signals on the left or on the right.
Both cases are briefly presented.

\RefFigure{fig:test-safety-fail-left} presents the scheme for failure of test on the left.
This happens if \SigTestLefti encounters anything: it is either  something that should not be involved and is too closed or just that \SigMaini is not the leftmost macro-signal involved, so that macro-collision has to be aborted to be started by the rightful macro-signal.
Since probe signal are not concerned, the signals that can be met on the left are: \SigMainl, \SigReadyl, \SigBorderRightl or \SigShrinkTopRl (for any \SpeedIndexL).

On meeting, \SigTestLefti bounces back as \SigTestLeftFaili.
It arrives back to \SigMaini and marks it as \SigMainTestFailLi.
When \SigTestRightOKij meets \SigMainTestFailLi, it is destroyed and \SigMaini is restored; nothing is emitted so that the macro-collision is aborted.
The signal \SigTestLeftUpik has to be disposed of; this is done either on \SigTestLeftFaili or on  \SigMainTestFailLi.

\begin{figure}[hbt]
  \centering\SetUnitlength{14em}\scalebox{.75}{\SafeZonePictureParameter\begin{tikzpicture}[y=1.5\unitlength]
  \path[use as bounding box,clip] (\SpeedIval+-2+\EpsilonVal-.05,-1.4em) rectangle (\SpeedIval+2-\EpsilonVal+.05,.59) ;
  \fill (0,0) \CoorNode{0-i} node[below left=.4ex]{\SafeBot${=}(\SpeedI,-1)$} circle (.015\unitlength) ;
  \fill (-\SpeedKval+\SpeedIval+.1,0) \CoorNode{0-k} ;
  \fill (-\SpeedJval+\SpeedIval,0) \CoorNode{0-j} ;
  \fill (\SpeedIval,1) \CoorNode{top} node[anchor=south west]{$(\SpeedI,1)$} circle (.015\unitlength) ; 
  \DrawSigMainiLUAbove(0-i)(top)
  \DrawSigMainkLUBelowParam[pos=.25](0-k)(top)
  \DrawSigMainjLUAbove(0-j)(top)
  \fill (\SpeedIval+1.75-\EpsilonVal,\EpsilonVal) \CoorNode{SZ-right} node[anchor=south east,inner xsep=0,outer xsep=0]{\SafeRight}  circle (.015\unitlength) ; 
  \path (\SpeedIval+1.75-\EpsilonVal,\EpsilonVal/2) \CoorNode{SZ-right-half};
  \fill (\SpeedIval-1.75+\EpsilonVal,\EpsilonVal) \CoorNode{SZ-left} ; \fill (2*\SpeedIval*\EpsilonVal,2*\EpsilonVal) \CoorNode{up-i-4} node[left=1ex]{$\left(
      \begin{array}{@{}c@{}}
        2\varepsilon\SpeedI{-}\SpeedI\\
        2\varepsilon{-}1
      \end{array}
    \right)$} circle (.015\unitlength) ;
  \path[name path=main-j-half] (0-i) -- (SZ-right-half) ;
  \path[name path=main-j] (0-j) -- (top) ;
  \path[name path=main-k] (0-k) -- (top) ;
  \SetIntersect{main-k}{main-j-half}{U}
  \path[name path=test-right] (U) -- (SZ-right) ;
  \SetIntersect{main-k}{test-right}{r-k}
  \DrawSigTestRightiLUBelow(0-i)(U)
  \SetIntersect{main-j}{test-right}{r-j}
  \DrawSigTestRightikLUBelow(r-k)(r-j)
  \DrawSigTestRightikjLUBelow(r-j)(SZ-right)
  \DrawSigTestRightUpiLUAbove(0-i)(SZ-right)
  \DrawSigTestRightOKijLUAbove(up-i-4)(SZ-right)
  \DrawSigTestStarti(.3,-.06)(0,0)
  \path[name path=e] (-.75,0) \CoorNode{E0} -- +(.1,.7) \CoorNode{Et} ;
  \DrawSigBorderRightlLUAbove(E0)(Et)
  \path[name path=test-left] (0-i) -- (SZ-left) ;
  \SetIntersect{test-left}{e}{E1}
  \path[name path=test-left-fail] (E1) -- (.9*\EpsilonVal*\SpeedIval,.9*\EpsilonVal)  \CoorNode{I1};
  \path[name path=test-left-up] (U) -- (SZ-left) ;
  \SetIntersect{test-left-fail}{test-left-up}{EI1}
  \DrawSigTestLeftiLUBelow(0-i)(E1)
  \DrawSigTestLeftUpikLUAboveRight(EI1)(U)
  \DrawSigTestLeftFailiLUAbove(E1)(I1) ;
  \DrawSigMainTestFailLiLLeft(I1)(up-i-4)
  \fill (U) node[anchor=south west]{\Uik} circle (.015\unitlength) ; 
\end{tikzpicture}
} 
  \caption{Testing for the safety zone, fail on left.}
  \label{fig:test-safety-fail-left}
\end{figure}

\RefFigure{fig:test-safety-fail-right} presents the scheme for failure of test on the right.
The alternation of \SigTestRightijl and \SigTestRightWaitijl is not indicated for clarity although they are used to detect failure.
The failure might come from some \SigBorderLeftl with $l$ too small or from \SigTestRightUpi meeting \SigTestRightWaitijl, i.e. before \SigMainl is met.
In any failure case, \SigTestRightijl (or \SigTestRightWaitijl) bounces back as \SigTestRightFaili.
When \SigTestRightFaili meets \SigMainTestOKi, it is destroyed and \SigMaini is restored; nothing is emitted so that the macro-collision is aborted.
The signal \SigTestRightUpi has to be disposed, this happens on meeting \SigTestRightFaili.

It might happen that the \SigTestRightFaili arrives before \SigTestLeftOKi onto \SigMaini.
It might also happen that there is failure on both side and arrival onto  \SigMaini can be in any order.
The listed rules do take this into account.

\begin{figure}[hbt]
  \centering\SetUnitlength{14em}\scalebox{.75}{\SafeZonePictureParameter\begin{tikzpicture}[y=1.5\unitlength]
  \path[use as bounding box,clip] (\SpeedIval+-2+\EpsilonVal-.05,-1.7em) rectangle (\SpeedIval+2-\EpsilonVal+.05,.59) ;
  \fill (0,0) \CoorNode{0-i} node[below left=.4ex]{\SafeBot${=}(\SpeedI,-1)$} circle (.015\unitlength) ;
  \fill (-\SpeedKval+\SpeedIval+.1,0) \CoorNode{0-k} ;
  \fill (-\SpeedJval+\SpeedIval,0) \CoorNode{0-j} ;
  \fill (\SpeedIval,1) \CoorNode{top} node[anchor=south west]{$(\SpeedI,1)$} circle (.015\unitlength) ; 
  \DrawSigMainiLUAbove(0-i)(top)
  \DrawSigMainkLUBelowParam[pos=.25](0-k)(top)
  \DrawSigMainjLUAbove(0-j)(top)
  \fill (\SpeedIval+1.75-\EpsilonVal,\EpsilonVal) \CoorNode{SZ-right} ; 
  \path (\SpeedIval+1.75-\EpsilonVal,\EpsilonVal/2) \CoorNode{SZ-right-half};
  \fill (\SpeedIval-1.75+\EpsilonVal,\EpsilonVal) \CoorNode{SZ-left} node[anchor=south west,inner xsep=0,outer xsep=0]{\SafeLeft} circle (.015\unitlength) ; 
  \fill (1.5*\SpeedIval*\EpsilonVal,1.5*\EpsilonVal) \CoorNode{up-i-3} node[below right=-.2em,inner sep=0,outer sep=0]{$\left(
      \begin{array}{@{}c@{}}
        1.5\TestEpsilon\SpeedI{-}\SpeedI\\
        1.5\TestEpsilon{-}1
      \end{array}
    \right)$} circle (.015\unitlength) ; 
  \DrawSigTestLeftiLUBelow(0-i)(SZ-left)
  \path[name path=main-j-half] (0-i) -- (SZ-right-half) ;
  \path[name path=main-j] (0-j) -- (top) ;
  \path[name path=main-k] (0-k) -- (top) ;
  \SetIntersect{main-k}{main-j-half}{U}
  \path[name path=test-right] (U) -- (SZ-right) ;
  \SetIntersect{main-k}{test-right}{r-k}
  \DrawSigTestRightiLUBelow(0-i)(U)
  \SetIntersect{main-j}{test-right}{r-j}
  \DrawSigTestRightijkLUBelow(r-k)(r-j)
  \DrawSigTestLeftUpikLUAbove(U)(SZ-left)
  \DrawSigTestLeftOKiLUAboveParam[pos=.6](SZ-left)(up-i-3) ;
  \DrawSigTestStarti(.3,-.06)(0,0)
  \path[name path=e] (2,0) \CoorNode{E0} -- +(.1,.7) \CoorNode{Et} ;
  \DrawSigBorderRightlLUBelow(E0)(Et)
  \path[name path=test-right] (0-i) -- (SZ-right) ;
  \SetIntersect{test-right}{e}{E1}
  \path[name path=test-right-fail] (E1) -- (1.7*\EpsilonVal*\SpeedIval,1.7*\EpsilonVal)  \CoorNode{I1};
  \path[name path=test-right-up] (U) -- (SZ-left) ;
  \SetIntersect{test-right-fail}{test-right-up}{EI1}
  \DrawSigTestLeftiLUBelow(0-i)(SZ-left)
  \DrawSigTestRightijLUBelow(r-j)(E1)
  \DrawSigTestRightUpiLUAbove(0-i)(EI1)
  \DrawSigMainTestOKiLLeft(up-i-3)(I1)
  \DrawSigTestRightFailiLUAbove(I1)(E1)
  \fill (U) node[anchor=south west]{\Uik} circle (.015\unitlength) ; 
\end{tikzpicture}
}
  \caption{Testing for the safety zone, fail on right.}
  \label{fig:test-safety-fail-right}
\end{figure}

The used meta-signals and (failure) collision rules are defined in Figs.\,\ref{fig:ms:test} and \ref{fig:rule:test-fail}.
The rules in \RefFigure{fig:rule:test-fail} are divided in three part: fail on left only, fail on right only and additional rules in case fail on both left and right.

\begin{figure}[hbt]
  \centerline{
    \begin{CRlist}
      \CRForAllSpeedIL{\SigBorderRightl, \SigTestLefti}{\SigBorderRightl, \SigTestLeftFaili}\CRForAllSpeedIL{\SigShrinkTopRl, \SigTestLefti}{\SigShrinkTopRl, \SigTestLeftFaili}\CRForAllSpeedIL{\SigReadyl, \SigTestLefti}{\SigReadyl, \SigTestLeftFaili}\CRForAllSpeedIL{\SigMainl, \SigTestLefti}{\SigMainl, \SigTestLeftFaili}\CRForAllSpeedI{\SigTestLeftFaili, \SigTestLeftUpik}{\SigTestLeftFaili}\CRForAllSpeedI{\SigTestLeftFaili, \SigMaini}{\SigMainTestFailLi}\CRForAllSpeedI{\SigMainTestFailLi, \SigTestLeftUpik}{\SigMainTestFailLi}\CRForAllSpeedI{\SigMainTestFailLi, \SigTestRightOKij}{\SigMaini}\hline \CRForAllSpeedKltILleK{\SigTestRightijk, \SigBorderLeftl}{\SigTestRightFaili, \SigBorderLeftl}\CRForAllSpeedKltILleK{\SigTestRightijk, \SigShrinkTopLl}{\SigTestRightFaili, \SigShrinkTopLl}\CRForAllSpeedJltKltI{\SigTestRightUpi, \SigTestRightWaitij}{\SigTestRightFaili}\CRForAllSpeedKltI{\SigTestRightFaili, \SigTestRightUpi}{\SigTestRightFaili}\CRForAllSpeedI{\SigMaini, \SigTestRightFaili}{\SigMainTestFailRi}\CRForAllSpeedI{\SigTestLeftOKi, \SigMainTestFailRi}{\SigMaini}\CRForAllSpeedI{\SigMainTestOKi, \SigTestRightFaili}{\SigMaini}\hline  \CRForAllSpeedI{\SigTestLeftFaili, \SigMainTestFailRi}{\SigMaini}\CRForAllSpeedI{\SigMainTestFailLi, \SigTestRightFaili}{\SigMaini}\end{CRlist}}
  \caption{Collision rules for testing, failure cases.}
  \label{fig:rule:test-fail}
\end{figure}

The function \AGCsimGroup can be trivially extended for this section by ignoring all of its signals, as they don't affect the identity of macro-signals. Again, the number of additional signals accounted for by this section for one collision is bounded, therefore \AGCsimGroup is indeed defined locally.

\subsection{Check Participating Signals}
\label{subsec:check}

From this point, it is known that the index of involved macro-signals ranges from \SpeedIndexJ to \SpeedIndexI (included).
But it is not known whether they actually participate in one single macro-collision (the situations in \RefFig{fig:shrink-role} are not yet distinguished).

\begin{figure}[hbt]
  \centering\newcommand{\Hei}{3.5cm}
  \SubFigure[scheme\label{fig:checking-position-of-mains:scheme}]{\footnotesize\SetUnitlength{1.4em}\newcommand{\ClipTop}{2.7}\newcommand{\ClipBot}{-1.2}\renewcommand{\ClipTop}{2.5}\renewcommand{\ClipBot}{-1}\begin{tikzpicture}[y=\Hei/(\ClipTop-\ClipBot)]
\ifDebugPicture\else
  \path[clip] (-5.5,\ClipBot) rectangle (2,\ClipTop+.5) ;
  \fi
  \path (0,10) \CoorNode{top} ;
  \path (-5,\ClipBot) \CoorNode{a0} ;
  \path (-2.5,\ClipBot) \CoorNode{a2} ;
  \path (-1,\ClipBot) \CoorNode{a3} ;
  \path (1,\ClipBot) \CoorNode{aj} ;
  \path (-5.5,-.2) \CoorNode{z0} ;
  \path (10,3) \CoorNode{z1} ;
  \begin{scope}[inner sep=.2em]
    \path[name path=top clip] (-10,\ClipTop) -- (7,\ClipTop);
    \path[name path=m0] (top) -- (a0) ;
    \path[name path=m2] (top) -- (a2) ;
    \path[name path=m3] (top) -- (a3) ;
    \path[name path=mj] (top) -- (aj) ;
    \SetIntersect{top clip}{m0}{d0} 
    \SetIntersect{top clip}{mj}{dj}
    \DrawSigMainiLUAboveRight(a0)(d0) 
    \DrawSigMainjLUAbove(dj)(aj) 
    \path[name path=lower] (z0) -- (z1) ;
    \path[name path=up] (z0) -- ([shift={(-6,0)}]z1) ;
    \SetIntersect{up}{mj}{c}
    \SetIntersect{top clip}{m2}{d2} 
    \SetIntersect{top clip}{m3}{d3} 
    \SetIntersect{lower}{m0}{b0} 
    \SetIntersect{lower}{m2}{b2} 
    \SetIntersect{lower}{m3}{b3} 
    \SetIntersect{lower}{mj}{b4} 
    \DrawSigMainTestOKiLUAbove(a0)(b0)
    \begin{scope}[inner sep=0,outer sep=0] 
      \DrawSigTestRightOKijLUBelowLeft(b0)(1,\ClipBot)
    \end{scope}
    \DrawSigCheckMaybeijLUBelow(b0)(b4)
    \DrawSigCheckUpijLUAbove(b0)(c)
    \DrawSigCheckOKijLUAbove(-2.5,\ClipTop)(c)
  \end{scope}
\end{tikzpicture}\begin{tikzpicture}[y=\Hei/(\ClipTop-\ClipBot)]
  \ifDebugPicture\else
  \path[clip] (-5.5,\ClipBot) rectangle (7,\ClipTop+.5) ;
  \fi
  \path (0,10) \CoorNode{top} ;
  \path (-5,\ClipBot) \CoorNode{a0} ;
  \path (-1.7,\ClipBot) \CoorNode{a1} ;
  \path (2.5,\ClipBot) \CoorNode{a2} ;
  \path (4,\ClipBot) \CoorNode{a3} ;
  \path (7,\ClipBot) \CoorNode{aj} ;
  \path (-5.5,-.2) \CoorNode{z0} ;
  \path (10,3) \CoorNode{z1} ;
  \begin{scope}[inner sep=.2em]
    \path[name path=top clip] (-10,\ClipTop) -- (7,\ClipTop);
    \path[name path=m0] (top) -- (a0) ;
    \path[name path=m1] (top) -- (a1) ;
    \path[name path=m2] (top) -- (a2) ;
    \path[name path=m3] (top) -- (a3) ;
    \path[name path=mj] (top) -- (aj) ;
    \SetIntersect{top clip}{m0}{d0} 
    \SetIntersect{top clip}{m1}{d1} 
    \SetIntersect{top clip}{mj}{dj}
    \DrawSigMainiLUAboveRight(a0)(d0) 
    \DrawSigMainkLUAboveRight(a1)(d1) 
    \DrawSigMainjLUBelow(dj)(aj) 
    \path[name path=lower] (z0) -- (z1) ;
    \path[name path=upper,yshift=.2\unitlength] (-5.5,.3) -- ([yshift=.2\unitlength]z1);
    \SetIntersect{top clip}{m2}{d2} 
    \SetIntersect{top clip}{m3}{d3} 
    \begin{scope}[dashed]
      \draw (d2) -- (a2) ;
      \draw (d3) -- (a3) ;
    \end{scope}
    \SetIntersect{lower}{m0}{b0} 
    \SetIntersect{lower}{m1}{b1} 
    \SetIntersect{lower}{m2}{b2} 
    \SetIntersect{lower}{m3}{b3} 
    \SetIntersect{lower}{mj}{b4} 
    \SetIntersect{upper}{m1}{c} 
    \DrawSigMainTestOKiLUAbove(a0)(b0)
    \begin{scope}[inner sep=0,outer sep=0] 
      \DrawSigTestRightOKijLUBelowLeft(b0)(1,\ClipBot)
    \end{scope}
    \begin{scope}[StyleCheck]
      \draw (b0) -- (c) ; 
      \draw (c) -- (b2) ; 
      \draw (c) -- (b3) ;  
    \end{scope}
    \DrawSigCheckMaybeijLUBelow(b0)(b1)
    \DrawSigCheckijLUBelow(b1)(b4)
    \DrawSigCheckUpijLUAbove(b0)(c)
    \DrawSigCheckInterceptikjLUAboveLeft(c)(b4) 
    \DrawSigCheckOKijLUAbove(1,\ClipTop)(b4)
    \draw (c) node[above left] {$A$};
    \draw (b4) node[above right=.1em] {$B$};
  \end{scope}
\end{tikzpicture}}\SubFigure[example\label{fig:checking-position-of-mains:example}]{\includegraphics[width=.35\textwidth,height=\Hei]{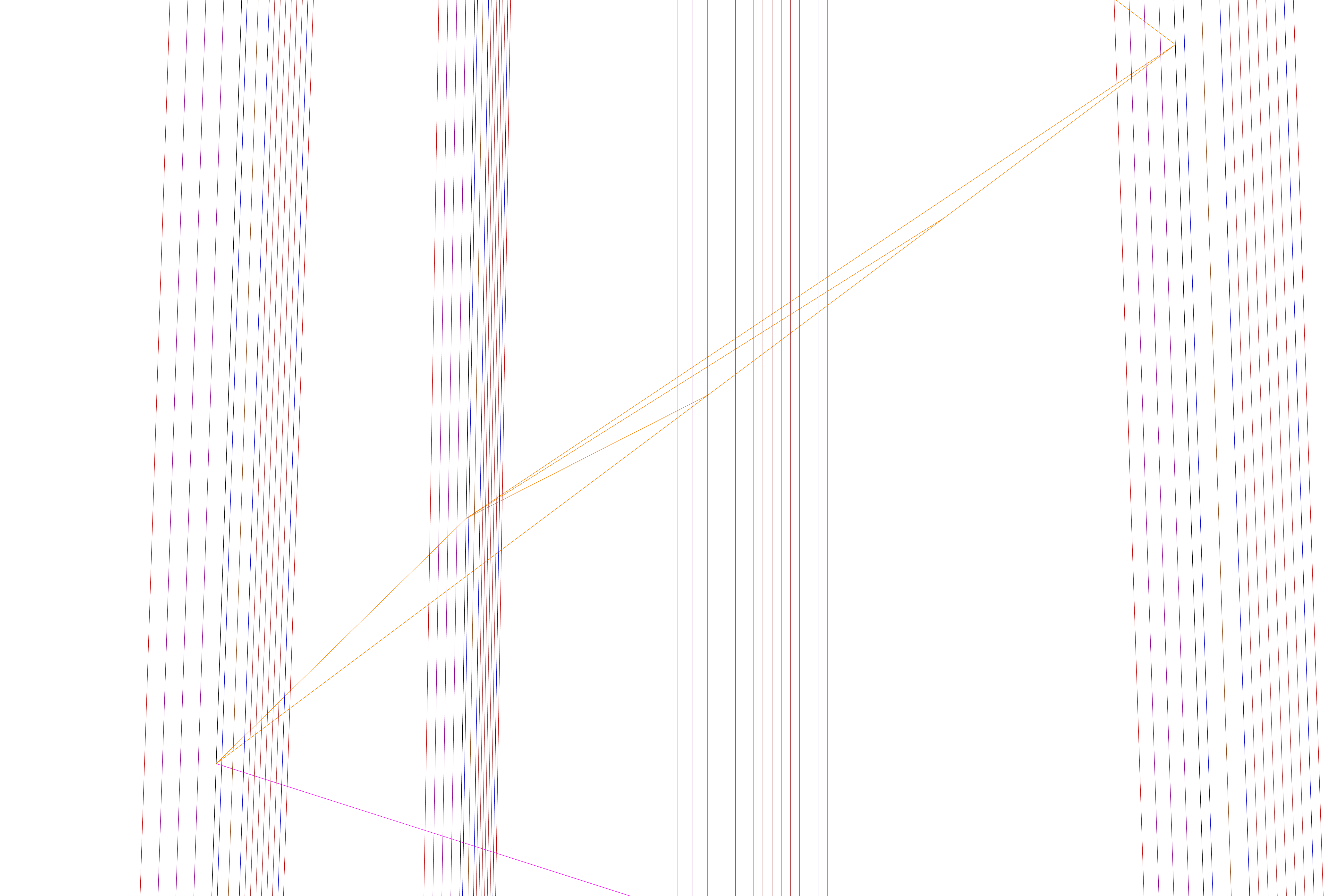}
  }
  \caption{Testing the other \SigMain signals.}
  \label{fig:checking-position-of-mains}
\end{figure}

To check this, the two first \SigMain (\SigMaini and \SigMaink) are used to organise meeting points with the all potential \SigMainl ($\SpeedIndexL\in\IntegerInterval{\SpeedIndexJ}{\SpeedIndexK-1}$).
If any appear anywhere except at their assigned meeting point, then it is known that it will not pass where \SigMaini and \SigMaink intersect (and the macro-collision aborts).
The meeting points are computed according to the speeds (like in \RefFig{fig:test-safety-identify}).
This constructions is presented on \RefFig{fig:checking-position-of-mains} with potential \SigMainl dashed.
The equation \RefEq{eq:intersection} is used again to compute the intersection points and to deduce the speeds.

Signal \SigCheckMaybeij and slower signal \SigCheckUpij go on the right.
If the first \SigMain \SigCheckMaybeij meets is \SigMainj, then there are only two macro-signals involved, it disappears and lets \SigCheckUpij starts the next stage.

Otherwise, \SigCheckMaybeij turns to \SigCheckij which crosses the configuration until it meets \SigMainj or a mismatch.
\SigCheckUpij cross \SigMaink at $A$ and branches into several \SigCheckInterceptikm.
Each time \SigCheckij meets some \SigMainl, then there should also be the corresponding \SigCheckInterceptikl. 
And the resolution is started.

If any \SigMainl is met with the wrong \SigCheckInterceptikm ($\SpeedIndexM\neq\SpeedIndexL$) or without any, \SigCheckij turns into \SigCheckFailij.
This latter cancels all remaining \SigCheckInterceptikl and disappears with \SigCheckInterceptikj.
The resolution is not started.

The used meta-signals and collision rules are defined in Figs.\,\ref{fig:ms:check} and \ref{fig:rule:check}. The speed of \SigCheckUpij has to be fast enough so that $A$ occurs before \SigCheckij intersects any \SigMaink.  The position of $A$ and of such intersections are again obtained using \RefEq{eq:intersection}, yielding the necessary speeds of the signals \SigCheckInterceptikj.

Once again, \AGCsimGroup simply ignores the new meta-signals and collisions, which are in bounded amount for each collision.

Altogether, the devices presented in this section yield the following lemma, which states that U correctly simulates one collision from clean inputs.

\begin{lemma}
  \label{lem:preparation}
  Let $\AGCrule$ be a collision rule of $\AGCmachine$, and let $\AGCconfigurationSMSymbol$ be a configuration of $\AGCmachine$ whose signals are exactly $\AGCruleIn$ and the positions of these signals are such that they all meet at some point $(\AGCspacialPosition, \AGCtemporalPosition)$. Let $\AGCspaceTimeDiagramSM$ be the associated space-time diagram.
  
  Let $\AGCconfigurationSMotherSymbol$ be a configuration such that $\AGCsimGroup(\AGCconfigurationSMotherSymbol) = \AGCconfigurationSMSymbol$, and which is clean at every position of a signal in $\AGCconfigurationSMSymbol$. Then there is a $\AGCsimStartingMSwidth$ and $t' < \AGCtemporalPosition$ such that $\AGCspaceTimeDiagramSMother(t')$ is a $\AGCsimStartingMSwidth$-checked configuration for \AGCconfigurationSMSymbol.
\end{lemma}

\begin{figure}[hbt]
  \centerline{\begin{ParameterList}
      \CoefCheck & \frac{101}{100} \\[.6em]
      $\forall 1\leq\SpeedIndexK\leq\SpeedIndexI$, $h_{\SpeedIndexI,\SpeedIndexK}$ & \frac{\SpeedI-\SpeedRapid}{\SpeedRapid-\BaseSpeed{\SpeedIndexK-1}} \CoefCheck\\[.6em]
      \multicolumn{2}{c}{$\SpeedCheckUp {=} \max\left(\frac{\SpeedRapid}{2},\max_{1\leq\SpeedIndexK\leq\SpeedIndexI}\left(
            \frac{h_{\SpeedIndexI,\SpeedIndexK}\SpeedK+ \SpeedI}{h_{\SpeedIndexI,\SpeedIndexK} + 1}
          \right)\right)$}
      \\[.6em]
      $\ForAllSpeedKltI,\,t_{i,k}^{\text{\sf chk-start}}$ & \frac{\SpeedI - \SpeedRapid}{\SpeedRapid-\SpeedK}  \\[.6em]
      $\ForAllSpeedKltI,\,t_{i,k}^{\text{\sf intersect}}$ & \frac{\SpeedI - \SpeedCheckUp}{\SpeedCheckUp-\SpeedK} \\[.6em]
    \end{ParameterList}
    \begin{MSlist}
      \MSForAllSpeedJltI[\SpeedRapid]{\SigCheckMaybeij}\MSForAllSpeedJltI[\SpeedRapid]{\SigCheckij}\MSForAllSpeedJltI[\SpeedCheckUp]{\SigCheckUpij}\MSForAllSpeedJltI[\frac{\SpeedL.t_{i,k}^{\text{\sf intersect}} - \SpeedK.t_{i,k}^{\text{\sf chk-start}}}{t_{i,k}^{\text{\sf intersect}} - t_{i,k}^{\text{\sf chk-start}}}]{\SigCheckInterceptikl}\MSForAllSpeedJltI[-\SpeedRapid]{\SigCheckFailij}\end{MSlist}
  }\caption{Meta-signals for checking.}
  \label{fig:ms:check}
\end{figure}

\begin{figure}[hbt]
  \centerline{\begin{CRlist}
      \CRForAllSpeedJltI{\SigCheckMaybeij, \SigMainj}{\SigMainj}\CRForAllSpeedJltI{\SigCheckUpij, \SigMainj}{\SigCheckOKij, \SigMainj}\CRForAllSpeedJltKltI{\SigCheckMaybeij, \SigMaink}{\SigMaink, \SigCheckij}\CRForAllSpeedJltKltI{\SigCheckUpij, \SigMaink}{\SigCheckInterceptikl\}$_{\SpeedIndexJ\leq\SpeedIndexL<\SpeedIndexK}\cup$ \{\SigMaink}\CRForAllSpeedJltLltKltI{\SigCheckInterceptikl, \SigCheckij}{\SigCheckij}\CRForAllSpeedJltLltKltI{\SigCheckInterceptikl, \SigCheckij, \SigMainl}{\SigMainl, \SigCheckij}\CRForAllSpeedJltKltI{\SigCheckInterceptikj, \SigCheckij, \SigMainj}{\SigCheckOKij, \SigMainj}\hline
      \CRForAllSpeedJltLltI{\SigCheckij, \SigMainl}{\SigCheckFailij, \SigMainl}\CRForAllSpeedJleLltKltIMneL{\SigCheckij, \SigMainl, \SigCheckInterceptikm}{\SigMainl, \SigCheckFailij}
      \CRForAllSpeedJltLltKltI{\SigCheckFailij, \SigCheckInterceptikl}{\SigCheckFailij}\CRForAllSpeedJltLltKltI{\SigCheckFailij, \SigCheckInterceptikj}{}\CRForAllSpeedJltKltI{\SigCheckij, \SigCheckInterceptikj}{\SigCheckFailij}\end{CRlist}}
  \caption{Collision rules for checking.}
  \label{fig:rule:check}
\end{figure}

\section{Simulation Examples}
\label{sec:example}

\begin{figure}[hbt]
  \centering\includegraphics[width=.85\textwidth]{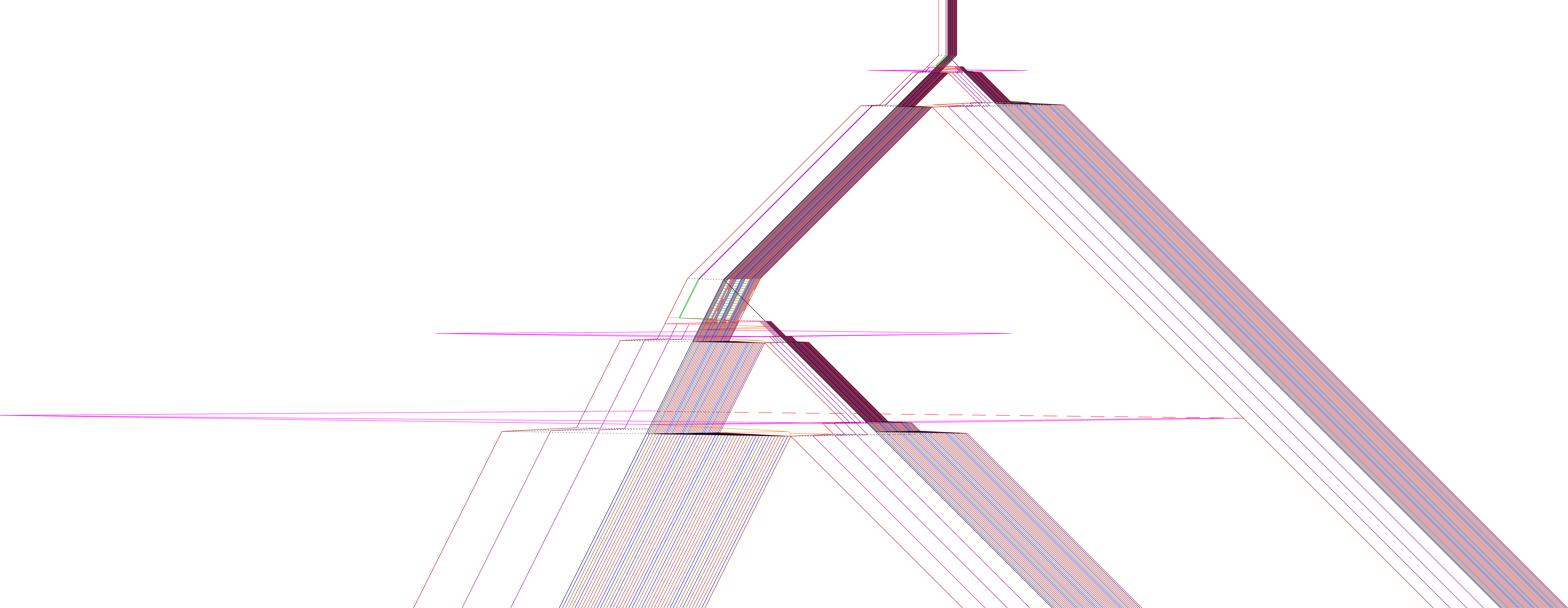}
  \caption{Space-time diagram of simulation of \RefFig{fig:schematic-simulation:ed}.}\label{fig:final-simulation}
\end{figure}

The presented construction works and has been implemented.
It has been entirely programmed in an ad hoc language for signal machines.
Given a signal machine, the library generate the corresponding \UniversalMSSpeed together with a function to translate initial configurations.
This has been used to generate all the pictures.
\RefFigure{fig:final-simulation} presents a simulation of the dynamics in \RefFig{fig:schematic-simulation:ed}.

\RefFigure{fig:annexe:example:5} provides different test and check failures before resolving the correct macro-collisions as well as a 3 macro-signal collision.

\PicExample{5}{.045}

\RefFigure{fig:annexe:example:2} represents a space-time diagram with finitely many collisions with no special regularity.

\PicExample{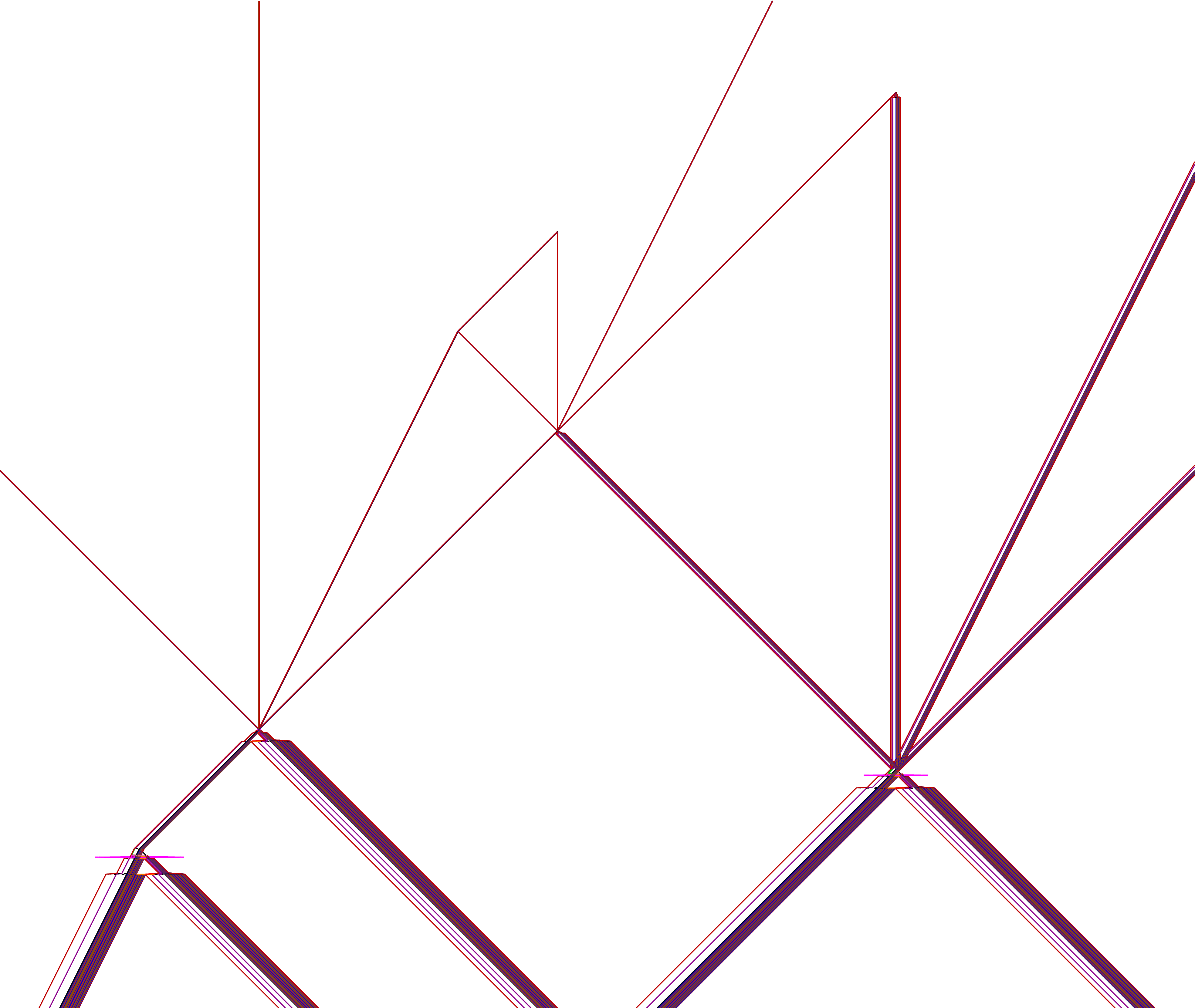}{.14}

\RefFigure{fig:annexe:example:7} represents a space-time diagram with a simple accumulation on top and its simulation.

\PicExample{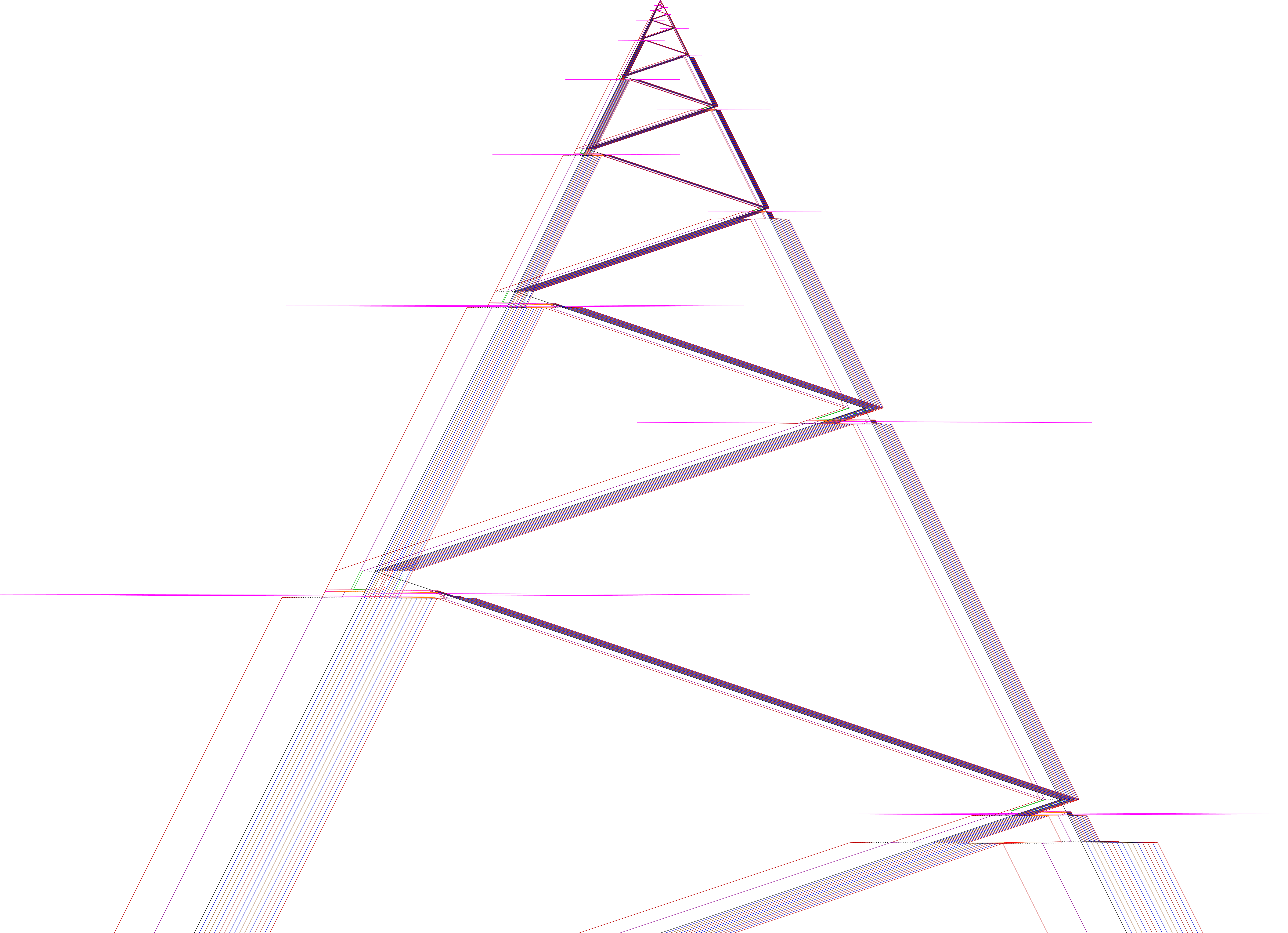}{.045}

\RefFigure{fig:annexe:example:3} is the basis for firing squad synchronisation on Cellular Automaton.
On signal machines, since space and time are continuous, it generates a fractal.
The simulation contains more than 100,000 signals.
It has been use as a test for robustness.

\PicExample{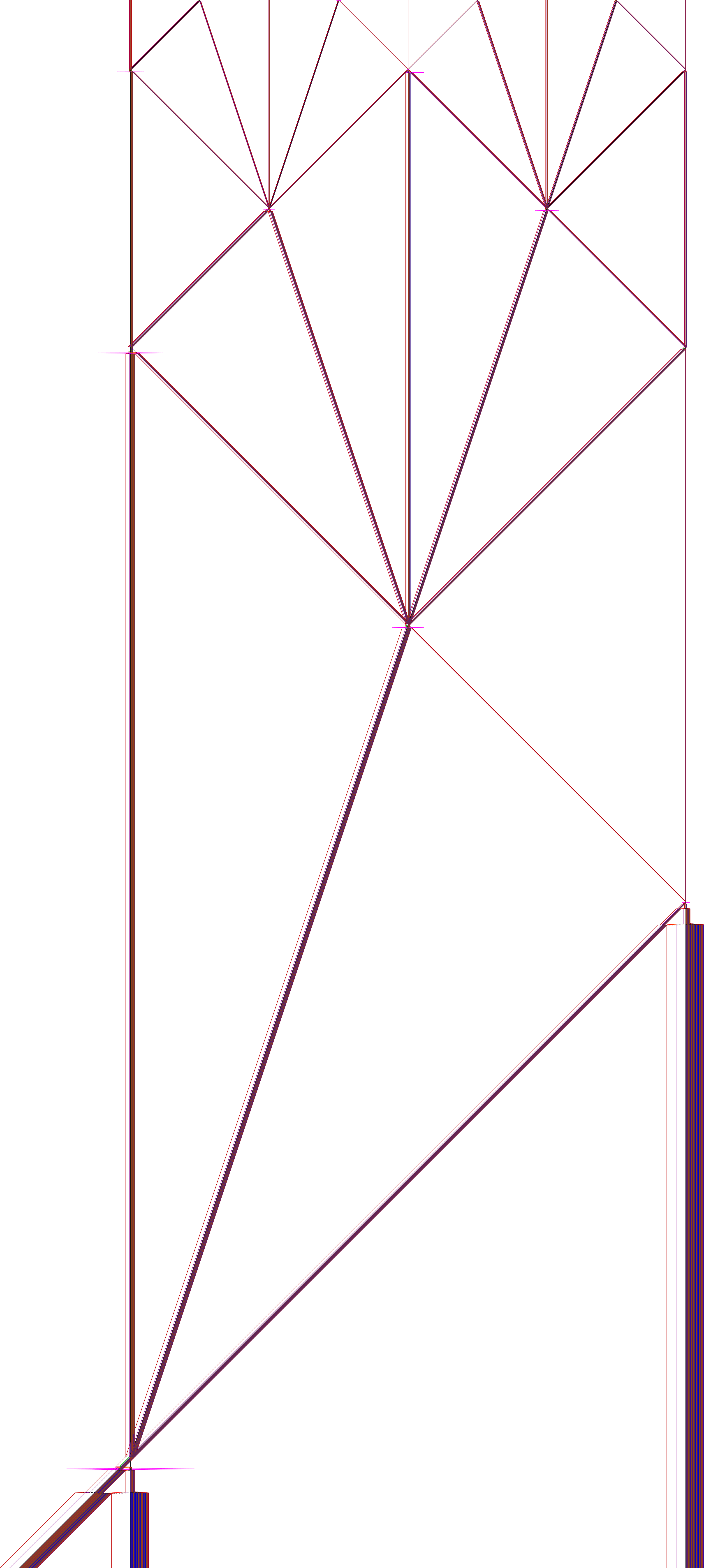}{.08}

\section{Conclusion}
\label{sec:conc}

Altogether, the construction proves the following result.

\begin{theorem}
  For any finite set of real numbers \SpeedSet, there is a \SpeedSet-universal signal machine.
  The set of \UniversalMSSpeed where \SpeedSet ranges over finite sets of real numbers is an intrinsically universal family of signal machines.
\end{theorem}

With the definition of simulation used, there does not exist any intrinsically universal signal machine since having a signal exactly located where the simulated signals implies that the speed of the simulated signal must be available, but every signal machine has finitely many speeds.
On the other hand, it might work with some other reasonable definition of simulation, maybe considering some kind of approximation.

Signal machine may produce accumulations (infinitely many collisions in a bounded part of the space-time).
The simulation works up to the first accumulation (excluded) where the configuration of the simulated machine cease to be defined in \RefFigure{fig:annexe:example:3}.

Using macro-signals forces to deal with width.
Hopefully, macro-signals can be made as thin as needed.
Nevertheless, as seen through the paper, it requires a lot of technicalities to deal with that.

In the construction, each macro-signals carries the list of rules associates with it, thus its dynamics, like a cell carries its DNA.
We wonder what might happen and what kind of artefact could be created if some way to dynamically modify the table were introduced.

\small

\providecommand{\href}[2]{#2}\makeatletter
  \@ifundefined{mathbb}{\long\def\mathbb{\mathsf}} \makeatother

\appendix

\clearpage

\NotArxiv{\section*{This part on is meant for reviewing purpose only}}
\ArxivOnly{\section*{Extra Material}}

\bigskip

An archive is available with examples, simulation source and java programs to run it at \\
\href{http://www.univ-orleans.fr/lifo/Members/Jerome.Durand-Lose/Recherche/AGC_Intrinsic_Univ_SM__FILES.tgz}{http://www.univ-orleans.fr/lifo/Members/Jerome.Durand-Lose/Recherche/AGC\_Intrinsic\_Univ\_SM\_\_FILES.tgz}

\bigskip

\begin{minipage}{1.0\linewidth}
  \setcounter{tocdepth}{2}

  \tableofcontents  
\end{minipage}

\bigskip

\RefFigure{fig:annexe:example:4} represents the same initial configuration as \RefFig{fig:annexe:example:2}, but the speed are $5$ times faster. This is used to check that large speeds are handled correctly.

\PicExample{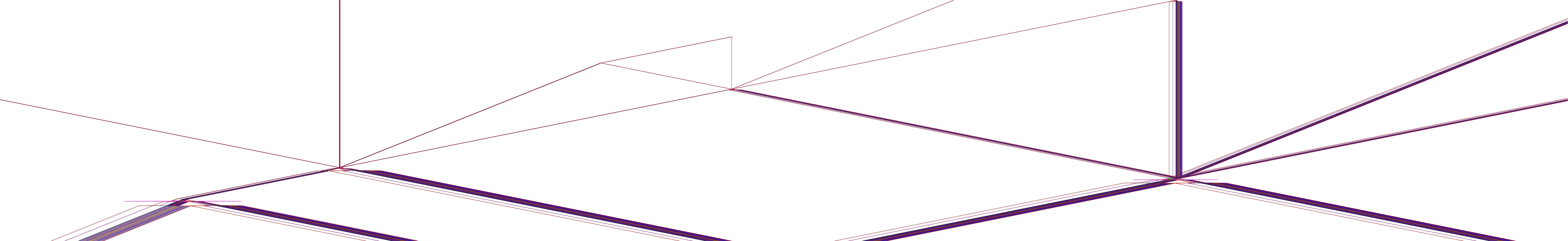}{.071}

\section{Table of Used Symbols}

\JDLvocabularyTableofsymbols

\end{document}